\documentclass[journal]{IEEEtran}
\usepackage{amsmath,amsfonts}
\usepackage{array}
\usepackage[linesnumbered,ruled,vlined]{algorithm2e}
\usepackage{textcomp}
\usepackage{stfloats}
\usepackage{url}
\usepackage{enumitem}
\usepackage{verbatim}
\usepackage{graphicx}
\usepackage{xcolor}
\usepackage{float}
\usepackage{hyperref}
\usepackage{subcaption}
\hypersetup{
    colorlinks=true,     
    linkcolor=blue,       
    citecolor=blue,     
    urlcolor=blue,       
    }
\usepackage{cite}

\makeatletter
\def\@cite#1#2{\textcolor{blue}{[{#1\if@tempswa , #2\fi}]}}
\makeatother

\begin{document}

\title{SpectrumFM: A Foundation Model for Intelligent Spectrum Management}

\author{Fuhui Zhou,~\IEEEmembership{Senior Member,~IEEE}, Chunyu Liu,~\IEEEmembership{Graduate Student Member,~IEEE},\\ Hao Zhang,~\IEEEmembership{Member,~IEEE,} Wei Wu,~\IEEEmembership{Member,~IEEE},  Qihui Wu,~\IEEEmembership{Fellow,~IEEE}, \\ Tony Q. S. Quek,~\IEEEmembership{Fellow,~IEEE}, and Chan-Byoung Chae,~\IEEEmembership{Fellow,~IEEE}
\thanks{This work was supported in part by the National Natural Science Foundation of China under Grant 62222107, 
in part by the National Key Research and Development Project under Grant 2023YFB2904500,
in part by the Yangtze River Delta Science and Technology Innovation Community Joint Research (Basic Research) Project under Grant 2024CSJZN00300,
in part by the Major Achievements Cultivation Project under Grant NC2025010, 
in part by the 407th Shuangqing Forum of the National Natural Science Foundation of China under Grant NQ2025005, and
in part by the Postgraduate Research \& Practice Innovation Program of Jiangsu Province under Grant SJCX25\_0151. 
The work of W. Wu was supported in part by the Open Fund of Anhui Province Key Laboratory of Cyberspace Security Situation Awareness and Evaluation under Grant CSSAE-2023-008. 
The work of T. Q. S. Quek was supported in part by the National Research Foundation, Singapore and Infocomm Media Development Authority under its Communications and Connectivity Bridging Funding Initiative. Any opinions, findings and conclusions or recommendations expressed in this material are those of the author(s) and do not reflect the views of National Research Foundation, Singapore.
The work of C.-B. Chae was supported in part by the Korea government under IITP/NRF Grants RS-2024-00428780 and 2022R1A5A1027646. (\textit{Corresponding author: Chunyu Liu}).}
\thanks{F. Zhou and H. Zhang are with the College of Artificial Intelligence and the Key Laboratory of Dynamic Cognitive System of Electromagnetic Spectrum Space, Nanjing University of Aeronautics and Astronautics, Nanjing 211106, China.
(E-mail: zhoufuhui@ieee.org, haozhangcn@nuaa.edu.cn)}
\thanks{C. Liu and Q. Wu are with the College of Electronic and Information Engineering, and with the Key Laboratory of Dynamic Cognitive System of Electromagnetic Spectrum Space, Nanjing University of Aeronautics and Astronautics, Nanjing 211106, China.
(E-mail: chunyu.liu@nuaa.edu.cn, wuqihui2014@sina.com)}
\thanks{W. Wu is with the College of Communication and Information Engineering, Nanjing University of Posts and Telecommunications, Nanjing 210003, China, and with the Anhui Province Key Laboratory of Cyberspace Security Situation Awareness and Evaluation, Hefei 230037, China.
(E-mail: weiwu@njupt.edu.cn)}
\thanks{T. Q. S. Quek is with the Singapore University of Technology and Design, Singapore 487372, and also with the Department of Electronic Engineering, Kyung Hee University, Yongin 17104, South Korea. (E-mail: tonyquek@sutd.edu.sg)}
\thanks{C.-B. Chae is with the School of Integrated Technology, Yonsei University, Seoul 03722, South Korea. (E-mail: cbchae@yonsei.ac.kr)}
}



\maketitle
\begin{abstract}
Intelligent spectrum management is crucial for improving spectrum efficiency and achieving secure utilization of spectrum resources. However, existing intelligent spectrum management methods, typically based on small-scale models, suffer from notable limitations in recognition accuracy, convergence speed, and generalization, particularly in the complex and dynamic spectrum environments. To address these challenges, this paper proposes a novel spectrum foundation model, termed SpectrumFM, establishing a new paradigm for spectrum management. SpectrumFM features an innovative encoder architecture that synergistically exploits the convolutional neural networks and the multi-head self-attention mechanisms to enhance feature extraction and enable robust representation learning. The model is pre-trained via two novel self-supervised learning tasks, namely masked reconstruction and next-slot signal prediction, which leverage large-scale in-phase and quadrature (IQ) data to achieve comprehensive and transferable spectrum representations. Furthermore, a parameter-efficient fine-tuning strategy is proposed to enable SpectrumFM to adapt to various downstream spectrum management tasks, including automatic modulation classification (AMC), wireless technology classification (WTC), spectrum sensing (SS), and anomaly detection (AD). Extensive experiments demonstrate that SpectrumFM achieves superior performance in terms of accuracy, robustness, adaptability, few-shot learning efficiency, and convergence speed, consistently outperforming conventional methods across multiple benchmarks. Specifically, SpectrumFM improves AMC accuracy by up to 12.1\% and WTC accuracy by 9.3\%, achieves an area under the curve (AUC) of 0.97 in SS at -4 dB signal-to-noise ratio (SNR), and enhances AD performance by over 10\%.
\footnote{The source code is available at \url{https://github.com/ChunyuLiu188/SpectrumFM.git}}
\end{abstract}


\section{Introduction}
\IEEEPARstart{T}{he} radio frequency (RF) spectrum is a fundamental yet increasingly scarce resource in modern wireless communication networks~\cite{meaning}. As the demand for ubiquitous connectivity continues to rise, driven by the proliferation of Internet-of-Things (IoT) devices, non-terrestrial networks (NTNs), and the sixth-generation (6G) wireless paradigms, spectrum management is of crucial importance for improving spectrum efficiency, mitigating interference, detecting the illegal spectrum utilization, achieving reliable and secure spectrum operation~\cite{10488747, 10685064}.

Spectrum management includes many tasks, such as spectrum sensing~\cite{10845212}, spectrum analysis~\cite{10602484}, spectrum resource decision-making~\cite{9479864, 10680609}, etc. Intelligent spectrum management aims to exploit machine learning to achieve those tasks. The current intelligent spectrum management methods are mainly oriented for the specific task and only adapt to one task. Although those methods have achieved considerable success, they often exhibit limited generalization to unseen environments and require substantial amounts of labeled data for training. Moreover, their performance deteriorates in low signal-to-noise ratio (SNR) settings or under non-stationary signal conditions, which are prevalent in practical wireless environments.

Inspired by the transformative success of foundation models in natural language processing (e.g., BERT, GPT) and computer vision (e.g., ViT, Segment Anything), there is growing interest in developing foundation models for wireless communications~\cite{10876763,10841938}. Those models are trained on large-scale heterogeneous data to learn universal representations that can be adapted to diverse downstream tasks with minimal supervision. Recent efforts, such as SpectralGPT~\cite{10490262} and WirelessGPT~\cite{yang2025wirelessgptgenerativepretrainedmultitask}, have demonstrated the potential of this paradigm in remote sensing and integrated sensing and communication (ISAC), respectively. However, the spectrum environment is inherently dynamic, with interference, signal fading, and noise introducing significant variability. This variability complicates the model ability to generalize across different environmental conditions and spectrum usage patterns. Moreover, the spectrum encompasses a broad range of frequency bands, from licensed to unlicensed, and the unpredictable nature of cognitive radio systems further amplifies the challenges. As a result, the application of foundation models to spectrum management remains largely unexplored.

To address this gap, we propose a novel spectrum foundation model, termed SpectrumFM, a foundation model tailored for generalizable spectrum management across a wide range of tasks and environments. At its core, SpectrumFM employs a hybrid encoder architecture that exploits convolutional neural networks (CNNs) for localized feature extraction and multi-head self-attention (MHSA) modules for capturing long-range dependencies,
enabling robust feature extraction and representation learning from large-scale in-phase and quadrature (IQ) data. Pre-training is conducted via two novel self-supervised learning tasks, namely masked reconstruction and next-slot signal prediction. The masked reconstruction task encourages the model to infer missing spectrum components, enhancing its ability to learn latent structures and intrinsic correlations while simultaneously improving its robustness to signal impairments such as noise and interference. 

In contrast, the next-slot signal prediction task focuses on temporal continuity by training the model to anticipate future spectrum behavior, thereby improving its predictive capabilities and adaptability to the dynamic nature of the spectrum environment. Furthermore, efficient parameter fine-tuning allows SpectrumFM to generalize effectively across a wide range of downstream tasks, including automatic modulation classification (AMC), wireless technology classification (WTC), spectrum sensing (SS), and anomaly detection (AD). 
The main contributions of our work are summarized as follows.
\begin{itemize}[left=0pt, labelsep=0.5em, itemsep=2pt]
\item {To the best of our knowledge, SpectrumFM is the first foundation model specifically designed for spectrum management, representing a promising new paradigm in this domain. In contrast to conventional methods that rely on hand-crafted features or task-specific architectures, SpectrumFM establishes an adaptable and highly generalizable framework capable of effectively addressing a wide range of spectrum-related tasks, including AMC, WTC, SS, and AD. Furthermore, it demonstrates strong robustness and adaptability across diverse environmental conditions and data distributions, effectively operating in real-world scenarios characterized by complex and dynamic spectrum data.}
\item {SpectrumFM introduces a novel architectural and learning framework that significantly advances spectrum management by seamlessly exploiting CNNs and MHSA mechanisms to achieve an enhanced feature extraction from IQ signals. The CNN component excels at identifying fine-grained, localized spectral patterns, while the MHSA module models long-range dependencies, enabling the model to capture comprehensive, high-level representations of spectrum signals.
To facilitate efficient pre-training without requiring extensive labeled data, we introduce two innovative self-supervised tasks, namely masked reconstruction and next-slot signal prediction. The masked reconstruction task enables the model to learn the underlying structure of the spectrum signals by predicting missing parts of the input, while the next-slot signal prediction task enhances the model temporal awareness by training it to predict future spectrum values based on historical data. As a result, these two tasks allow SpectrumFM to develop rich and transferable representations that can generalize well across complex and dynamic spectrum environments.
}
\item {We conduct comprehensive experiments across four key downstream tasks, including AMC, WTC, SS, and AD, to rigorously evaluate the performance of our proposed model in comparison with state-of-the-art methods. The results highlight its effectiveness in diverse spectrum scenarios. Specifically,
in the AMC task, SpectrumFM achieves notable accuracy improvements over existing methods, demonstrating enhanced robustness across datasets with varying complexities. In particular, it outperforms previous methods by 7.5\% on the RML2016.10A dataset, 1.9\% on the RML2016.10B dataset, and 12.1\% on the RML2016.04C dataset.
In the WTC task, our model exhibits a substantial classification accuracy gain, surpassing the best-performing baseline by 9.3\%, highlighting its strong generalization ability in wireless technology identification.
In the SS task, SpectrumFM achieves an area under the curve (AUC) of 0.97 even at a challenging SNR level of -4 dB, demonstrating exceptional robustness in low SNR environments.
In the AD task, our model significantly enhances anomaly detection performance, achieving an AUC improvement of over 10\%, underscoring its ability to identify complex spectrum anomalies with high precision.
Overall, these results strongly validate the effectiveness, adaptability, and superior generalization ability of SpectrumFM across a wide range of spectrum management tasks.
}
\end{itemize}
The remainder of this paper is organized as follows. In Section~\ref{sec:relatedwork}, we provide a comprehensive review of related works, focusing on task-specific spectrum management methods and recent advanced foundation models. Section~\ref{sec:model} presents an in-depth description of our proposed SpectrumFM, detailing its architecture, design principles, and two novel pre-training tasks, namely mask reconstruction and next-slot signal prediction. Section~\ref{sec:experiments} outlines the experimental results obtained from evaluating our model on various downstream tasks. Finally, we conclude the paper in Section~\ref{sec:conclusion}, summarizing the key findings and discussing potential directions for future research.

\section{Related Work}\label{sec:relatedwork}
\subsection{Task-Specific Spectrum Management Methods} Spectrum management has traditionally been approached through task-specific methods, where models are designed and optimized for individual tasks such as AMC, WTC, SS, and AD.  AMC, in particular, is a challenging task that requires accurate classification of the modulation scheme from a given IQ signal. Traditional methods have primarily relied on hand-crafted features, such as cyclic-moment-based features extracted through expert domain knowledge \cite{10667001}. A significant milestone was achieved in \cite{o2016convolutional}, where CNNs were introduced to learn features automatically, eliminating the dependence on expert-designed representations. This marked the advent of deep learning techniques in AMC.

While CNNs demonstrated strong performance in extracting spatial features from IQ signals, they are inherently limited in modeling temporal dependencies. To address this, GrrNet \cite{9079888} first employed a CNN-based module to extract representative features, followed by a gated recurrent unit (GRU) to capture temporal information. This hybrid design allowed GrrNet to balance spatial and temporal modeling, achieving notable improvements in AMC performance. However, GRUs still struggle with long-range dependencies and high-dimensional input data. To overcome these challenges, the authors in~\cite{9785878} proposed an attention-based AMC model. After initial feature extraction via a complex-valued CNN (CCNN), a modified multi-head attention mechanism was introduced, yielding a 1-10\% accuracy improvement over GRU-based models.

Deep learning-based AMC methods often suffer from reliance on large amounts of labeled data. To mitigate this, SSwsrNet~\cite{10496203} introduced a semi-supervised learning framework based on MixMatch, effectively leveraging both labeled and unlabeled data. This method significantly enhanced performance in few-shot learning scenarios. In low SNR environments, AMC accuracy typically degrades due to noise interference. AMC\_Net~\cite{10097070} addressed this issue by applying signal denoising in the frequency domain, combined with multi-scale feature extraction. This approach significantly improved classification accuracy under low SNR conditions. However, denoising-based methods often rely on paired training data (i.e., clean and noisy signal pairs). To relax this constraint,~\cite{10857965} proposed a contrastive learning objective that eliminates the need for paired inputs. This method enhances generalization by enabling the model to learn from a broader distribution of signal variations across diverse noise conditions.

WTC is another challenging task that requires the model to accurately classify the type of wireless transmission based on the input IQ signal. Numerous studies have demonstrated significant progress in this area. For example, the authors in \cite{FONTAINE2019101881} employed CNNs to automatically extract features from the input signal and used a feedforward neural network to classify the wireless technology. Data augmentation techniques were also incorporated, enabling the model to generalize well to unseen scenarios while maintaining high accuracy. In \cite{9637487}, a multiscale module was introduced to capture features at various resolutions. Additionally, a global average pooling layer was utilized to aggregate the extracted features, which reduced the number of parameters and accelerated the training process. Despite its simplified architecture, the model achieved strong classification performance and fast convergence. 

The authors in \cite{10.1016/j.vehcom.2022.100563} proposed a comprehensive WTC system designed to identify and analyze a wide range of wireless technologies coexisting within the 5.9 GHz band. The system specifically focused on LTE, Wi-Fi, 5G NR, C-V2X PC5, and ITS-G5. To ensure accurate identification, the study adopted a short time resolution window based on the minimum frame duration of the target technologies. Unlike most IQ-based approaches, MFBLN \cite{10555427} leveraged alternative features such as received signal strength indicator (RSSI), fast fourier transform (FFT), and spectrogram representations to extract multi-dimensional signal characteristics. This allowed the model to capture richer information from the input, thereby enhancing classification accuracy and robustness across diverse wireless environments.

SS is a fundamental task in cognitive radio networks, referring to the process of monitoring and analyzing the RF spectrum to determine its occupancy status \cite{10598350}. The primary goal of SS is to identify spectrum holes, which are unused or underutilized frequency bands in a specific geographical area. By detecting these holes, cognitive radios can dynamically access the available spectrum without interfering with primary users (PUs). SS is generally categorized into two types, namely narrowband SS and wideband SS. Narrowband SS analyzes one frequency band at a time, while wideband SS simultaneously evaluates multiple bands to identify vacant channels suitable for transmission~\cite{muzaffar2024review}. Common narrowband SS techniques include energy detection, cyclostationary detection, and matched filter detection~\cite{5723805}. Among these, energy detection is widely used due to its simplicity. It determines spectrum occupancy by comparing the measured signal energy against a predefined threshold.

However, energy detection suffers from the well-known SNR wall issue caused by noise uncertainty, which significantly degrades its performance in low SNR environments. To address this limitation, the authors in~\cite{8824091} proposed a deep learning-based signal detector. Unlike conventional energy detection, the proposed method does not require prior knowledge of channel state information or background noise, resulting in greater robustness and adaptability under varying conditions. For wideband SS, the authors in~\cite{10509639} introduced a Transformer-based architecture that effectively learns both intra-band spectral features and inter-band occupancy correlations in the wideband regime. To further address the challenges of wideband SS under few-shot or cross-domain few-shot scenarios, the authors in~\cite{10758375} proposed a pre-training and fine-tuning approach leveraging transfer learning. This method improves performance when labeled data is scarce, making it well-suited for real-world applications where large annotated datasets are unavailable.

AD plays a critical role in enhancing spectrum sharing security, particularly in cognitive radio networks (CRNs) \cite{10589480}. After spectrum sensing determines that the PU band is occupied, AD is employed to further assess whether the detected signal exhibits normal behavior or anomalous characteristics. The authors in \cite{9863875} proposed an improved deep support vector data description (Deep SVDD) method for extracting low-dimensional features of the samples represented
in the time-frequency domain. The method achieved excellent detection performance while maintaining real-time capability. To enable effective monitoring over a wide spectrum, a spectrum abnormal detection scheme was introduced in \cite{10035489}, which employed an adversarial autoencoder (AAE) to recover signals from sub-Nyquist samples. This design allowed for efficient processing of undersampled data while preserving the ability to detect spectral anomalies.

To address the challenge of implementing spectrum sensing and abnormal detection as separate modules, which increases system complexity and compromises real-time performance, the authors in \cite{10765508} proposed a unified deep learning framework that performs spectrum sensing and abnormal detection simultaneously. This integrated approach delivers high detection accuracy while ensuring low-latency operation, offering a more efficient and practical solution for cognitive radio systems.



\subsection{Foundation Models}
Foundation models are large-scale machine learning models that are pre-trained on vast amounts of data to learn general-purpose representations. The models serve as a foundational basis for a wide range of downstream tasks, making them highly versatile and adaptable across various domains. Foundation models are typically developed using deep learning techniques and are characterized by their massive scale, extensive training data, and strong transferability \cite{10834497, 10173618}. The authors in \cite{vaswani2017attention} introduced the attention mechanism and the Transformer architecture, which serve as the cornerstone of foundation models. BERT \cite{devlin-etal-2019-bert} pioneered the era of foundation models. By introducing a bidirectional Transformer-based architecture and innovative pre-training strategies such as masked language modeling (MLM) and next sentence prediction (NSP), BERT achieved state-of-the-art performance on a wide range of nature language processing (NLP) tasks. 

The BERT-like prediction loss is employed in HuBERT \cite{10.1109/TASLP.2021.3122291} for self-supervised speech representation learning, leading to the development of a speech foundation model. The authors in \cite{10.5555/3600270.3602542} present a foundational model that reformulates and unifies four core vision tasks, namely object detection, instance segmentation, keypoint prediction, and image captioning, into a single pixel-to-sequence framework. By framing multiple vision tasks within a shared pixel-to-sequence paradigm, this work demonstrated the potential of foundation models to generalize across tasks, reducing the need for specialized architectures and promoting greater flexibility in multimodal applications.
SpectralGPT \cite{10490262} represents a pioneering effort in creating a universal foundation model tailored for remote sensing (RS) applications. Trained on a large-scale dataset comprising one million spectral RS images, SpectralGPT achieved a parameter size exceeding 600 million. Evaluations across four downstream tasks reveal substantial performance improvements, highlighting the potential of foundation models to tackle a wide range of RS tasks.

WirelessGPT \cite{yang2025wirelessgptgenerativepretrainedmultitask}, a pioneering foundation model tailored for multi-task learning in ISAC, features an initial parameter size of approximately 80 million. Pre-trained on a large-scale wireless channel dataset, it delivers significant advancements in critical wireless tasks such as channel estimation, channel prediction, human activity recognition, and wireless reconstruction. WirelessGPT demonstrates that foundation models can achieve impressive performance improvements over conventional methods and smaller artificial intelligence (AI) models while requiring fewer labeled data, thereby enhancing efficiency and scalability in ISAC applications.

\section{Our Proposed Model}\label{sec:model}
\begin{figure*}[htbp]
    \centering
    \includegraphics[width=0.98\textwidth]{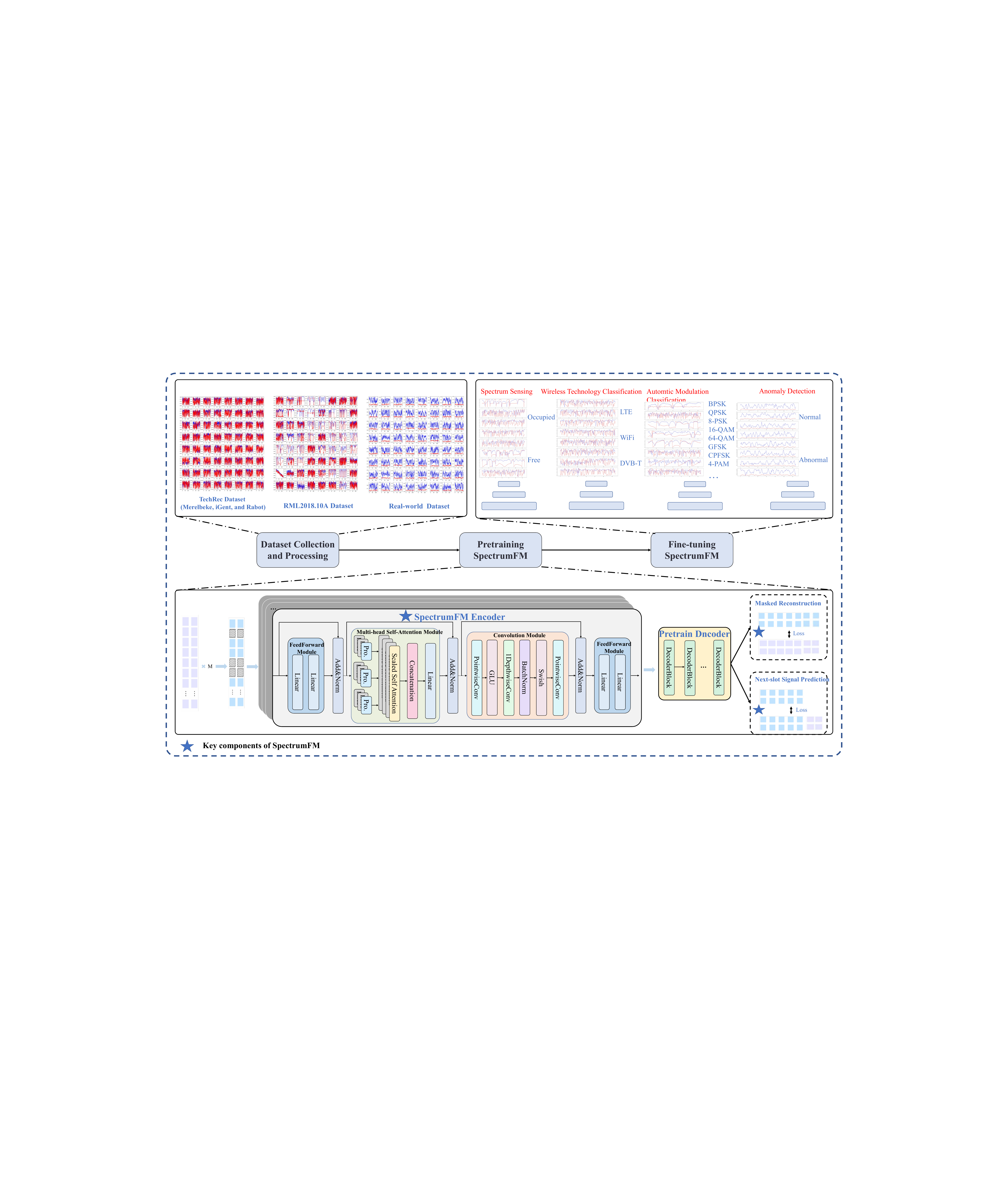} 
    \caption{The SpectrumFM framework comprises three key stages. First, in the data collection and processing stage, diverse spectrum data from multiple sources are gathered and preprocessed to ensure consistency and compatibility across datasets. Second, during the pre-training stage, the model learns fundamental spectrum representations through self-supervised learning tasks, namely, masked reconstruction and next-slot signal prediction. Finally, in the fine-tuning stage, the pre-trained model is adapted to specific downstream tasks, including AMC, WTC, SS, and AD.}
    \label{framework}
\end{figure*}

In this section, we present our proposed foundational model for spectrum management. The overall framework of SpectrumFM is illustrated in Fig. \ref{framework}, comprising three main stages, namely, data collection and processing, pre-training, and fine-tuning. In the data collection and processing stage, we aggregate a comprehensive dataset from diverse sources, including publicly available open-source datasets and signals collected in real-world practice scenarios. A crucial aspect of this stage is data preprocessing, where normalization techniques are exploited to ensure seamless integration of heterogeneous data into a unified training framework.
Furthermore, during the pre-training stage, SpectrumFM undergoes self-supervised learning on the integrated dataset exploiting two designed tasks, namely masked reconstruction and next-slot signal prediction. The two tasks enable the model to learn fundamental representations of spectrum data, capturing both generalizable patterns and environment-specific variations.
Finally, in the fine-tuning stage, the pre-trained model is adapted to specific downstream tasks, including AMC, WTC, SS, and AD. The step ensures that SpectrumFM effectively captures task-specific spectrum features, enhancing its adaptability across diverse spectrum management applications. The variables used throughout this paper are summarized in TABLE \ref{tab:notation}.
\begin{table}[t]
    \centering
    \caption{Key Notations Used in the Paper.}
    \begin{tabular}{c|c}
    \hline
    \textbf{Notation} & \textbf{Description}  \\
    \hline
    $\mathbf{x}_{\text{ap}}$ & The AP signal.\\
    \hline
    $\mathbf{x}^{\text{norm}}_{\text{ap}}$ & The normalized AP signal.\\ \hline
    $F$ & The channel fading coefficient.\\\hline
    $\text{abs}(x)$ & The amplitude function.\\\hline
    $\text{angle}(x)$ & The phase function.\\\hline
    $\Re(x)$ & The real part of the signal.\\\hline
    $\Im(x)$ & The imaginary part of the signal.\\\hline
    $\mathbf{A}$ & The amplitude component. \\\hline
    $\boldsymbol\varTheta$ & The phase component.\\\hline
    $RN$ & The residual connection and normalization operation.\\\hline
    $\text{concat}(\cdot)$ & The concatenation operation.\\ \hline
    $p$ & The position index.\\ \hline
    $d$ & The latent dimension.\\ \hline
    $d_\text{{feed}}$ & The feedforward dimension.\\ \hline
    $i$ & The index of the dimension.\\ \hline
    $N$ & The number of signal symbols.\\ \hline
    $k$ & The kernel size of the convolution layer.\\ \hline
    $H$ & The number of attention heads.\\ \hline
    $h$ & The attention head index.\\ \hline
    $g$ & The number of groups in the convolution layer.\\ \hline
    $L$ & The number of SpectrumFM encoder layers. \\ \hline
    $\ell$ & The SpectrumFM encoder layer index.\\ \hline
    $r$ & The mask ratio of signal symbols.\\ \hline
    $\mathbf{m}$ & The binary mask vector.\\ \hline
    $\mathbf{M}$ & The mask of attention scores.\\ \hline
    $\mathbf{\Theta}$ & All the parameters of the SpectrumFM.\\ \hline
    $E_p$ & The number of epochs for pre-training.\\ \hline
    $E_f$ & The number of epochs for fine-tuning.\\ \hline
    $\eta$ & The learning rate.\\ \hline
    \end{tabular}
    \label{tab:notation}
\end{table}
\subsection{Data Collection And Processing Stage}
\label{sec:data_collection_and_processing_stage}
In the data collection and processing stage, a diverse set of spectrum data is gathered from multiple sources, including publicly available open-source datasets and signals collected in real-world for this study, to facilitate comprehensive training of SpectrumFM. The open-source datasets utilized in this study include the RML2018.01A dataset \cite{8267032} and the technology recognition (TechRec) dataset\footnote{\url{https://ieee-dataport.org/documents/iq-signals-captured-lte-wifi-and-dvb-t}}.  
The RML2018.01A dataset comprises 24 distinct modulation types, including OOK, 4-ASK, 8-ASK, BPSK, QPSK, 8-PSK, 16-PSK, 32-PSK, 16-APSK, 32-APSK, 64-APSK, 128-APSK, 16-QAM, 32-QAM, 64-QAM, 128-QAM, 256-QAM, AM-SSB-WC, AM-SSB-SC, AM-DSB-WC, AM-DSB-SC, FM, GMSK, and OQPSK. This dataset spans a broad range of SNRs, ranging from -20 dB to 30 dB in increments of 2 dB, providing a comprehensive representation of diverse signal conditions.  
This dataset mainly reflects typical wireless communication scenarios across different modulation schemes under various channel impairments, such as fading, frequency offsets, and noise. 
The TechRec dataset consists of over-the-air IQ samples collected from six distinct locations in Gent, Belgium, including UZ, Reep, Rabot, Merelbeke, iGent, and Gentbrugge. Each location exhibits unique signal characteristics due to variations in signal strength, noise levels, and physical propagation conditions. The SNR in the dataset ranges from -6 dB to 12 dB, with increments of 2 dB. Specifically, IQ samples collected from Merelbeke, iGent, and Rabot are leveraged for pre-training, while data from UZ, Reep, and Gentbrugge are exploited for fine-tuning and testing.  
This dataset reflects real-world urban communication scenarios, capturing the effects of multipath propagation, background interference, and heterogeneous spectrum occupancy across different city environments.
In addition to the publicly available datasets, signals are also collected by using a Ceyear 1435B-V RF signal generator for transmission and a SAM~60~MK2 receiver for reception.  
This self-collected dataset covers both indoor scenarios, such as office-like multipath-rich environments, and outdoor scenarios, such as open-space propagation conditions, thereby complementing the public datasets with more diverse real-world signal characteristics.
\begin{table}[htbp]
\centering
\caption{Overview of The Pretraining Datasets Used in This Study.}
\label{tab:pretrain_datasets}
\begin{tabular}{lccc}
\hline
Dataset & \# Samples & Sample Shape & Approx. Contribution  \\
\hline
RML2018.01A & 2,555,904 & 1024 $\times$ 2 & 52.1\%  \\
TechRec & 9,747,800 & 128 $\times$ 2 & 24.9\%  \\
Self-collected & $\sim$9,000,000 & 128 $\times$ 2 & 23.0\%  \\
\hline
\end{tabular}
\end{table}
The total size of the pre-training dataset amounts to approximately 25 GB. A detailed breakdown of the datasets is shown in TABLE~\ref{tab:pretrain_datasets}. The received signal $H$ is formulated as
\begin{equation}
    H: x\left(n\right)= f\left(s\left(n\right), F\right)+\omega \left(n\right), n=1,2,3,...N, 
    \label{eq:1}
\end{equation}
where $x\left(n\right)$ and $s\left(n\right)$ are the received signal and the transmitted signal, respectively, $F$
represents the channel fading coefficient, $N$ is the number of signal symbols, and $\omega \left(n\right)$ is the additive white Gaussian noise with zero mean and variance $\sigma^2$.
The IQ signal samples can be represented as a vector by transforming the received signal \( x\left(n\right) \) into the vector \( \mathbf{x} \), given as 
\begin{subequations}
    \begin{align}
        \mathbf{x} &= \mathbf{I} + \mathbf{Q},  \\
                   &= \Re\left(x\right) + j\Im \left(x\right),
    \end{align}
\end{subequations} 
where $\mathbf{I}$ and $\mathbf{Q}$ represent the in-phase and quadrature components of the received signal, respectively, and $j=\sqrt{-1}$, and $\Re\left(x\right)$ and $\Im \left(x\right)$ correspond to the real and imaginary parts of the signal, respectively.
The raw data \( \mathbf{x} \) can be explicitly represented in matrix form, given as 
\begin{equation}
    \mathbf{x} = \left[ \begin{array}{l}
        \Re\left[ {x\left( 1 \right),x\left( 2 \right),\cdots,x\left( N \right)} \right]\\
        \Im\left[ {x\left( 1 \right),x\left( 2 \right),\cdots,x\left( N \right)} \right]
    \end{array} \right]. 
\end{equation}
To facilitate normalization, we first convert the IQ signals into amplitude-phase (AP) signals, since the value ranges of amplitude and phase components are better suited for standardization, specifically 
$[-\pi,\pi]$ for the phase component and $[0, +\infty]$ for the amplitude component. Mathematically, this transformation for the IQ signals to AP signals can be expressed as
\begin{equation}
    \label{eq:iq_to_ap}
    \mathbf{x}_{\text{ap}}=\mathbf{A}\exp(j\boldsymbol \varTheta ),
\end{equation}
where $\mathbf{A}$ is the amplitude component and $\boldsymbol\varTheta$ is the phase component, respectively. It can be further expressed as 
\begin{equation}
{\mathbf{x}_{\text{ap}}} = \left[ \begin{array}{l}
    \text{abs}\left[ {x\left( 1 \right),x\left( 2 \right),\cdots,x\left( N \right)} \right]\\
    \text{angle}\left[ {x\left( 1 \right),x\left( 2 \right),\cdots,x\left( N \right)} \right]
    \end{array} \right],
\end{equation}
where $\text{abs}\left[ \cdot \right]$ and $\text{angle}\left[ \cdot \right]$ are the amplitude and phase functions, respectively, given as 
\begin{subequations}
    \begin{align}
        \text{abs}(x)&=\sqrt{\Re^{2}(x)+\Im^{2}(x)},\\
        \text{angle}(x)&=\arctan\left( \frac{\Im(x)}{\Re(x)} \right).
    \end{align}
\end{subequations}
To ensure that the distribution of each sample in different dataset is consistent while preserving the intrinsic characteristics of the data, we propose a sample-level normalization method, where each sample is normalized independently based on its own maximum and minimum values by using min-max normalization, given as 
\begin{equation}
    \mathbf{x}^{\text{norm}}_{\text{ap}}=\frac{x_{\text{ap}}-\min(x_{\text{ap}})}{\max(x_{\text{ap}})-\min(x_{\text{ap}})},
    \label{eq:7}
\end{equation}
where $\min(\cdot)$ and $\max(\cdot)$ are the functions to obtain the minimum and maximum values of the sample $x_{\text{ap}}$, respectively.
\subsection{SpectrumFM Encoder}
After obtaining the processed signal $\mathbf{x}^{\text{norm}}_{\text{ap}}$, a 1D convolution projection layer with a kernel size $k = 1$ is leveraged to project the input signal into a higher-dimensional space, given as 
\begin{equation}
    \mathbf{x}_{\text{proj}}[p, :] = \sum_{c=0}^{1} \mathbf{x}^{\text{norm}}_{\text{ap}}[p, c] \cdot \mathbf{W}_{\text{proj}}[0, c, :],
    \label{eq:projection}
\end{equation}
where $\mathbf{x}_{\text{proj}} \in \mathbb{R}^{N\times d}$ is the projected signal, $\mathbf{W}_{\text{proj}} \in \mathbb{R}^{d \times 2 \times 1} $ is the convolution kernel, $d$ is the latent dimension, and $p$ represents the position index, ranging from 0 to $N-1$.
To capture the sequential order in a signal sequence, positional encoding is added to $\mathbf{x}_{\text{proj}}$. The positional encoding is given as 
\begin{subequations}
    \begin{align}
        \mathrm{PE}_{p, 2i} &= \sin\left(\frac{p}{10000^{2i/d}}\right),\\
        \mathrm{PE}_{p, 2i+1} &= \cos\left(\frac{p}{10000^{2i/d}}\right),
    \end{align}
\end{subequations}
where $i$ is the index of the dimension, ranging from 0 to $d-1$.
The signal with the positional encoding is given as 
\begin{equation}
    \mathbf{x}_{\text{position}}[p, :] = \mathbf{x}_{\text{proj}}[p, :] + \text{PE}[p, :].
\end{equation}

Subsequently, the position-encoded signal $\mathbf{x}_{\text{position}}$ is fed into the proposed SpectrumFM encoder to facilitate feature extraction. The SpectrumFM encoder follows a structured pipeline comprising an initial feedforward module, followed by a multi-head self-attention module, then a convolution module, and finally another feedforward module. Each module is connected with residual connections followed by normalization to stabilize the training process.
This design is particularly well-suited for wireless communication applications, where spectrum management requires efficient feature extraction, robust signal representation, and adaptive processing. The feedforward module captures low-level signal characteristics, while multi-head self-attention models spectral dependencies, enhancing feature discernment in dynamic environments. Convolutions refine local spectral details, improving resilience to distortions, and the final feedforward module consolidates learned representations. Residual connections and normalization stabilize training, ensuring effective feature propagation.

Initially, the input sequence undergoes preliminary processing through an initial feedforward module, which enriches the feature representation by applying non-linear transformations. Specifically, the feedforward module consists of two linear transformations separated by a non-linear activation function. The output of the module can be expressed as
\begin{equation}
\mathbf{x}_{\text{ffn}} = RN \left( GELU(\mathbf{x}_{\text{position}} \mathbf{W}_1 + \mathbf{b}_1) \mathbf{W}_2 + \mathbf{b}_2 \right),
    \label{eq:ffn}
\end{equation}
where $\mathbf{W}_1 \in \mathbb{R}^{d \times d_{\text{feed}}}$ and $\mathbf{W}_2 \in \mathbb{R}^{d_{\text{feed}} \times d}$ are the weight matrices, $\mathbf{b}_1 \in \mathbb{R}^{d_{\text{feed}}}$ and $\mathbf{b}_2 \in \mathbb{R}^{d}$ are the bias vectors, $d_{\mathrm{feed}}$ is the feedforward dimension, $GELU$ is the Gaussian error linear unit (GELU) activation function and $RN$ represents the residual connection and normalization operation.

The enriched representation $\mathbf{x}_{\text{ffn}}$ is then subjected to global dependency analysis via the multi-head self-attention mechanism, thereby enabling the model to capture long-range dependencies across the entire sequence. For each attention head $h$, the input $\mathbf{x}_{\text{ffn}} \in \mathbb{R}^{N \times d}$ is transformed into query ($\mathbf{Q}$), key ($\mathbf{K}$), and value ($\mathbf{V}$) matrices by using learnable weight matrices, given as
\begin{subequations}
    \begin{align}
        \mathbf{Q}_h &= \mathbf{x}_{\text{ffn}} \mathbf{W}_Q^h, \\
        \mathbf{K}_h &= \mathbf{x}_{\text{ffn}} \mathbf{W}_K^h, \\
        \mathbf{V}_h &= \mathbf{x}_{\text{ffn}} \mathbf{W}_V^h,
    \end{align}
\end{subequations}
where $\mathbf{W}_Q^h \in \mathbb{R}^{d \times d_{h}}$, $\mathbf{W}_K^h \in \mathbb{R}^{d \times d_{h}}$, and  $\mathbf{W}_V^h \in \mathbb{R}^{d \times d_{h}}$ are the learnable weight matrices, $d_{h} = \frac{d}{H}$ denoting dimension of each head, and $H$ is the number of heads.
The output of each head $h$ is computed by using a scaled dot-product attention, given as 
\begin{equation}
    \mathbf{x}_{h} = \text{softmax}\left(\frac{\mathbf{Q}_h \mathbf{K}_h^T}{\sqrt{d_h}}\right) \mathbf{V}_h,   
    \label{eq:attention}
\end{equation}
where $\text{softmax}(\cdot)$ applies the softmax function row-wise.
The outputs from all heads are concatenated and linearly transformed to produce the final output of the multi-head self-attention module, given as
\begin{equation}
    \mathbf{x}_{\text{attention}} = RN\left(\text{concat}(\mathbf{x}_{1}, \cdots, \mathbf{x}_{H}) \mathbf{W}^o\right),
\end{equation}
where $\text{concat}$$(\cdot)$ is the concatenation operator, and $\mathbf{W}^o \in \mathbb{R}^{d \times d}$ is the weight matrix.

Following the multi-head self attention module, the convolution module identifies and extracts local dependencies within localized regions of the sequence, enhancing the model ability to detect fine-grained patterns. Specifically, a 1D convolution layer with a kernel size $k = 1$ and a gated linear unit (GLU) activation function is applied to the output of the multi-head self-attention module, given as 
\begin{equation}
    \mathbf{x}_{\text{inconv}}[p, :] = GLU\left(\sum_{c=0}^{1} \mathbf{x}_{\text{attention}}[p, c] \cdot \mathbf{W}_{k1}[0, c, :]\right),
\end{equation}
where $\mathbf{W}_{k1} \in \mathbb{R}^{2d \times d \times 1} $ is the convolution kernel.
And then, a 1D convolution layer with a kernel size $k = 3$, groups $g = d$ and a swish activation function is leveraged to extract local dependencies, given as
\begin{equation}
    \begin{split}
    \mathbf{x}_{\text{\text{localconv}}}[p, :] = \text{swish}\bigg( 
    & \sum_{c=0}^{d-1} \bigg( 
    \mathbf{x}_{\text{inconv}}\underbrace{[p-1:p+1,}_{\text{kernel size}=3}\\ c]
    & \ast \mathbf{W}_{k2}[c, :] \bigg) + \mathbf{b}_{k2} \bigg),
    \end{split}
\end{equation}
where \text{swish}($\cdot$) represents the swish activation function, $\mathbf{W}_{k2} \in \mathbb{R}^{d \times 1 \times 3} $ is the convolution kernel, and $\mathbf{b}_{k2} \in \mathbb{R}^{d}$ is the bias.
Another 1D convolution layer with kernel size $k = 1$ is leveraged to further refine the extracted local features, given as
\begin{equation}
    \mathbf{x}_{\text{outconv}}[p, :] = RN\left(\sum_{c=0}^{1} \mathbf{x}_{\text{localconv}}\left[p, c\right]  \cdot \mathbf{W}_{k3}[0, c, :]\right),
\end{equation}
where $\mathbf{W}_{k3} \in \mathbb{R}^{d \times d \times 1} $ is the convolution kernel.

Finally, a subsequent feedforward module integrates the insights obtained from both the attention and convolution modules, given as
\begin{equation}
    \mathbf{x}_{\text{latent}} = (GELU(\mathbf{x}_{\text{outconv}}\mathbf{W}_3 + \mathbf{b}_3))\mathbf{W}_4 + \mathbf{b}_4,
\end{equation} 
where $\mathbf{W}_3 \in \mathbb{R}^{d \times d_\text{feed}}$ and $\mathbf{W}_4 \in \mathbb{R}^{d_\text{feed} \times d}$ are the weight matrices, $\mathbf{b}_3 \in \mathbb{R}^{d_\text{feed}}$ and $\mathbf{b}_4 \in \mathbb{R}^{d}$ are the bias vectors. 

To enhance the model representational capacity and capture more complex patterns within the input data, a multi-layer encoder architecture is exploited. Specifically, each layer of the encoder progressively transforms the latent representation from the previous layer until the $L$-th layer. The process can be expressed as
\begin{equation}
    \mathbf{x}_\text{latent}^{(\ell)} = \text{encoder}(\mathbf{x}_{\text{latent}}^{(\ell-1)}),
    \label{eq:encoder}
\end{equation} 
where $\text{encoder}(\cdot)$ is the SpectrumFM encoder, $\mathbf{x}_{\text{latent}}^{(\ell)}$ is the latent representation of the $\ell$-th layer, and $\mathbf{x}_{\text{latent}}^{(\ell-1)}$ is the latent representation of the previous layer. And the final output after 
$L$ layers of the encoder is denoted as $\mathbf{x}_{\text{latent}}^{(L)}$.
\subsection{Pre-training Stage}
To pre-train SpectrumFM without labeled data, two self-supervised pre-training tasks, namely mask reconstruction and next-slot signal prediction are proposed. These tasks are well-suited for spectrum data due to the dynamic and noisy nature of spectrum environments.
Mask reconstruction helps the model learn robust signal representations by reconstructing missing or degraded signal segments, simulating real-world scenarios where signals are often impaired by noise and interference.
Next-slot signal prediction is crucial for understanding the dynamic nature of spectrum usage. As the spectrum environment continuously evolves, predicting future spectrum behavior enables the model to adapt to these changes, ensuring effective management of the constantly shifting spectrum.
As a result, these tasks enable SpectrumFM to learn the key characteristics of spectrum data, preparing it for real-world spectrum management challenges.
\subsubsection{The Masked Reconstruction Task} 
In the masked reconstruction task, the objective is to mask a portion of the original AP sequence $\mathbf{x}_{\text{ap}}^{\text{norm}}$ and train the model to reconstruct the masked parts. The task facilitates learning general representations in an unsupervised manner by training the model to infer and reconstruct masked portions of the input data, thereby enhancing its robustness and generalization capabilities.
Specifically, given the sequence $\mathbf{x}_{\text{ap}}^{\text{norm}}$, we first randomly mask a subset of the sequence according to a predefined masking ratio $r$. The binary mask vector $\mathbf{m}$ is defined as $\mathbf{m} \sim \text{Bernoulli}(1-r)$, which consists of elements $\left\{m_1, m_2, \cdots, m_N\right\}$. Here $m_{p} = 0$ if the $p$-th symbol is masked out, and $m_{p} = 1$ otherwise. The masked sequence is then given as 
\begin{equation}
    \mathbf{x}_{\text{ap}}^{\text{mask}} = \mathbf{x}_{\text{ap}}^{\text{norm}} \odot \mathbf{m},
    \label{eq:20}
\end{equation}
where $\odot$ represents the element-wise multiplication. The masked AP sequence is then fed into the SpectrumFM encoders to obtain $\mathbf{x}_{\text{latent}}^{(L)}$. It is important to note that an attention mask $\mathbf{M}$ is also applied during the computation of attention weights in the SpectrumFM encoders. Specifically, in the pre-training phase, equation (\ref{eq:attention}) is updated as 
\begin{equation}
    \mathbf{x}_{h} = \text{softmax}\left(\frac{\mathbf{Q}_h \mathbf{K}_h^T}{\sqrt{d_h}} + \mathbf{M}\right) \mathbf{V}_h,   
\end{equation}
where $\mathbf{M}$ assigns extremely small values (e.g., $-\infty$) to the positions that are masked out, thereby ensuring that the positions do not contribute to the computed attention weights.
After obtaining $\mathbf{x}_\text{{latent}}^{(L)}$, a lightweight decoder is applied to the latent representation to reconstruct the original input sequence, given as 
\begin{equation}
\mathbf{x}_\text{{recon}} = \mathbf{x}_\text{{latent}}^{(L)} \mathbf{W}_\text{{recon}} + \mathbf{b}_\text{{recon}},
\label{eq:22}
\end{equation}
where $\mathbf{W}_\text{{recon}} \in \mathbb{R}^{d\times 2}$ and $\mathbf{b}_\text{{recon}} \in \mathbb{R}^{2}$ are the learnable weights and biases of the decoder, respectively.
Finally, the mean squared error (MSE) loss is calculated by comparing the reconstructed sequence with the original input sequence in the masked region, given as 
\begin{equation}
    \mathcal{L}_\text{{recon}} = \frac{1}{\sum_{p=1}^{N} (1 - m_{p})} \sum_{i=p}^{N} (1 - m_{p}) \left(\mathbf{x}^{\text{norm}}_{\text{ap}}  -\mathbf{x}_\text{{recon}}\right)^2.
    \label{eq:23}
\end{equation}
\subsubsection{Next-slot Signal Prediction Task}
The next-slot signal prediction task aims to predict the signal symbol at the $N$-th time slot based on the historical signal observations from the previous $N-1$ slots. The task facilitates SpectrumFM to understand the temporal characteristics of signals. Specifically, $\mathbf{x}_{\text{ap}}^{\text{norm}}[:N-1]$ is input to the SpectrumFM encoders to obtain the latent representation $\mathbf{x}_\text{{latent}}^{(L)}$, and then an long short-term memory (LSTM) layer is leveraged to aggregate the latent representation, and finally a linear layer is exploited to predict the $N$-th signal symbol. Mathematically, the predicted signal symbol is given as 
\begin{align}
    \hat{x}_{\text{ap}}^{\text{norm}}[N] = \sigma\left(LSTM\left(\mathbf{x}_\text{{latent}}^{(L)}\right)\mathbf{W}_{\text{pre}} + \mathbf{b}_{\text{pre}}\right),
    \label{eq:24}
\end{align}
where $\sigma(\cdot)$ is the sigmoid function, $\mathbf{W}_{\text{pre}} \in \mathbb{R}^{d\times 2}$ and $\mathbf{b}_{\text{pre}}\in \mathbb{R}^{2}$ are the weights and biases of the linear layer, respectively.
The MSE loss is adopted to train the task, given as 
\begin{align}
    \mathcal{L}_{\text{pre}} = \frac{1}{2}  \left( \hat{x}_{\text{ap}}^{\text{norm}}[N] - \mathbf{x}_{\text{ap}}^{\text{norm}}[N] \right)^2.
    \label{eq:25}
\end{align}
The Pre-training algorithm is summarized in Algorithm \ref{alg:1}.
\begin{algorithm}
    
    \SetAlgoLined 
    \caption{Pre-training Algorithm For SpectrumFM} 
    \label{alg:1}
    \KwIn{Pre-training dataset} 
    \KwOut{All the parameters $\mathbf{\Theta}$ of SpectrumFM} 

    Initialize all the parameters $\mathbf{\Theta}$ of SpectrumFM\; 
    
    Set $\mathrm{best\_val\_loss} \leftarrow \infty$\; 
    Set $\mathrm{no\_improvement\_count} \leftarrow 0$\; 
    
    \For{epoch = 1 to $E_p$}{ 
        \For{batch data in the pretraining dataset}{ 
            Preprocess the batch data according to equations (\ref{eq:1}) to (\ref{eq:7})\; 
            
            Feed processed data into SpectrumFM encoders and obtain the latent representation $\mathbf{x}_{\text{latent}}^{(L)}$ according to equations (\ref{eq:projection}) to (\ref{eq:encoder})\; 
            
            Obtain reconstructed signal symbols according to equations (\ref{eq:20}) to (\ref{eq:22})\;
            
            Compute reconstruction loss $\mathcal{L}_{\text{recon}}$ according to equation (\ref{eq:23})\;
            
            Predict next-slot signal symbols according to equation (\ref{eq:24})\;
            
            Compute predicted loss $\mathcal{L}_{\text{pre}}$ according to equation (\ref{eq:25})\;
            
            Obtain pre-training loss according to $\mathcal{L} \leftarrow \alpha\mathcal{L}_{\text{recon}}+ \beta \mathcal{L}_{\text{pre}}$\;
            
            Update $\mathbf{\Theta}$ using the pre-training loss according to $\mathbf{\Theta} \leftarrow \mathbf{\Theta} - \eta \nabla_{\mathbf{\Theta}} \mathcal{L}$\; 
        }
        
        Evaluate the model on the validation set and compute $\mathrm{val\_loss}$\;
        
        \If{$\mathrm{val\_loss < best\_val\_loss}$}{
            $\mathrm{best\_val\_loss \leftarrow val\_loss}$\; 
            $\mathrm{no\_improvement\_count \leftarrow 0}$\; 
        }
        \Else{
            $\mathrm{no\_improvement\_count \leftarrow no\_improvement}$ \newline $\mathrm{\_count + 1}$\; 
        }
        
        \If{$\mathrm{no\_improvement\_count \geq 2}$}{
            \textbf{break}  \; 
        }
    }

    Return the pretrained parameters $\mathbf{\Theta}$\; 
    
\end{algorithm}
\begin{algorithm}[htbp]
    
    \SetAlgoLined 
    \caption{Fine-tuning Algorithm For SpectrumFM} 
    \label{alg:2}
    \KwIn{Fine-tuning dataset, pre-trained parameters $\mathbf{\Theta}$} 
    \KwOut{All the fine-tuned parameters $\mathbf{\Theta}$ of SpectrumFM} 
    \For{$\mathrm{epoch}$ = 1 to $E_f$}{ 
        \For{batch data in the fine-tuning dataset}{ 
            Preprocess the batch data according to equations (\ref{eq:1}) to (\ref{eq:7})\; 
            
            Feed processed data into the SpectrumFM encoder to obtain the latent representation $\mathbf{x}_{\text{latent}}^{(L)}$ according to equations (\ref{eq:projection}) to (\ref{eq:encoder})\; 
            
            Compute the classification loss for AMC, WTC, and SS tasks according to equations (\ref{eq:finetune}) to (\ref{eq:finetune_loss}) or the AD task loss according to equation (\ref{eq:finetune_ad})\;
            Update $\mathbf{\Theta}$ by using the classification loss or the AD task loss according to $\mathbf{\Theta} \leftarrow \mathbf{\Theta} - \eta \nabla_{\mathbf{\Theta}} \mathcal{L}$\; 
        }
    }

    Return finetuned parameters $\mathbf{\Theta}$\; 
    
\end{algorithm}
\subsection{Fine-tuning Stage}
\begin{figure}[htbp]
    \centering
    \begin{subfigure}[b]{0.4\textwidth}
        \centering
        \includegraphics[width=\linewidth]{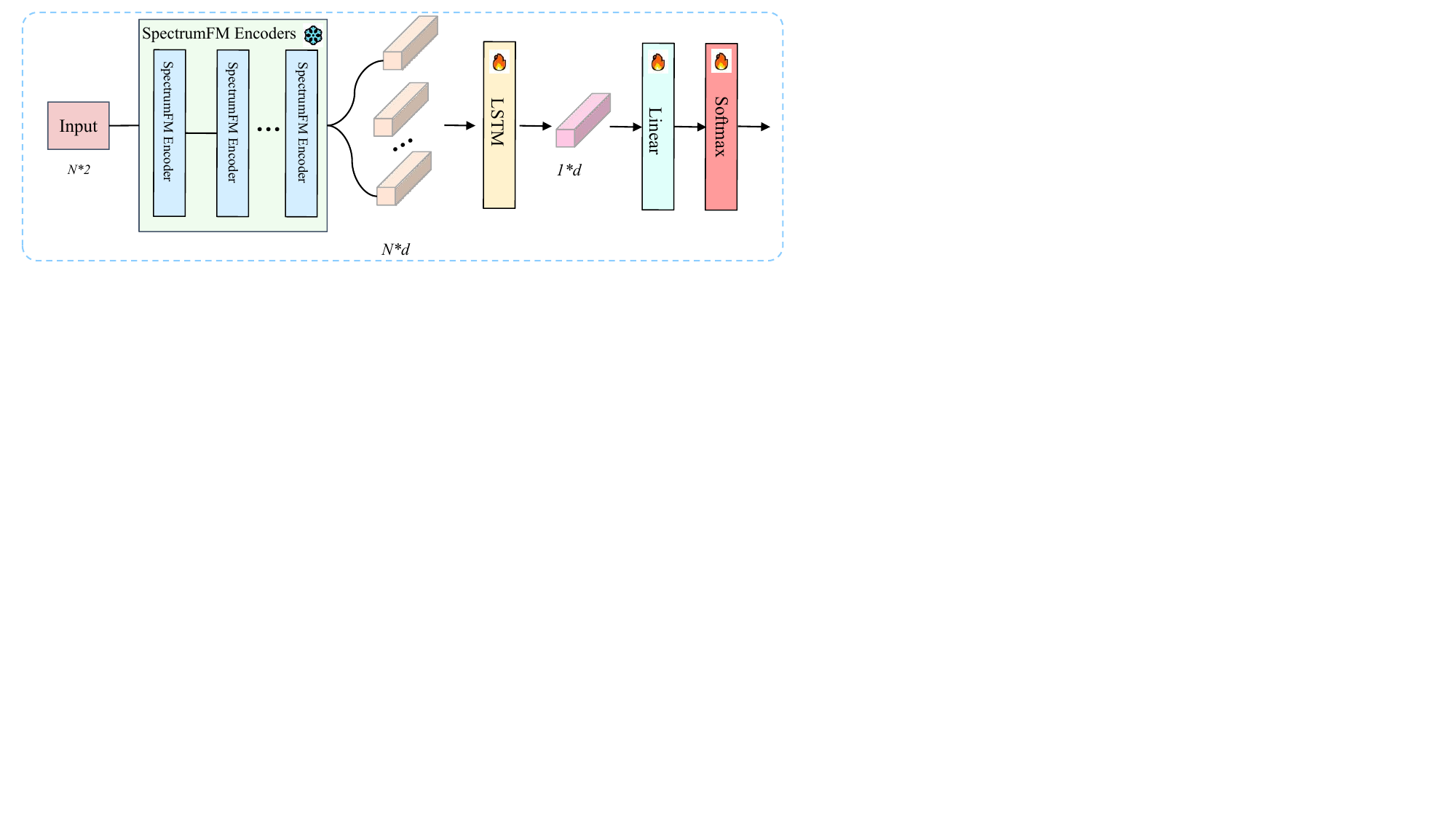}
        \caption{The fine-tune framework for AMC, WTC, and SS tasks.}
        \label{fig:finetune1}
    \end{subfigure}
    
    \begin{subfigure}[b]{0.4\textwidth}
        \centering
        \includegraphics[width=\linewidth]{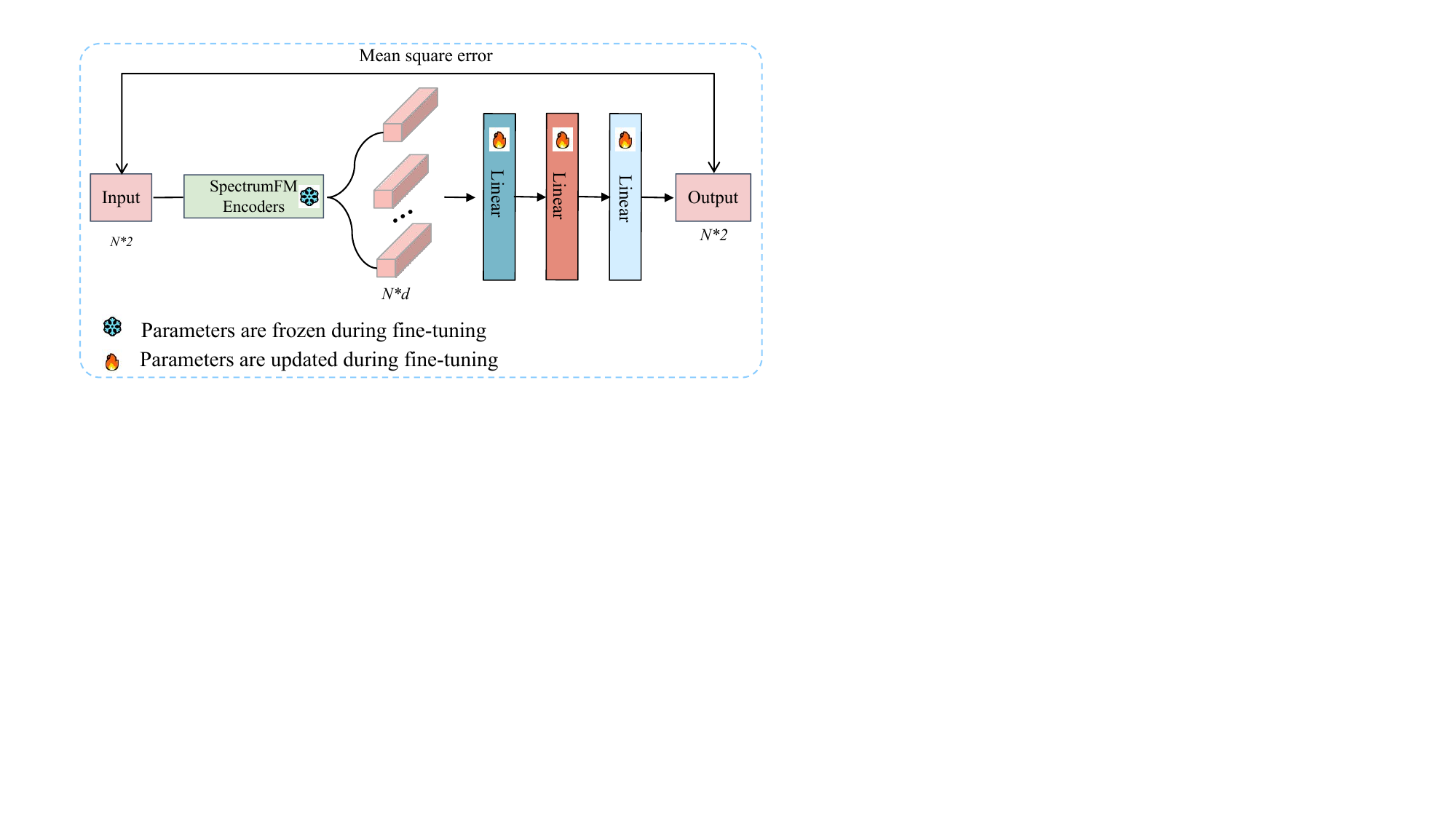}
        \caption{The fine-tune framework for AD task.}
        \label{finetune2}
    \end{subfigure}
    \caption{The fine-tune framework for AMC, WTC, SS, and AD tasks.}
    \label{finetune}
\end{figure}

After completing the pre-training stage, which enables the model to acquire generalizable representations from large volumes of unlabeled data, the next critical step is the fine-tuning stage. During this phase, the pre-trained model is further refined for a specific downstream task exploiting a small but task-relevant labeled dataset. The primary goal of fine-tuning is to adjust a limited set of model parameters, enabling the model to effectively capture task-specific characteristics while preserving the general knowledge acquired during pre-training. The fine-tune framework for AMC, WTC, and SS tasks is shown in Fig. \ref{fig:finetune1}. 
In particular, the input sequence is first processed through the SpectrumFM encoders to generate a sequence of hidden states as described in equations (\ref{eq:projection}) to (\ref{eq:encoder}). These hidden states are subsequently aggregated leveraging an LSTM network. Finally, the aggregated hidden state is passed through a linear layer, followed by a softmax function, to perform the classification task, given as 
\begin{equation}
    \label{eq:finetune}
    \begin{aligned}
        \hat{\mathbf{y}} = \text{softmax}\left(LSTM\left(\mathbf{x}_\text{{latent}}^{(L)}\right)\mathbf{W}_{f} + \mathbf{b}_{f}\right),
    \end{aligned}
\end{equation}
where $\mathbf{W}_{f} \in \mathbb{R}^{d\times C}$ and $\mathbf{b}_{f}\in \mathbb{R}^{C}$ are the weights and biases of the linear layer, respectively, and $C$ is the number of classes.
The cross-entropy loss is leveraged to fine-tune for the AMC, WTC, and SS tasks, given as 
\begin{equation}
    \label{eq:finetune_loss}
    \begin{aligned}
        \mathcal{L}_\text{{classification}} =  - \sum_{a=1}^{C} y_a \log(\hat{y}_a).
    \end{aligned}
\end{equation}

The fine-tune framework for AD task is shown in Fig.~\ref{finetune2}. The input sequence is first processed through the SpectrumFM encoders to obtain a series of hidden states (as described in equations (\ref{eq:projection}) to (\ref{eq:encoder})). Contrary to the methodology employed in classification tasks, wherein hidden states undergo aggregation via an LSTM network, the proposed framework directly feeds the hidden states into a decoder layer to generate a reconstructed signal, given as 
\begin{equation}
    \label{eq:finetune_ad}
    \begin{aligned}
        \hat{\mathbf{x}} = \text{tanh}\left(\mathbf{x}_\text{{latent}}^{(L)}\mathbf{W}_{a} + \mathbf{b}_{a}\right),
    \end{aligned}
\end{equation}
where $\mathbf{W}_{a} \in \mathbb{R}^{d\times 2}$ and $\mathbf{b}_{a}\in \mathbb{R}^{2}$ are the weights and biases of the decoder layer, respectively, and $\text{tanh}(\cdot)$ is the tanh activation function.
The reconstructed signal is then compared with the original input signal to compute the MSE reconstruction loss. The MSE loss serves as the criterion for distinguishing normal from anomalous signals. A low reconstruction loss indicates that the signal is normal, whereas a high loss suggests the presence of an anomaly, given as 
\begin{equation}
    \label{eq:finetune_loss_ad}
    \begin{aligned}
        \mathcal{L}_{AD} =  \frac{1}{N} \sum_{p=1}^{N} \left(\hat{x}_{p} - x_{p}\right)^{2}.
    \end{aligned}
\end{equation}
The fine-tuning algorithm is summarized in Algorithm \ref{alg:2}.
\subsection{Theoretical Computational Complexity}
To further clarify the computational footprint of SpectrumFM, we provide a theoretical analysis of the per-layer complexity. Each encoder layer consists of a multi-head self-attention (MHSA) module, two feed-forward (FFN) modules, and a convolution (Conv) module. Let $N$ denote the input sequence length, $d$ be the latent dimension, $h$ be the number of attention heads, $d_h = d / h$ be the dimension per head, $d_{\text{feed}}$ be the feed-forward hidden dimension, $k$ be the kernel size of the depthwise convolution, and $L$ be the number of layers. The complexity of the MHSA module can be expressed as $\mathcal{O}_{\text{MHSA}} = 4 N d^2 + 2 h N^2 d_h$, where the first term accounts for the linear projections of queries, keys, and values, and the second term corresponds to the scaled dot-product attention. Each feed-forward module contributes $\mathcal{O}_{\text{FFN}} = 2 N d d_{\text{feed}}$, and the convolution module contributes $\mathcal{O}_{\text{Conv}} = N d k$. Therefore, the total complexity per layer is approximately $\mathcal{O}_{\text{layer}} = \mathcal{O}_{\text{MHSA}} + 2 \, \mathcal{O}_{\text{FF}} + \mathcal{O}_{\text{Conv}}$, and for a $L$-layer encoder, the overall forward pass complexity is $\mathcal{O}_{\text{encoder}} = L\mathcal{O}_{\text{layer}}$. Substituting our model parameters ($N=128$, $d=256$, $h=4$, $d_{\text{feed}}=512$, $k=3$, $L=16$) yields a theoretical per-sample FLOPs estimate of approximately $2.03 \times 10^9$ FLOPs. This analysis provides a clear understanding of the computational demands associated with SpectrumFM, facilitating informed decisions regarding its deployment in various applications.

\section{Experiments}
\label{sec:experiments}
In this section, we first present the experimental settings, followed by the experiments on four downstream spectrum management tasks, including AMC, WTC, SS, and AD.
\subsection{Experimental Settings}
The SpectrumFM is pre-trained on a setup comprising four NVIDIA RTX 4090 GPUs, utilizing PyTorch and DeepSpeed for efficient training. The hyperparameters for SpectrumFM are as follows. The number of signal symbols is set to 128, the number of attention heads $H$ is set to 4, the latent dimension $d$ is set to 256, the feedforward dimension $d_{\text{feed}}$ is set to 512, and the number of SpectrumFM encoder layers $L$ is set to 16~\cite{gulati2020conformer}. The mask ratio $r$ is set to 15\%, the pre-training phase consists of 10 epochs with a batch size of 256 and a learning rate of 0.001~\cite{devlin-etal-2019-bert}. The AdamW optimizer is employed for optimization, and early stopping is utilized to prevent overfitting. During the fine-tuning stage, the same learning rate of 0.001 and the AdamW optimizer are used to further adapt the model to specific downstream tasks.

As shown in Fig.~\ref{fig:loss_plot}, the pre-training loss decreases smoothly without oscillations across different objectives, indicating stable convergence of the training process.

\subsection{Automatic Modulation Classification Task}
\begin{table*}[htbp]
    \centering
    \caption{The Overall Performance in the AMC Task.}
    \label{tab:model_performance}
    \begin{tabular}{c|c|c|c|c|c|c|c|c|c}
        \hline
        Dataset & \multicolumn{3}{c|}{RML2016.10A} & \multicolumn{3}{c|}{RML2016.10B} & \multicolumn{3}{c}{RML2016.04C} \\\hline
        Model & Precision (\%) & Recall (\%) & F1 (\%) & Precision (\%) & Recall (\%) & F1 (\%) & Precision (\%) & Recall (\%) & F1 (\%) \\\hline
        ResNet & 54.27 & 50.04 & 48.94 & 57.16 & 54.88 & 52.80 & 64.43 & 55.97 & 57.04 \\\hline
        MCNet & 60.19 & 53.52 & 54.79 & 60.78 & 59.22 & 59.61 & 65.97 & 59.57 & 61.11 \\\hline
        VGG & 67.99 & 54.72 & 54.06 & 71.04 & 58.65 & 59.31 & \textbf{74.80} & 61.88 & 63.85 \\\hline
        CNN2 & 68.67 & 53.25 & 55.32 & 64.61 & 57.14 & 57.27 & 74.69 & 59.45 & 62.33 \\\hline
        GRU2 & 68.19 & 58.80 & 60.22 & 66.15 & \underline{64.11} & 63.97 & 71.46 & 63.13 & 64.81 \\\hline
        DAE & 69.96 & 58.97 & 60.72 & 64.88 & 61.46 & 61.17 & 53.61 & 55.91 & 53.85 \\\hline
        CGDNN & \underline{73.03} & 56.57 & 59.36 & 64.23 & 58.26 & 59.53 & 75.21 & 60.34 & 63.37 \\\hline
        Transformer & 68.26 & \underline{59.27} & 61.08 & 68.75 & 63.10 & 63.28 & 71.31 & \underline{65.41} & \underline{66.70} \\\hline
        MSNet & 69.41 & 58.33 & 60.60 & 65.29 & 63.49 & 63.50 & 71.02 & 63.66 & 65.36 \\\hline
        AMC\_Net & 71.41 & 59.10 & \underline{61.11} & \underline{69.25} & 63.38 & \underline{64.05} & 68.03 & 63.01 & 64.35 \\\hline
        Our & \textbf{75.29} & \textbf{63.72} & \textbf{66.20} & \textbf{70.07} & \textbf{65.35} & \textbf{65.93} & \underline{73.93} & \textbf{73.37} & \textbf{73.46} \\\hline
        \end{tabular}
\end{table*}
\subsubsection{Dataset}
We leverage the RML2016 datasets~\footnote{\url{https://www.deepsig.ai/datasets/}} for AMC task. The RML2016 datasets, including RML2016.10A, RML2016.10B, and RML2016.04C, are benchmark datasets widely used for AMC task in wireless communications. RML2016.10A contains 11 modulation types under various SNR conditions, providing a foundational testbed for evaluating AMC task. RML2016.10B emphasizes real-world RF signals captured in complex propagation environments, offering a more challenging scenario. RML2016.04C further extends the challenges by incorporating effects such as multipath fading and Doppler shifts, rigorously testing the robustness of classifiers in dynamic channel conditions. 
\subsubsection{Baselines}
\label{sec:baselines}
We leverage 10 state-of-the-art baseline models for the AMC task. The outlines of the models are as follows.
\begin{itemize}[left=0pt, labelsep=0.5em, itemsep=2pt]
    \item AMC\_Net \cite{10097070} employs a frequency-domain denoising module and a multi-scale feature extraction mechanism, leveraging attention operations to enhance modulation classification accuracy.
    \item Transformer \cite{vaswani2017attention} leverages the self-attention mechanism to extract features from the input signal.
    \item MSNet \cite{9463441} leverages a multi-scale feature extraction and fusion mechanism to classify the modulation schemes.
    \item CGDNN \cite{cgdnn} composes of a shallow convolutional network, a gated recurrent unit, and a deep neural network, for robust automatic modulation recognition.
    \item DAE \cite{dae} is an LSTM denoising auto-encoder framework that efficiently extracts robust features from noisy radio signals for the classification of communication technologies and modulation schemes.
    \item MCNet \cite{8963964} employs specialized convolutional blocks with asymmetric kernels to concurrently capture spatio-temporal signal correlations, enhancing classification accuracy and efficiency.
    \item ResNet \cite{resnet} is a residual network, a widely used deep learning model.
    \item VGG \cite{8267032} is a 1D CNN network that adapt the VGG architecture principles.
    \item CNN2 \cite{8267032} is a CNN classifier model for classifying modulation schemes according to their families. 
    \item GRU2 \cite{gru2} is a two layers of the GRU classifier model.
\end{itemize}
\subsubsection{Overall Performance}
To evaluate the overall performance of our proposed method for the AMC task, we conduct extensive experiments on three datasets, namely RML2016.10A, RML2016.10B, and RML2016.04C. The results are summarized in TABLE \ref{tab:model_performance}, which presents the precision, recall, and F1-score for various models. The results underscore the superior capability of our model in the AMC task, achieving notable improvements in precision, recall, and F1-score over other baseline models. Specifically, our model attains F1-scores of 66.20\% on RML2016.10A, 65.93\% on RML2016.10B, and 73.46\% on RML2016.04C, outperforming baselines like Transformer and AMC\_Net by 2.65-9.61 percentage points. Notably, it demonstrates robust precision (75.29\% on RML2016.10A) and recall (73.37\% on RML2016.04C), addressing challenges in class imbalance.
\subsubsection{The Performance on Various SNR Conditions}
\begin{figure*}[htbp]
    \centering
    \begin{minipage}[t]{0.23\textwidth}
        \centering
        \includegraphics[width=\linewidth, height=1.35in]{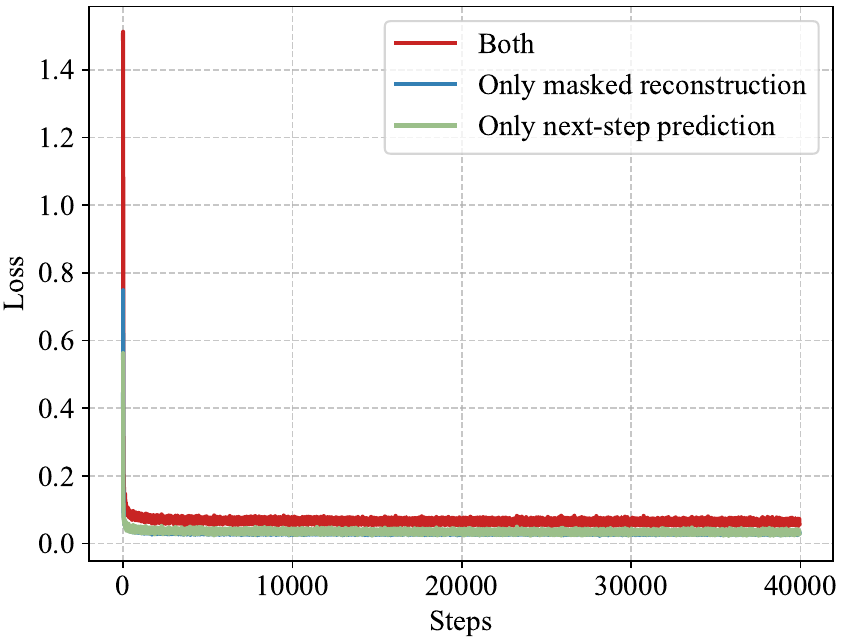}  
        \caption{Loss curves during the pre-training stage.}
        \label{fig:loss_plot}
    \end{minipage}
    \begin{minipage}[t]{0.23\textwidth}
        \centering
        \includegraphics[width=\linewidth]{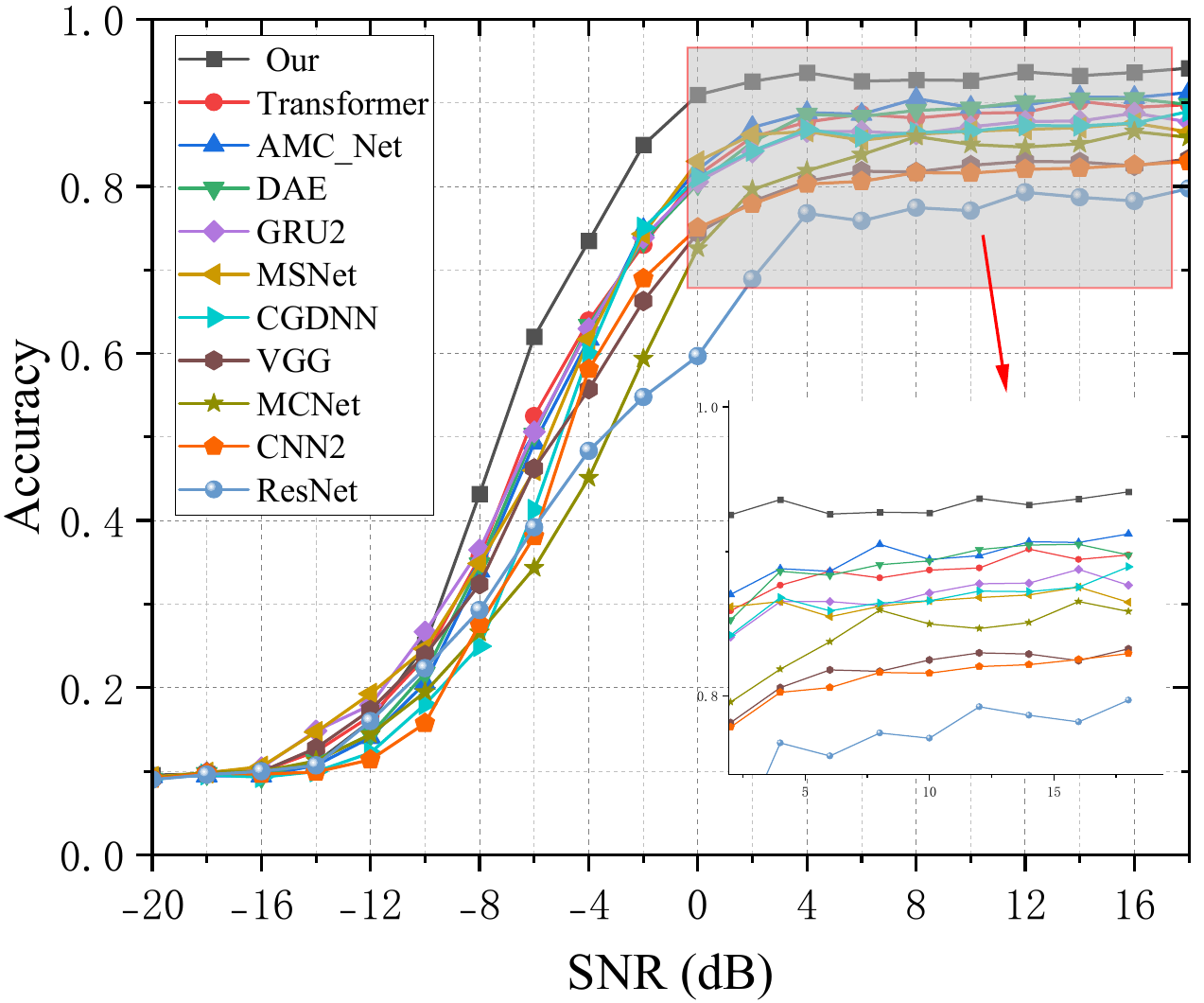}  
        \caption{The accuracy of models on different SNR conditions on RML2016.10A dataset.}
        \label{fig:3a}
    \end{minipage}
    \begin{minipage}[t]{0.23\textwidth}
        \centering
        \includegraphics[width=\linewidth]{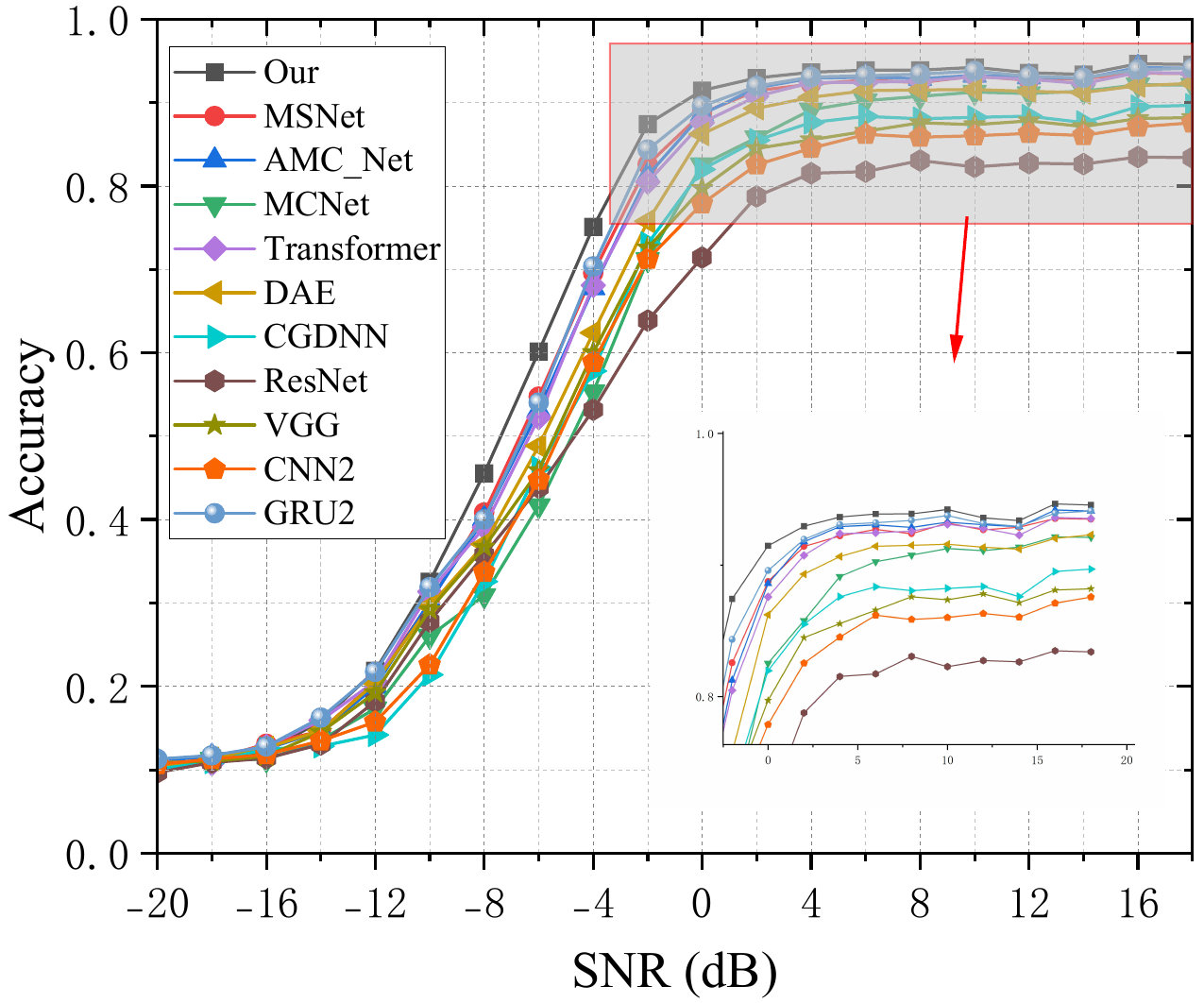}  
        \caption{The accuracy of models on different SNR conditions on RML2016.10B dataset.}
        \label{fig:3b}
    \end{minipage}
    \begin{minipage}[t]{0.23\textwidth}
        \centering
        \includegraphics[width=\linewidth]{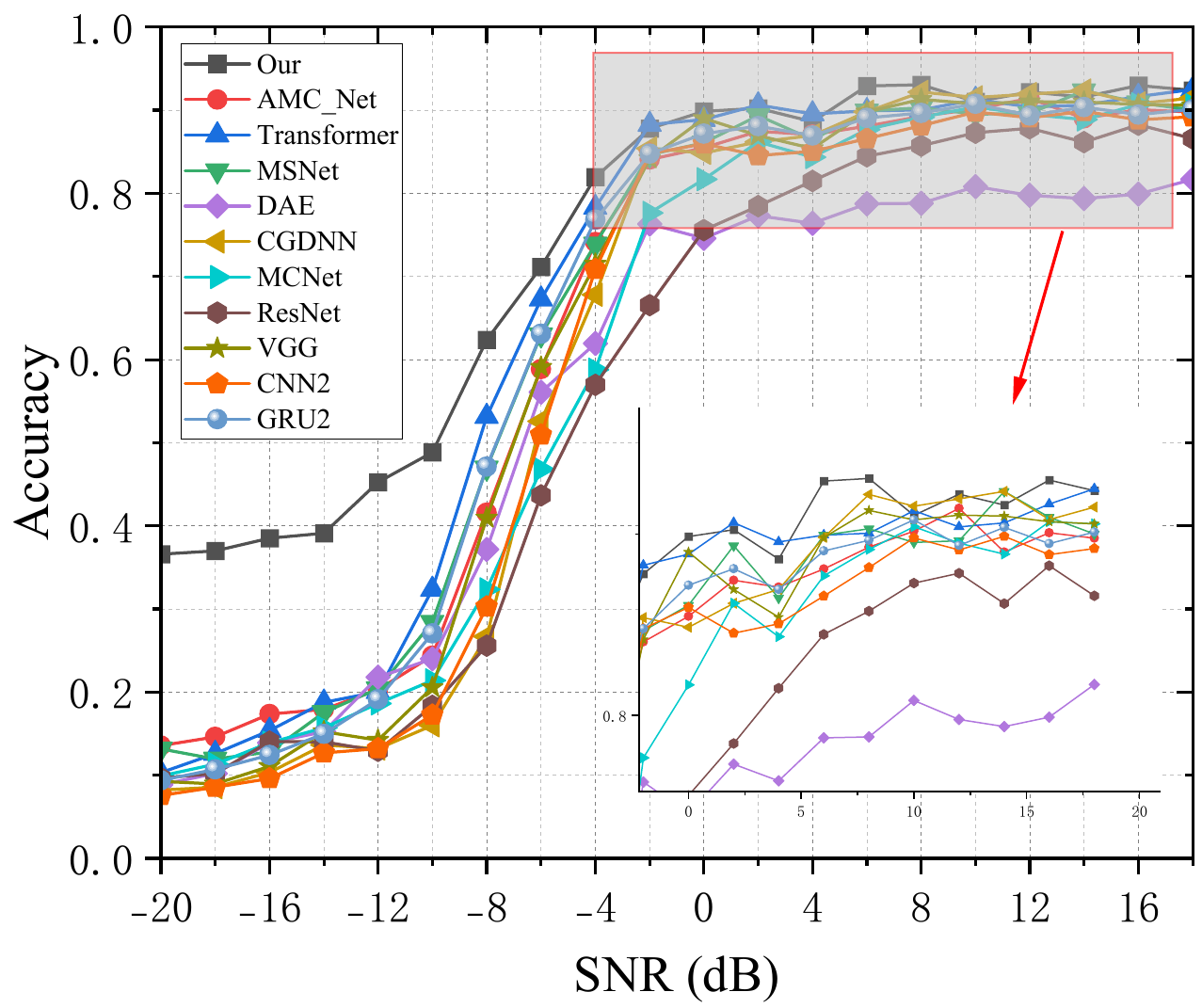}  
        \caption{The accuracy of models on different SNR conditions on RML2016.04C dataset.}
        \label{fig:3c}
    \end{minipage}
\end{figure*}
To evaluate the robustness of our proposed method, its performance is assessed and contrasted with several baseline models across different SNR values, as illustrated in Fig. \ref{fig:3a}-\ref{fig:3c}. The findings highlight the exceptional capability of our method across varying SNR conditions on three distinct datasets, namely, RML2016.10A, RML2016.10B, and RML2016.04C.
Particularly, on the RML2016.10A dataset (Fig. \ref{fig:3a}), our method showcases a marked improvement in accuracy at low SNR levels (-10 to 0 dB), significantly outperforming all other models. The trend continues on the RML2016.10B dataset (Fig. \ref{fig:3b}), where our method maintains its superior performance across the entire SNR range, demonstrating enhanced reliability. Furthermore, the robustness of our method is particularly evident on the RML2016.04C dataset (Fig. \ref{fig:3c}), our method not only consistently outperforms other models but also shows a particularly significant advantage under challenging low SNR conditions (-10 to -5 dB). 
These results collectively emphasize the efficacy of our proposed method in achieving high accuracy across diverse SNR conditions, underscoring its superior performance and resilience in various communication environments. 

\begin{figure*}[htbp]
    \centering
    \begin{subfigure}{0.3\textwidth}
        \centering
        \includegraphics[width=\linewidth]{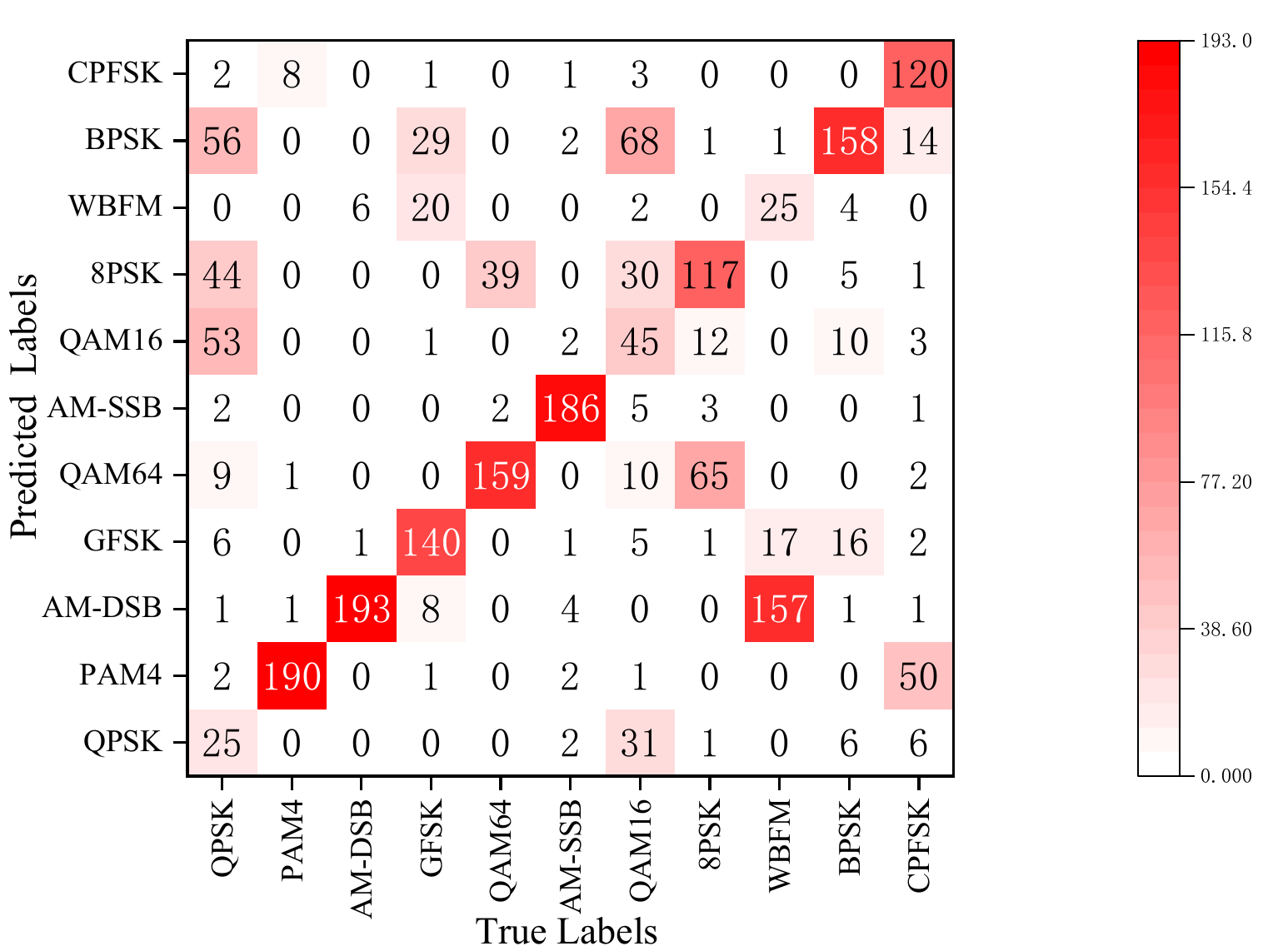}  
        \caption{AMC\_Net confusion matrix at -4 dB SNR.}
    \end{subfigure}%
    \begin{subfigure}{0.3\textwidth}
        \centering
        \includegraphics[width=\linewidth]{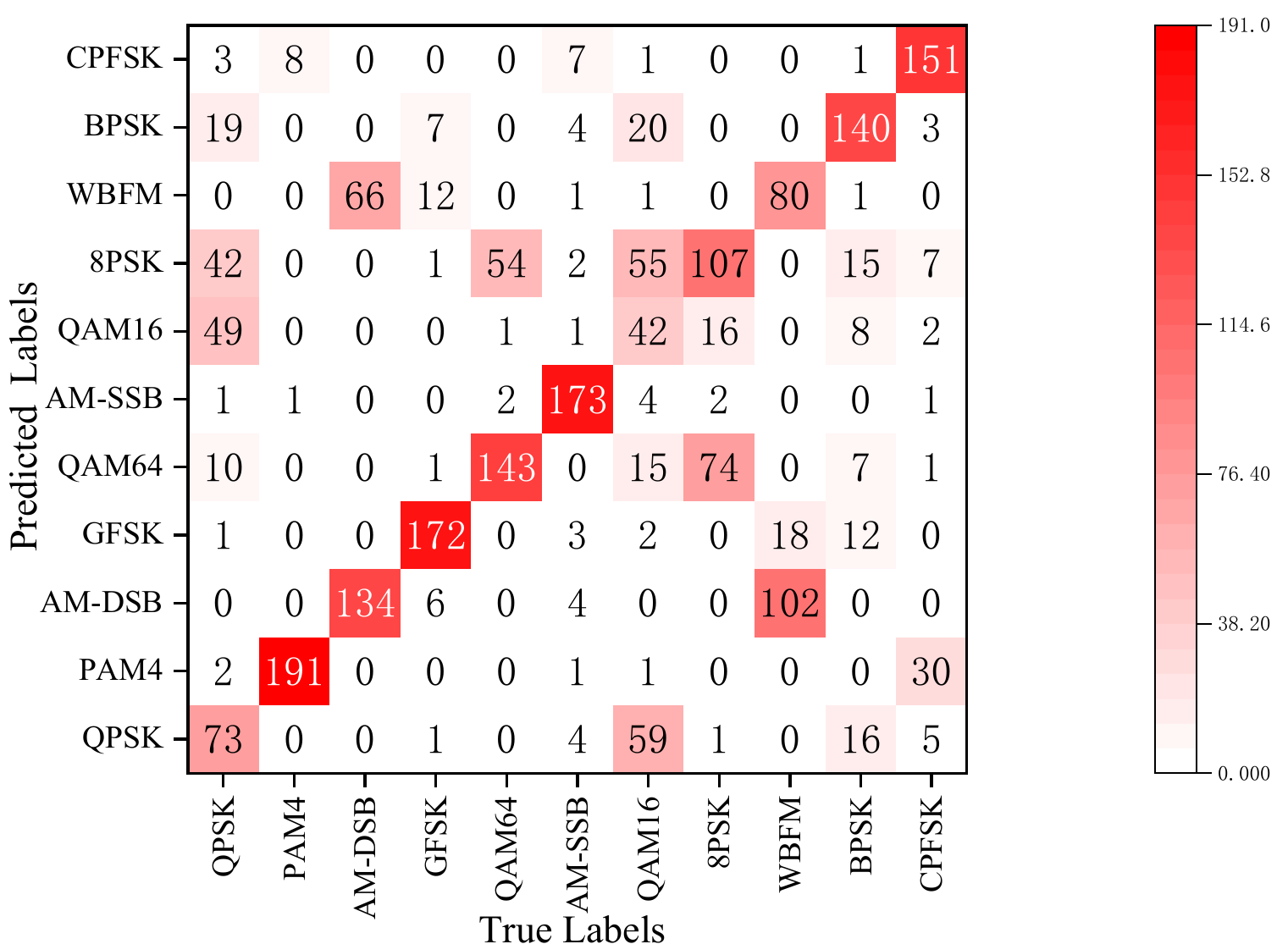}  
        \caption{Transformer confusion matrix at -4 dB SNR.}
    \end{subfigure}%
    \begin{subfigure}{0.3\textwidth}
        \centering
        \includegraphics[width=\linewidth]{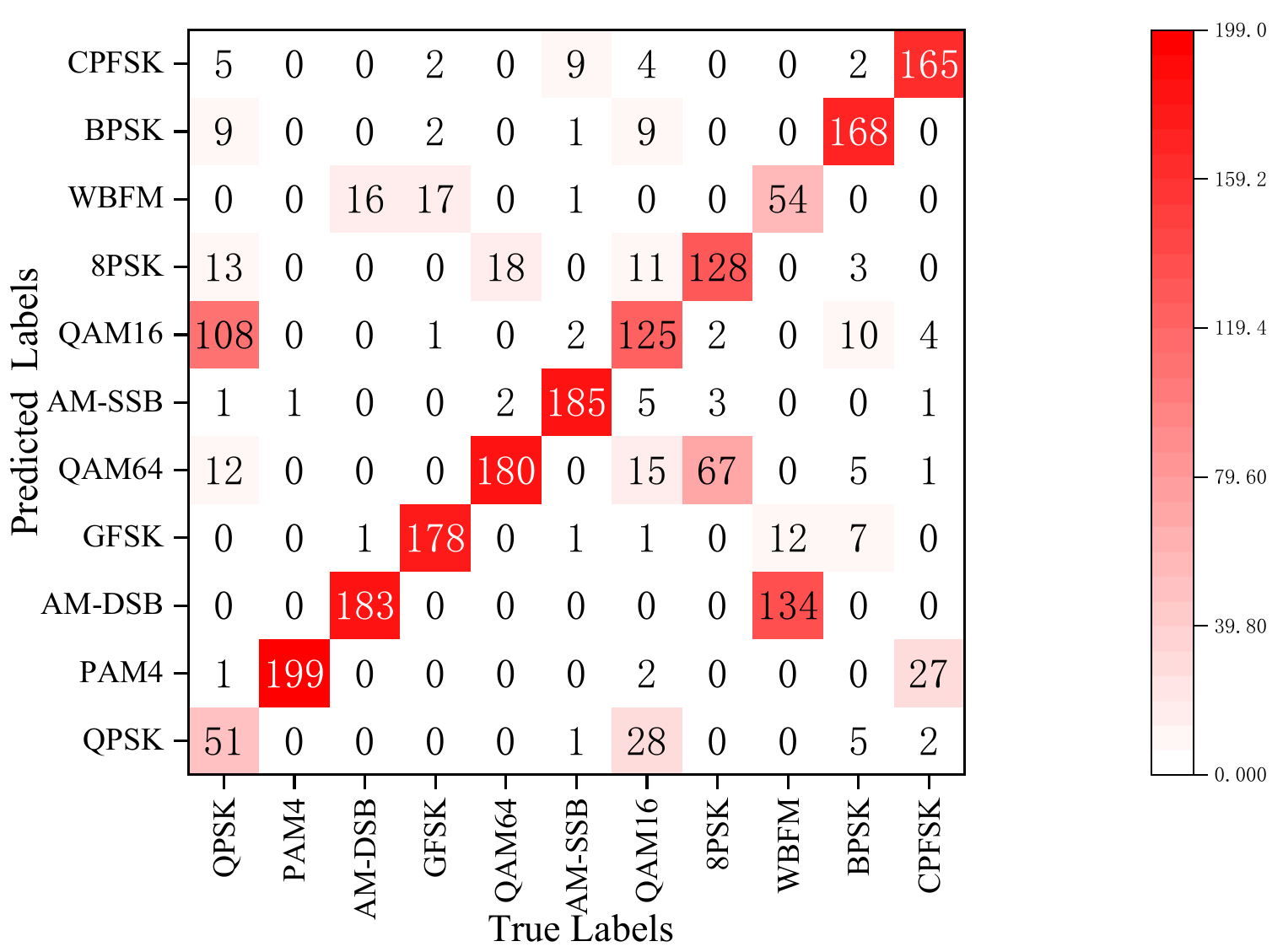}  
        \caption{SpectrumFM confusion matrix at -4 dB SNR.}
    \end{subfigure}

    \begin{subfigure}{0.3\textwidth}
        \centering
        \includegraphics[width=\linewidth]{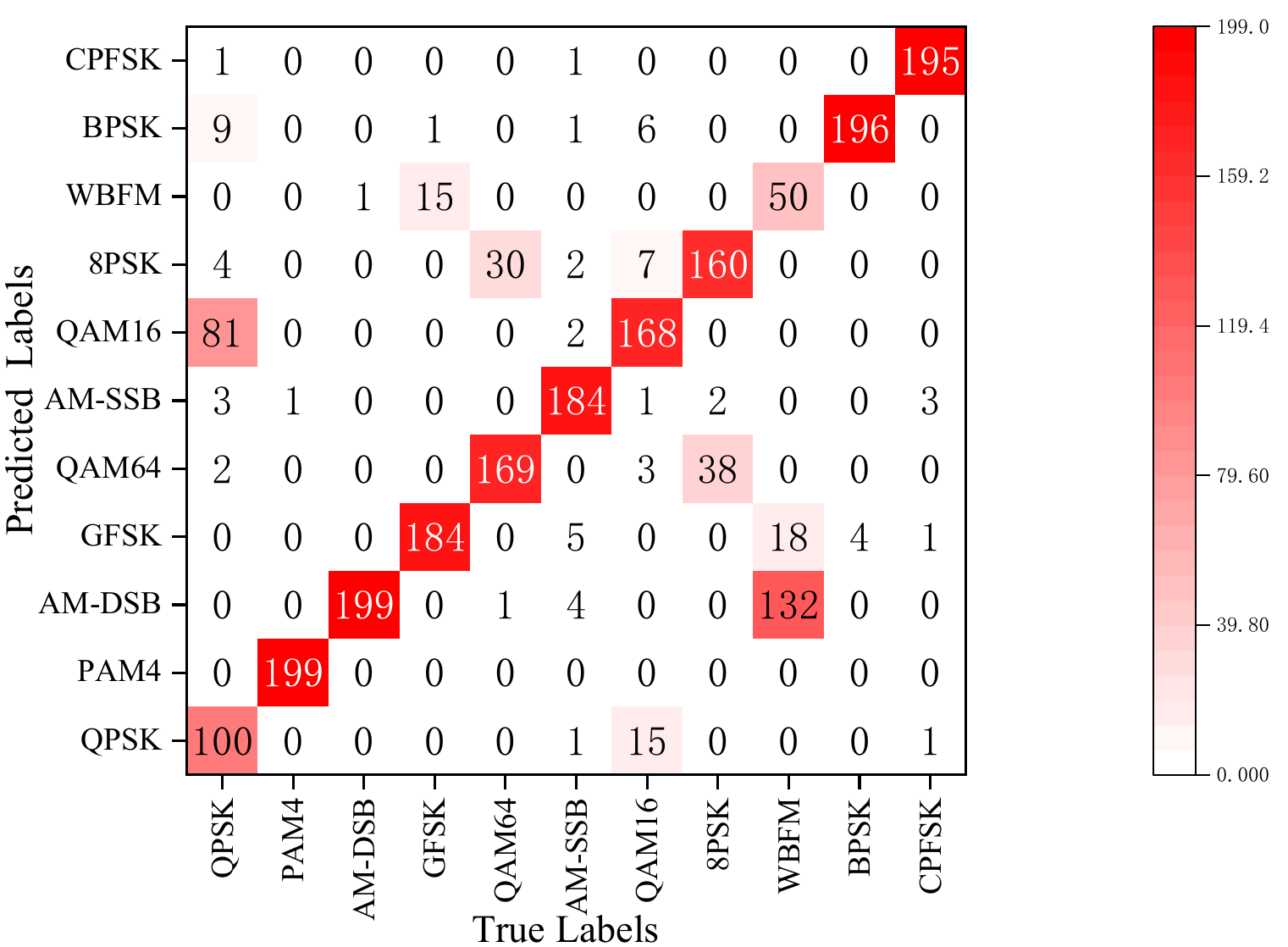}  
        \caption{AMC\_Net confusion matrix at 0 dB SNR.}
    \end{subfigure}%
    \begin{subfigure}{0.3\textwidth}
        \centering
        \includegraphics[width=\linewidth]{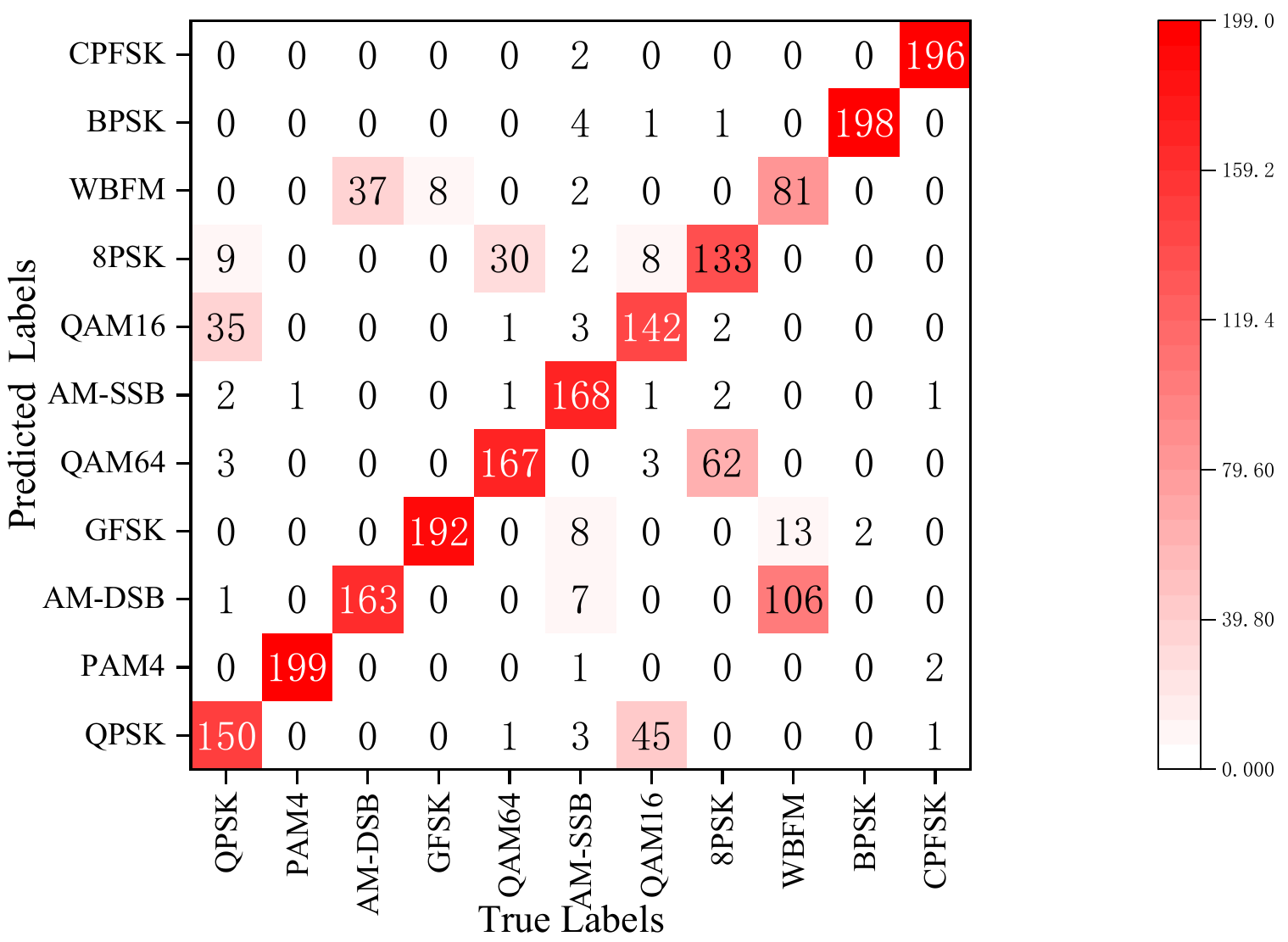}  
        \caption{Transformer confusion matrix at 0 dB SNR.}
    \end{subfigure}%
    \begin{subfigure}{0.3\textwidth}
        \centering
        \includegraphics[width=\linewidth]{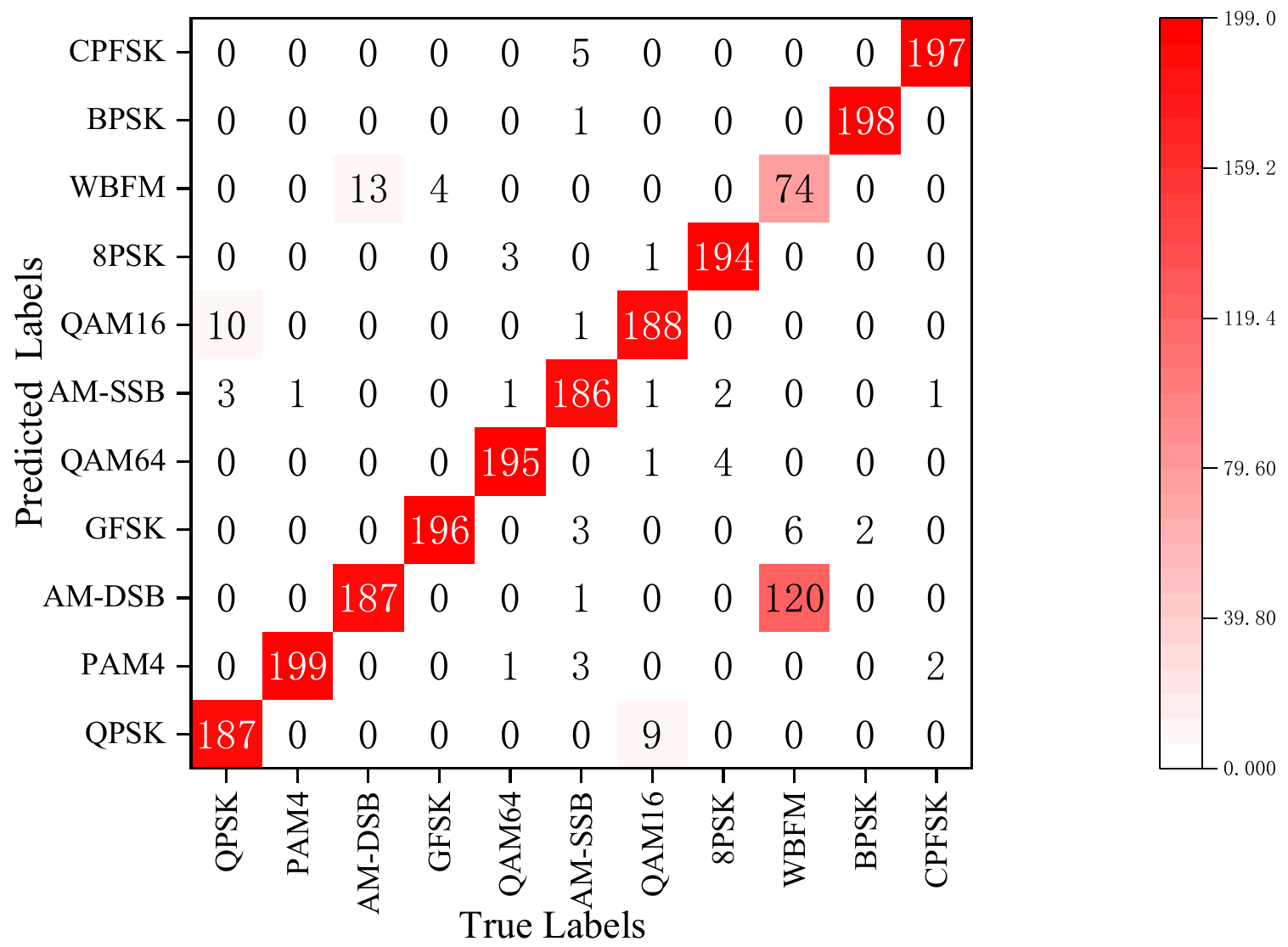}  
        \caption{SpectrumFM confusion matrix at 0 dB SNR.}
    \end{subfigure}


    \begin{subfigure}{0.3\textwidth}
        \centering
        \includegraphics[width=\linewidth]{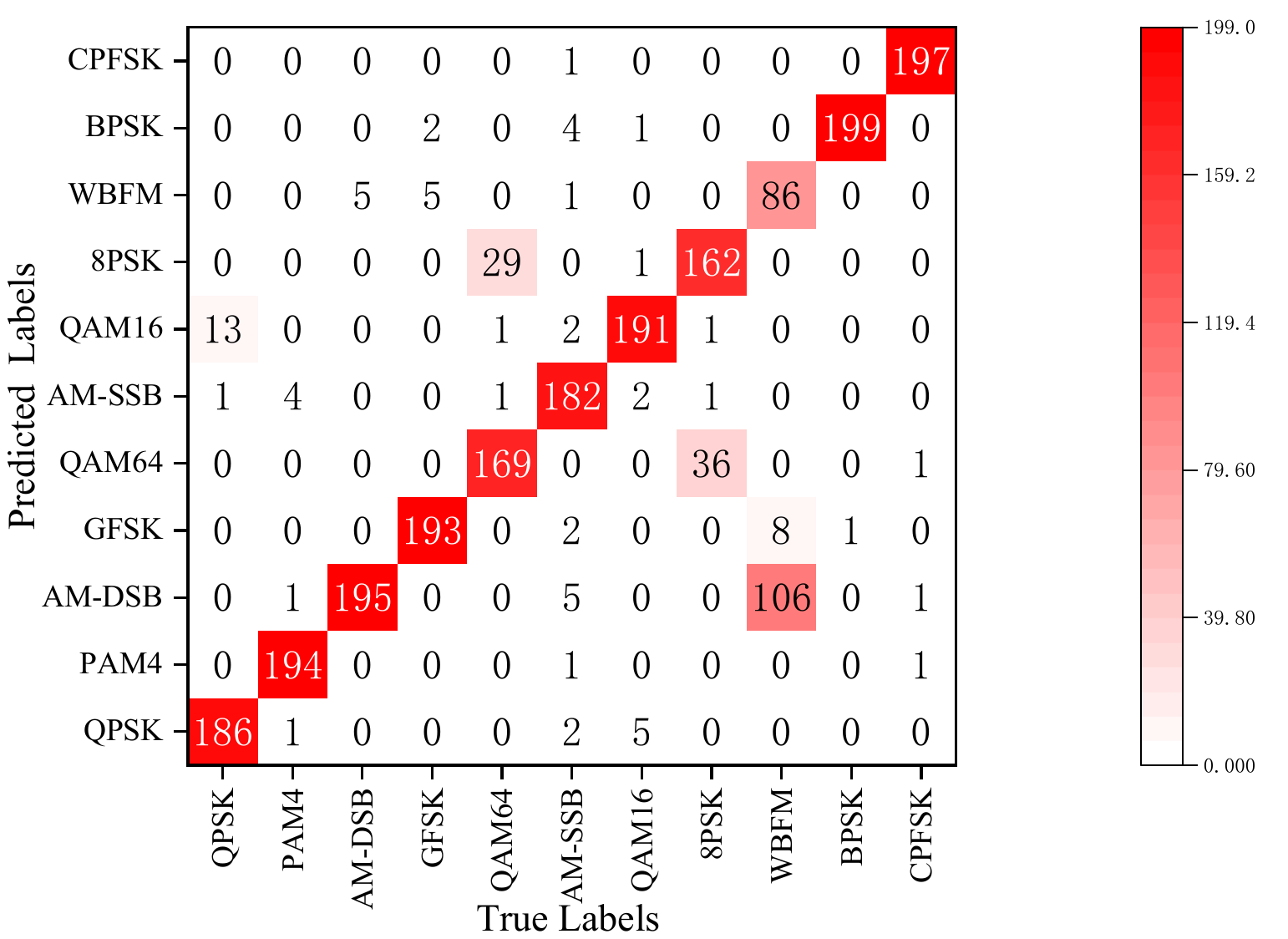}  
        \caption{AMC\_Net confusion matrix at 4 dB SNR.}
    \end{subfigure}%
    \begin{subfigure}{0.3\textwidth}
        \centering
        \includegraphics[width=\linewidth]{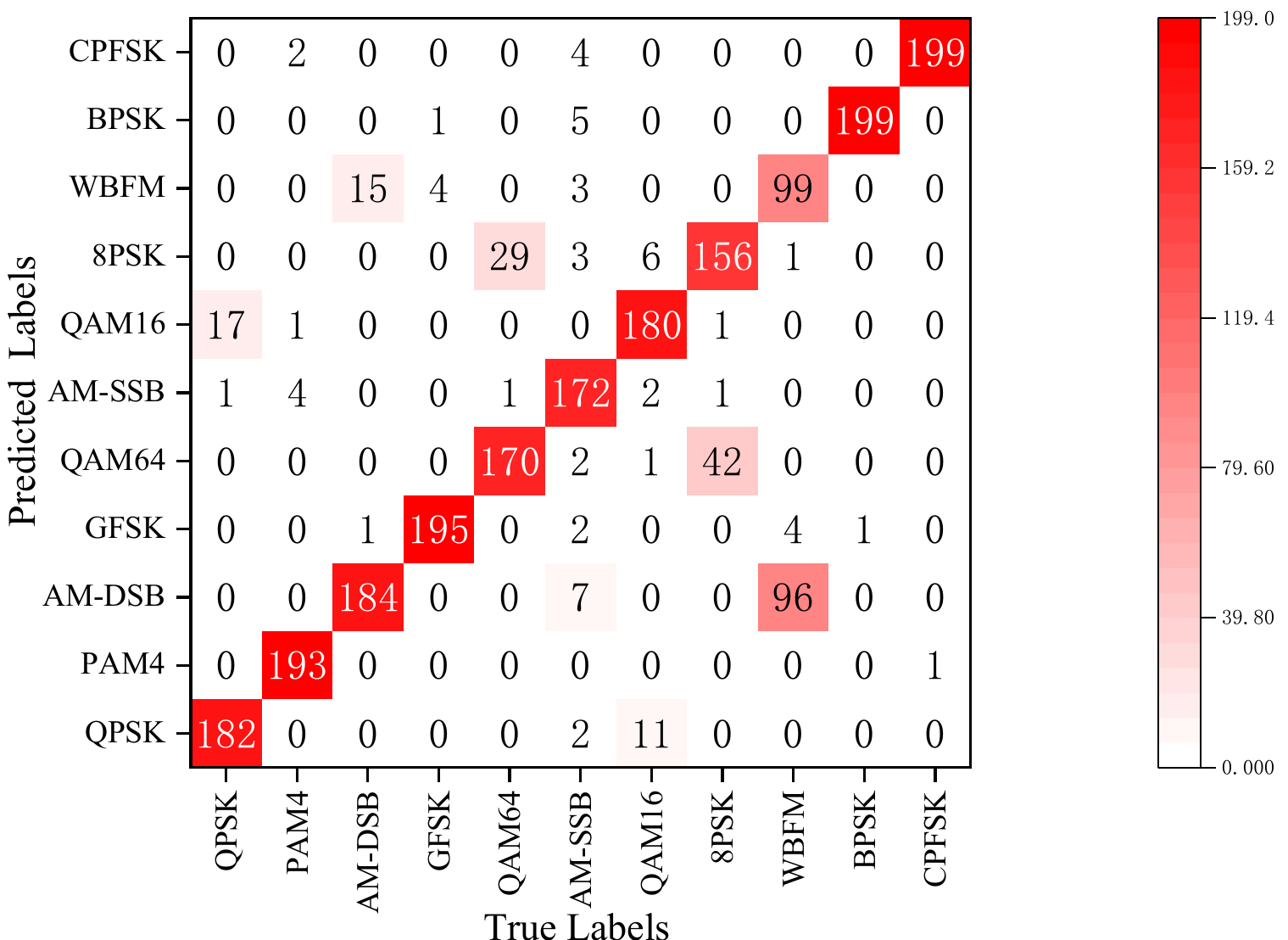}  
        \caption{Transformer confusion matrix at 4 dB SNR.}
    \end{subfigure}%
    \begin{subfigure}{0.3\textwidth}
        \centering
        \includegraphics[width=\linewidth]{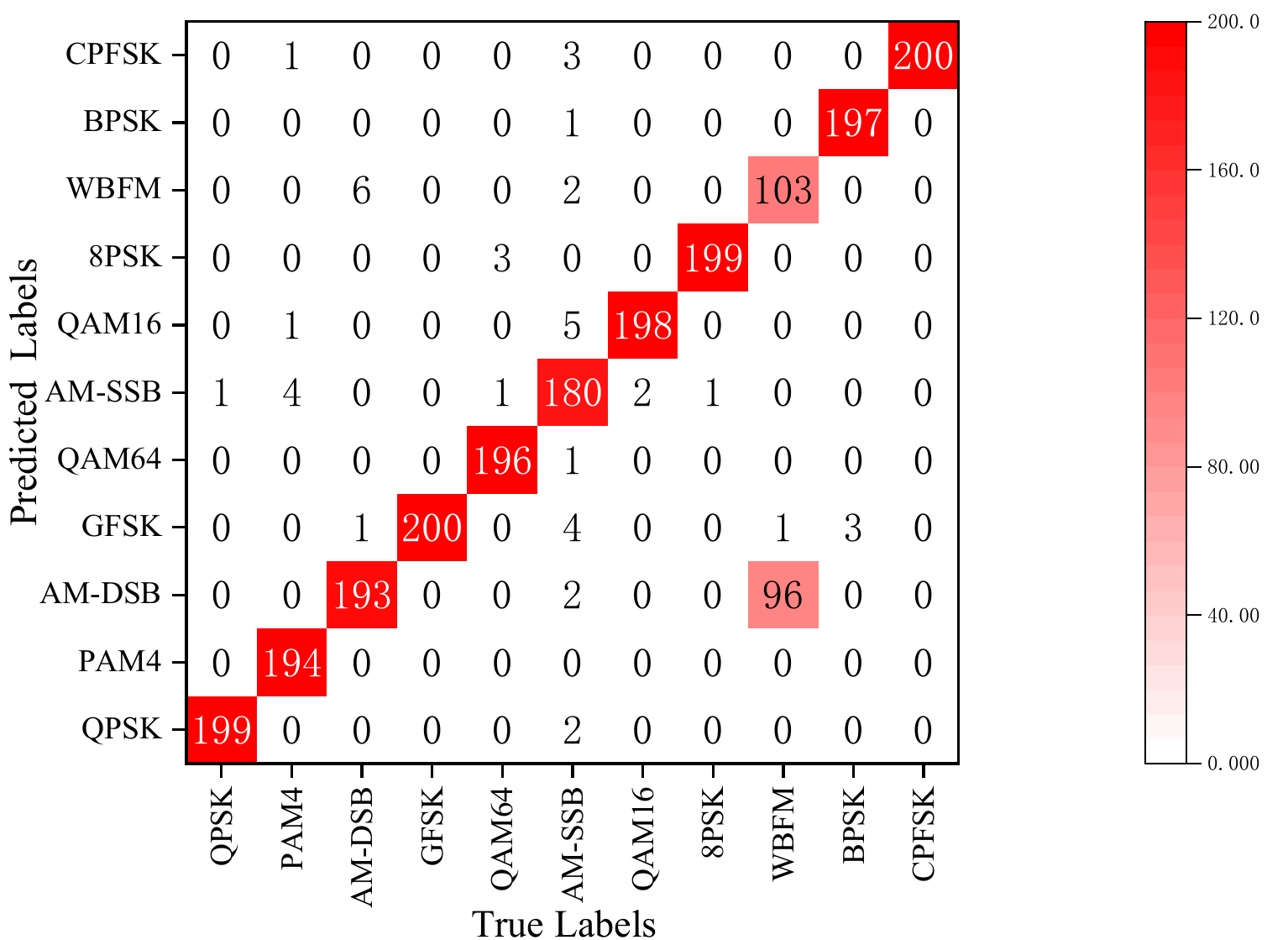}  
        \caption{SpectrumFM confusion matrix at 4 dB SNR.}
    \end{subfigure}
    \caption{The confusion matrices for the AMC\_Net, Transformer, and our proposed model under SNR conditions of -4 dB, 0 dB, and 4 dB.}
    \label{fig:confusion_matrix}
\end{figure*}

The confusion matrices for our proposed model and the two most competitive baseline models, namely  AMC\_Net and Transformer under SNR conditions of -4 dB, 0 dB, and 4 dB are shown in Fig. \ref{fig:confusion_matrix}, revealing interesting insights into their classification performance. At low SNR (-4 dB), all three models exhibit relatively poor classification performance, with significant confusion between modulation types such as AM-DSB, 16-QAM, and QPSK. However, it is clear that the proposed method achieves better diagonal clarity and fewer off-diagonal errors compared to AMC\_Net and Transformer, indicating stronger noise robustness.
As the SNR increases to 0 dB, the classification performance of all methods improves significantly. Notably, the confusion matrices of the Transformer and the proposed method show a clear diagonal trend with reduced misclassification. The proposed method consistently outperforms the other two, with higher correct classification counts and fewer false predictions, especially for easily confused categories such as QPSK and 64-QAM.
At high SNR (4 dB), the performance gap becomes more evident. The proposed method achieves near-perfect classification accuracy, with almost all samples correctly classified and the confusion matrix showing a clean diagonal structure. In comparison, AMC\_Net and Transformer still exhibit sporadic misclassification, particularly in closely related classes such as 64-QAM and 8-PSK.

\begin{figure*}[htbp]
    \centering
    \begin{subfigure}{0.3\textwidth}
        \centering
        \includegraphics[width=\linewidth]{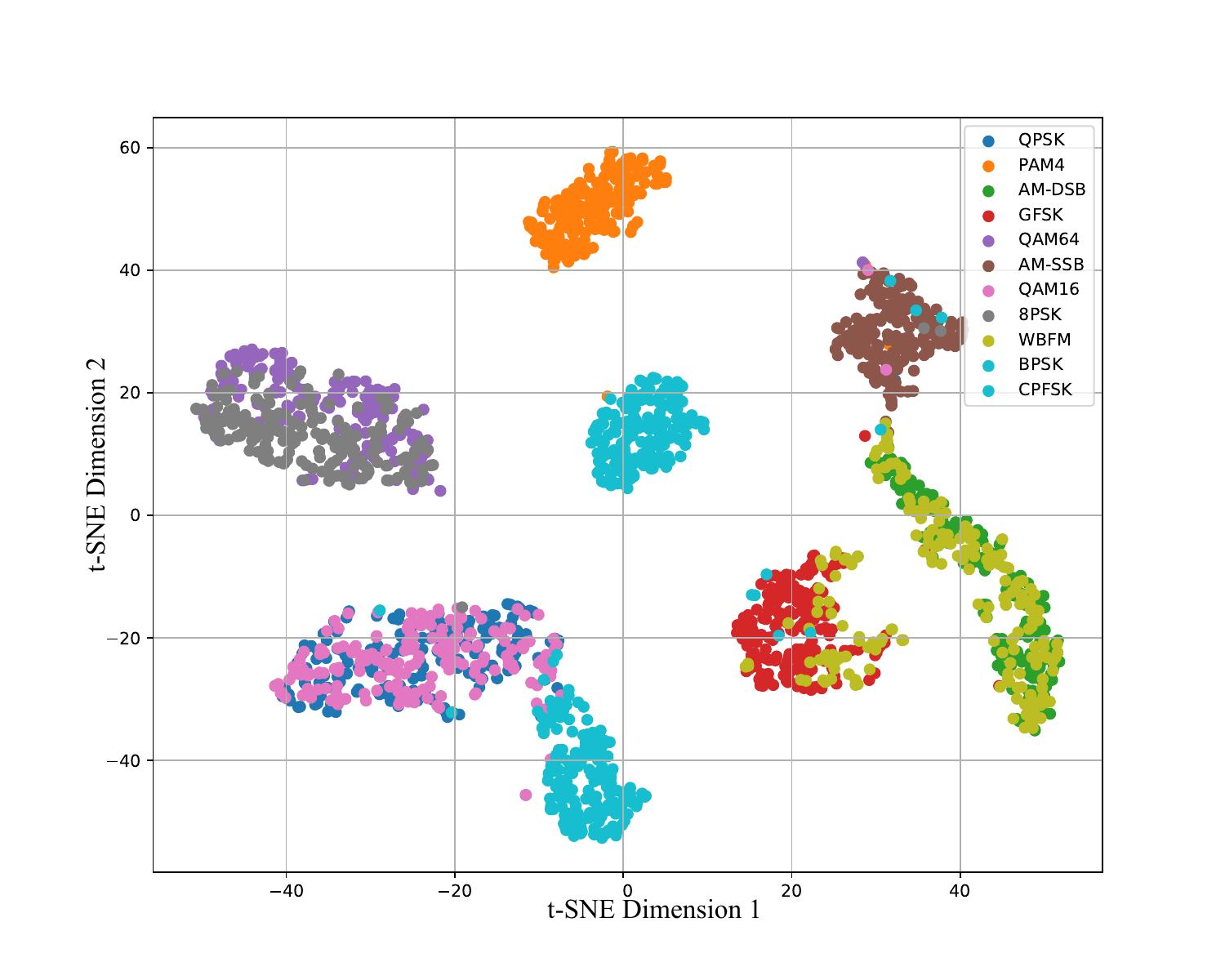}  
        \caption{The t-SNE visualizations of the RML2016.10A dataset for AMC\_Net.} 
        \label{fig:amc_net}
    \end{subfigure}
    \begin{subfigure}{0.3\textwidth}
        \centering
        \includegraphics[width=\linewidth]{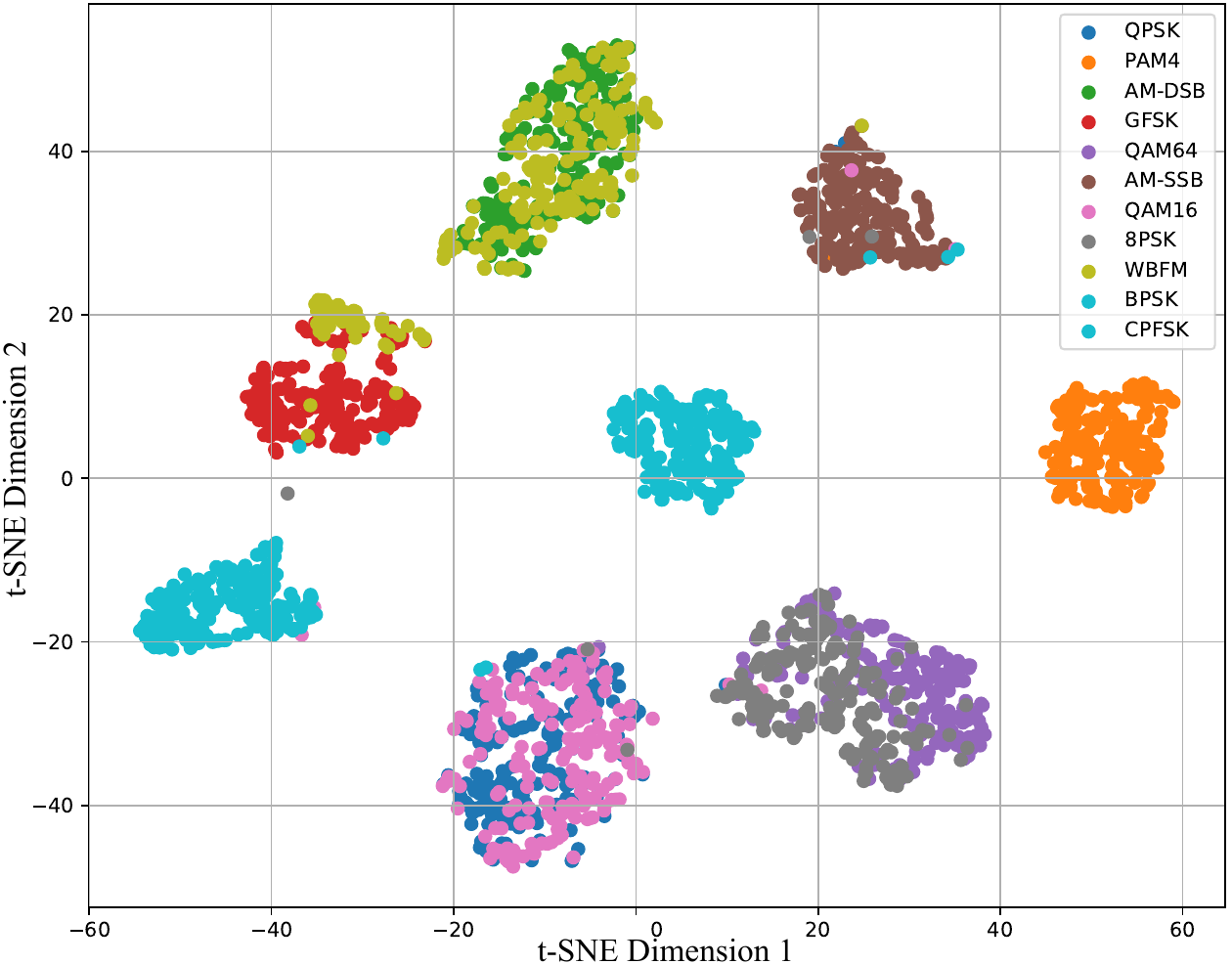}  
        \caption{The t-SNE visualizations of the RML2016.10A dataset for Transformer.}
        \label{fig:transformer}
    \end{subfigure}
    \begin{subfigure}{0.3\textwidth}
        \centering
        \includegraphics[width=\linewidth]{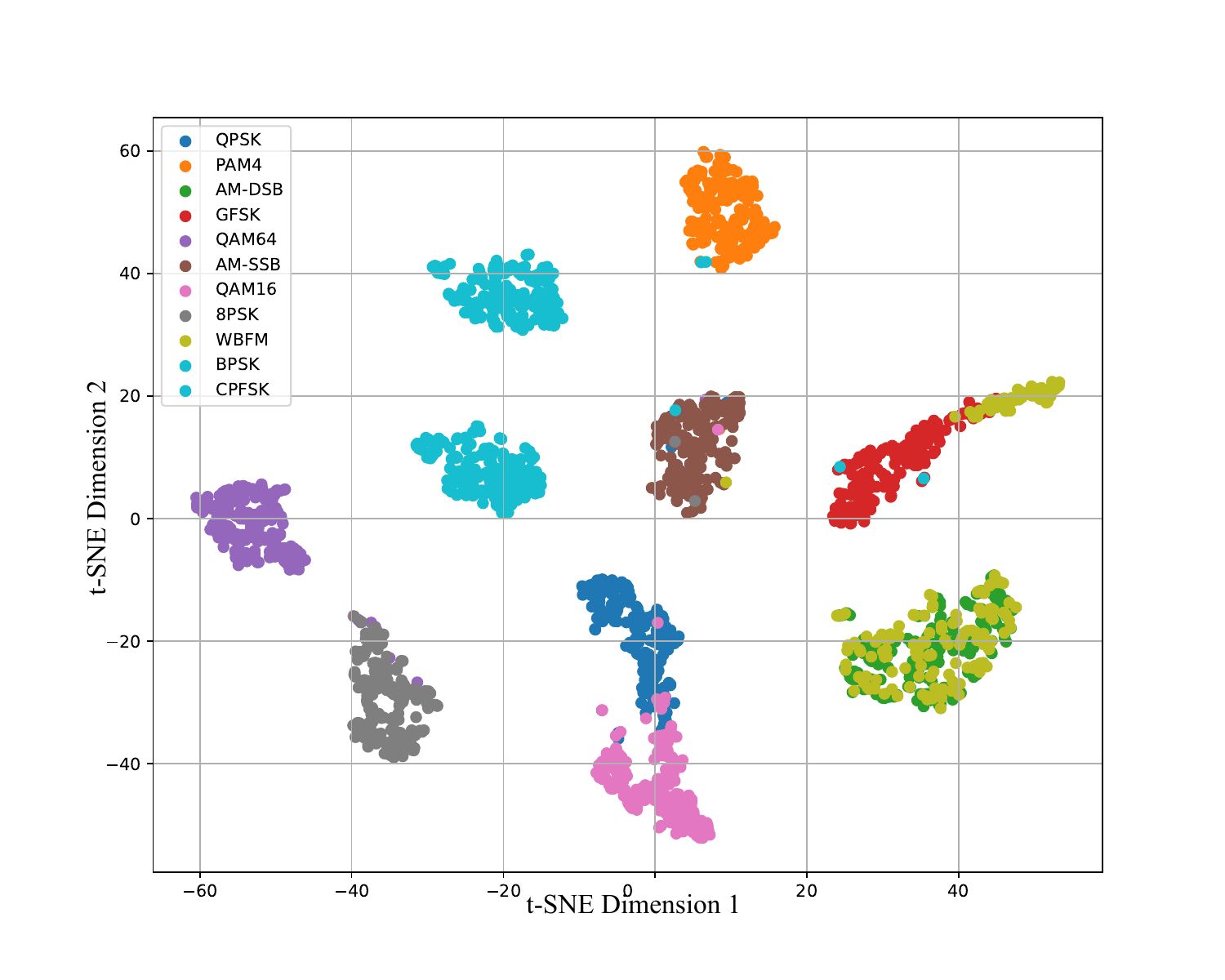}  
        \caption{The t-SNE visualizations of the RML2016.10A dataset for our model.}
        \label{fig:our_method}
    \end{subfigure}
    \caption{The t-SNE visualizations of the RML2016.10A dataset
    under 0 dB SNR condition for our proposed model and two most competitive baseline models.}
    \label{fig:tsne}
\end{figure*}
The t-SNE visualizations of the RML2016.10A dataset under 0 dB SNR condition for our proposed model, AMC\_Net and Transformer is presented in Fig. \ref{fig:tsne}. The t-SNE visualizations reveal notable clustering overlaps for the AMC\_Net model, specifically between 8-PSK and 64-QAM, 16-QAM and QPSK, WBFM and AM-DSB, as well as WBFM and GFSK. The overlaps indicate difficulties in distinguishing the modulation types using AMC\_Net.
Similarly, the Transformer model exhibits significant overlap between 16-QAM and QPSK, 8-PSK and 64-QAM, as well as WBFM and AM-DSB, suggesting similar challenges in classification accuracy for these pairs.
In contrast, our proposed method demonstrates clear boundaries between most modulation classes, with the exception of some overlap between WBFM and AM-DSB. It indicates a higher discriminative capability for our model across various modulation types, thereby showcasing its superior performance compared to the baseline models.
\subsubsection{Performance in Few-shot Conditions}
\begin{figure}[h]
    \centering
    \begin{subfigure}{0.21\textwidth}
        \centering
        \includegraphics[width=\linewidth]{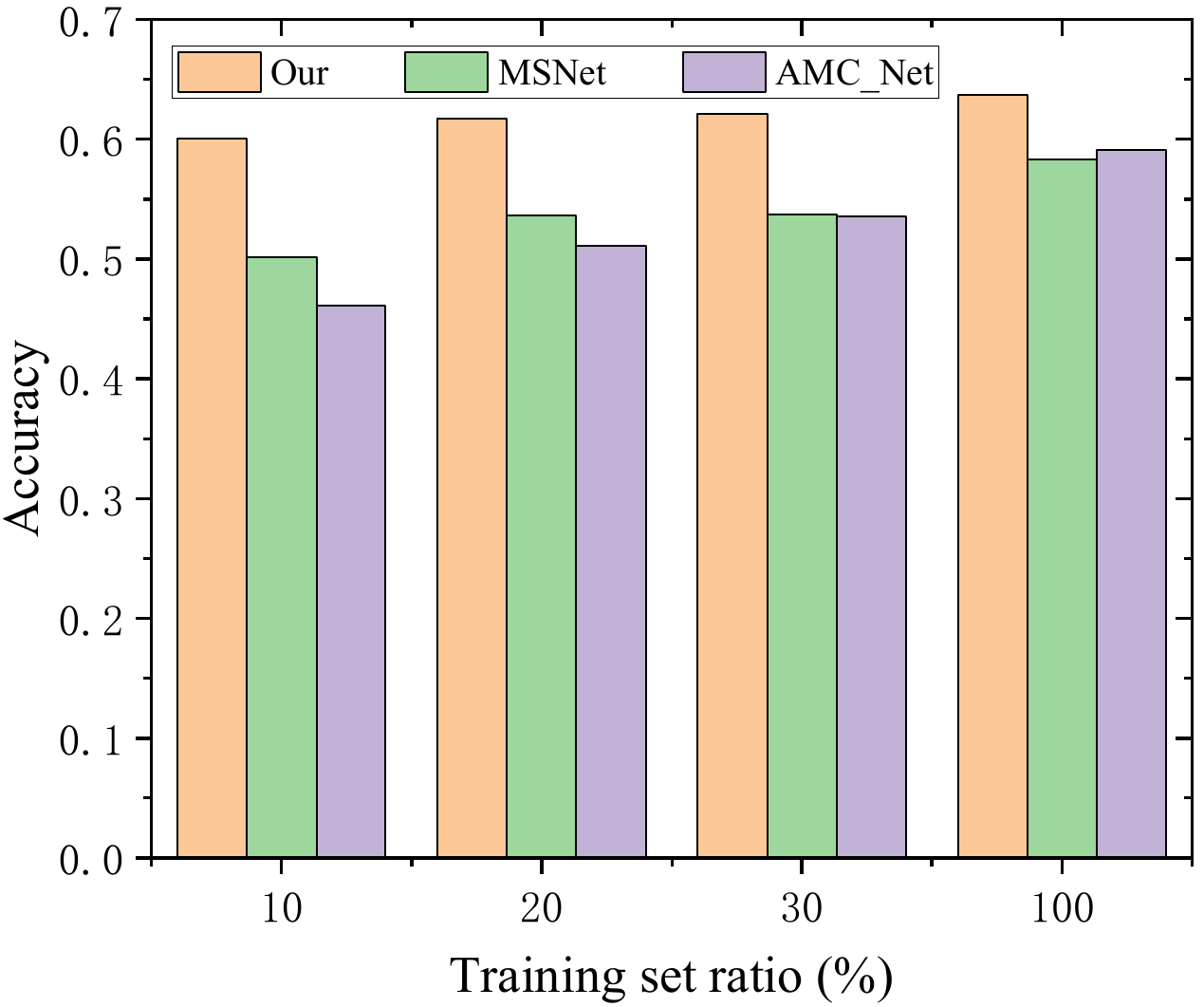}  
        \caption{Performance in few-shot conditions on RML2016.10A dataset.}
        \label{fig:small_sample_amc}
    \end{subfigure}
    \begin{subfigure}{0.21\textwidth}
        \centering
        \includegraphics[width=\linewidth]{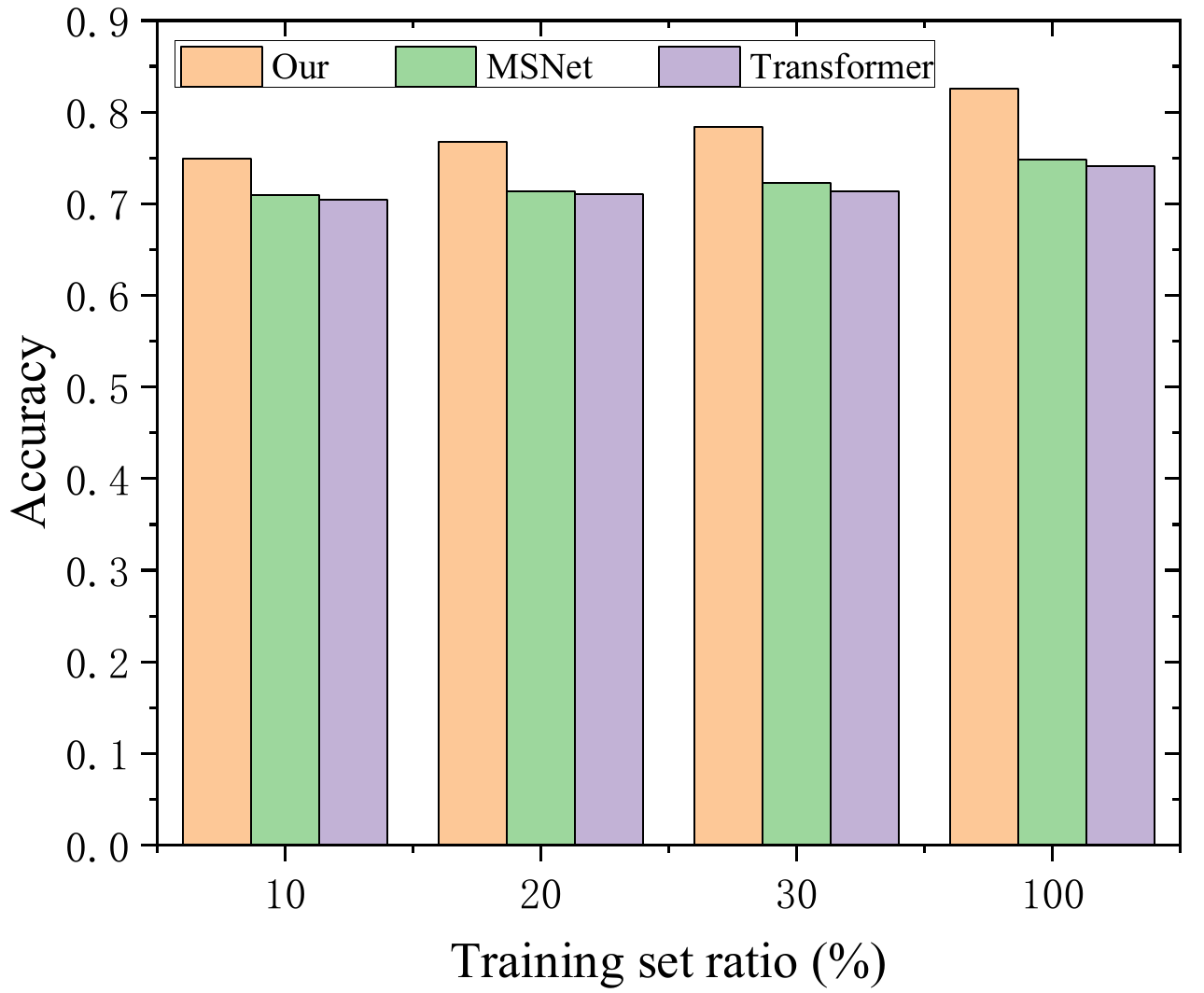}  
        \caption{Performance in few-shot conditions on TechRec dataset.}
        \label{fig:small_sample_wtc}
    \end{subfigure}
    \caption{Performance in few-shot conditions.}
    \label{fig:small_sample}
\end{figure}

\begin{figure}[h]
    \centering
    \begin{subfigure}{0.21\textwidth}
        \centering
        \includegraphics[width=\linewidth]{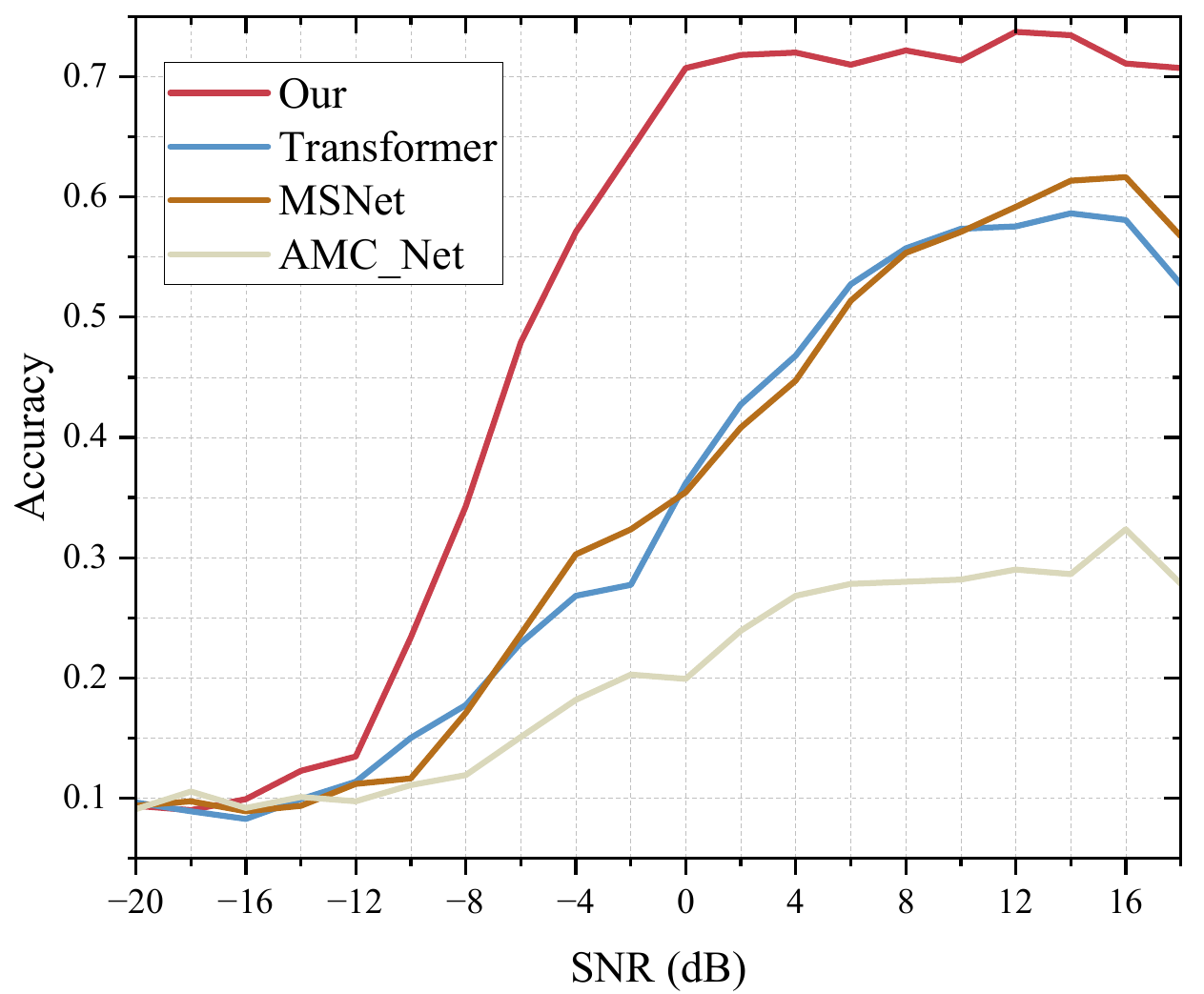}
        \caption{AMC.}
        \label{fig:100_amc}
    \end{subfigure}
    \begin{subfigure}{0.21\textwidth}
        \centering
        \includegraphics[width=\linewidth]{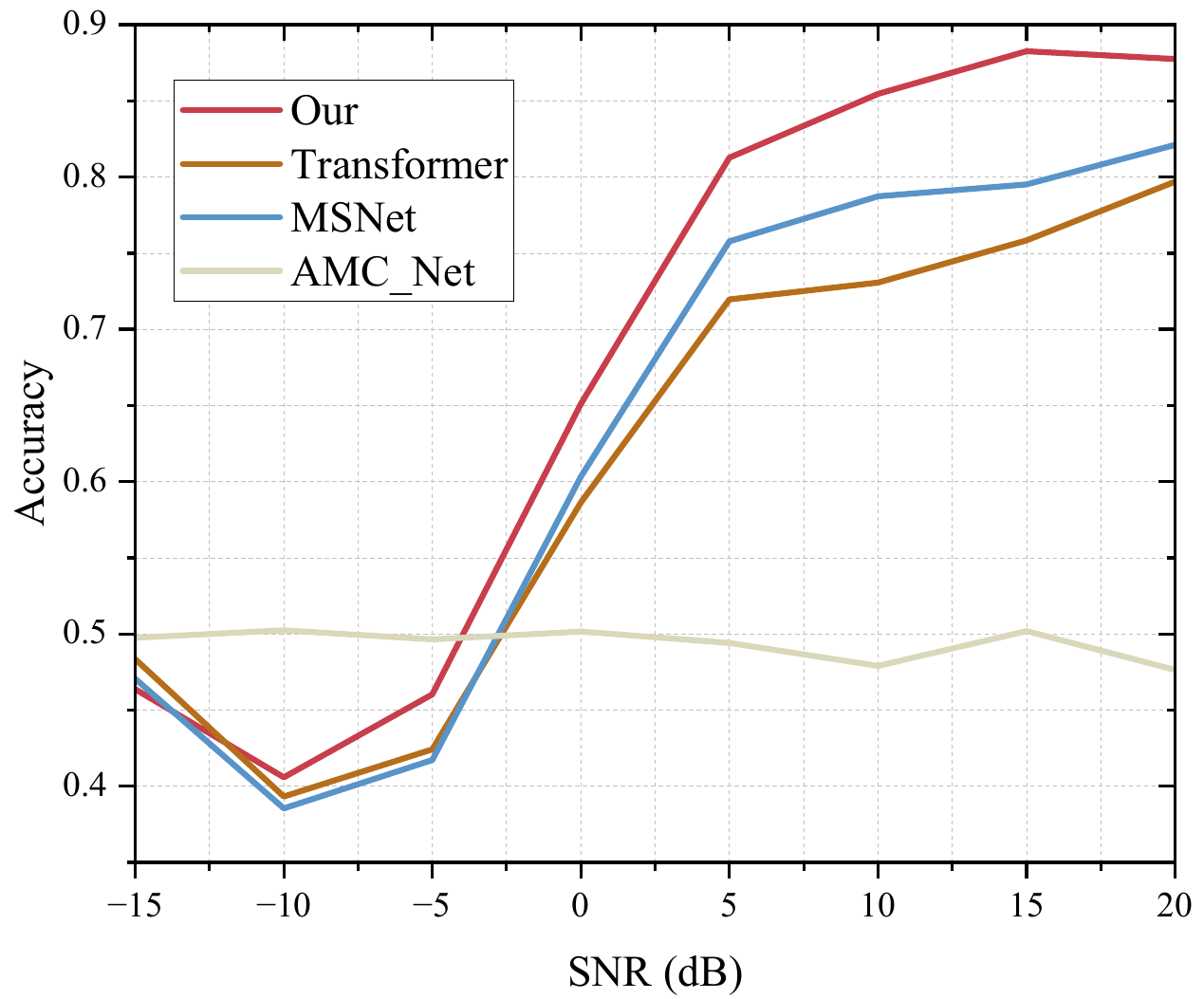}
        \caption{WTC.}
        \label{fig:100_wtc}
    \end{subfigure}
    \caption{Few-shot fine-tuning with 100 samples per class on AMC and WTC tasks in different SNR conditions.}
\end{figure}

The bar chart in Fig. \ref{fig:small_sample_amc} illustrates the classification accuracy of our method, MSNet, and AMC\_Net under different training set ratios (10\%, 20\%, 30\%, and 100\%). The vertical axis represents accuracy, while the horizontal axis denotes the proportion of the training set used for model training. From the results, it is evident that our proposed method consistently outperforms MSNet and AMC\_Net across all training set ratios. Notably, a key observation from the results is that as the training data ratio decreases, the accuracy of MSNet and AMC\_Net drops significantly, whereas our proposed method remains remarkably stable. Even when trained with only 10\% of the dataset, our model achieves an accuracy comparable to MSNet and AMC\_Net trained on 100\% of the data. The results highlight the strong few-shot learning capability of our method. Fig.~\ref{fig:100_amc} illustrates the few-shot fine-tuning performance with 100 samples per class on the AMC task using the RML2016.10A dataset under varying SNR conditions. The results show that our method achieves consistently better accuracy than MSNet, Transformer, and AMC\_Net across all SNR levels, demonstrating its robustness and good transferability in low-data regimes.
\subsubsection{Learning Efficiency}
\begin{figure}[h]
    \centering
    \begin{subfigure}{0.21\textwidth}
        \centering
        \includegraphics[width=\linewidth]{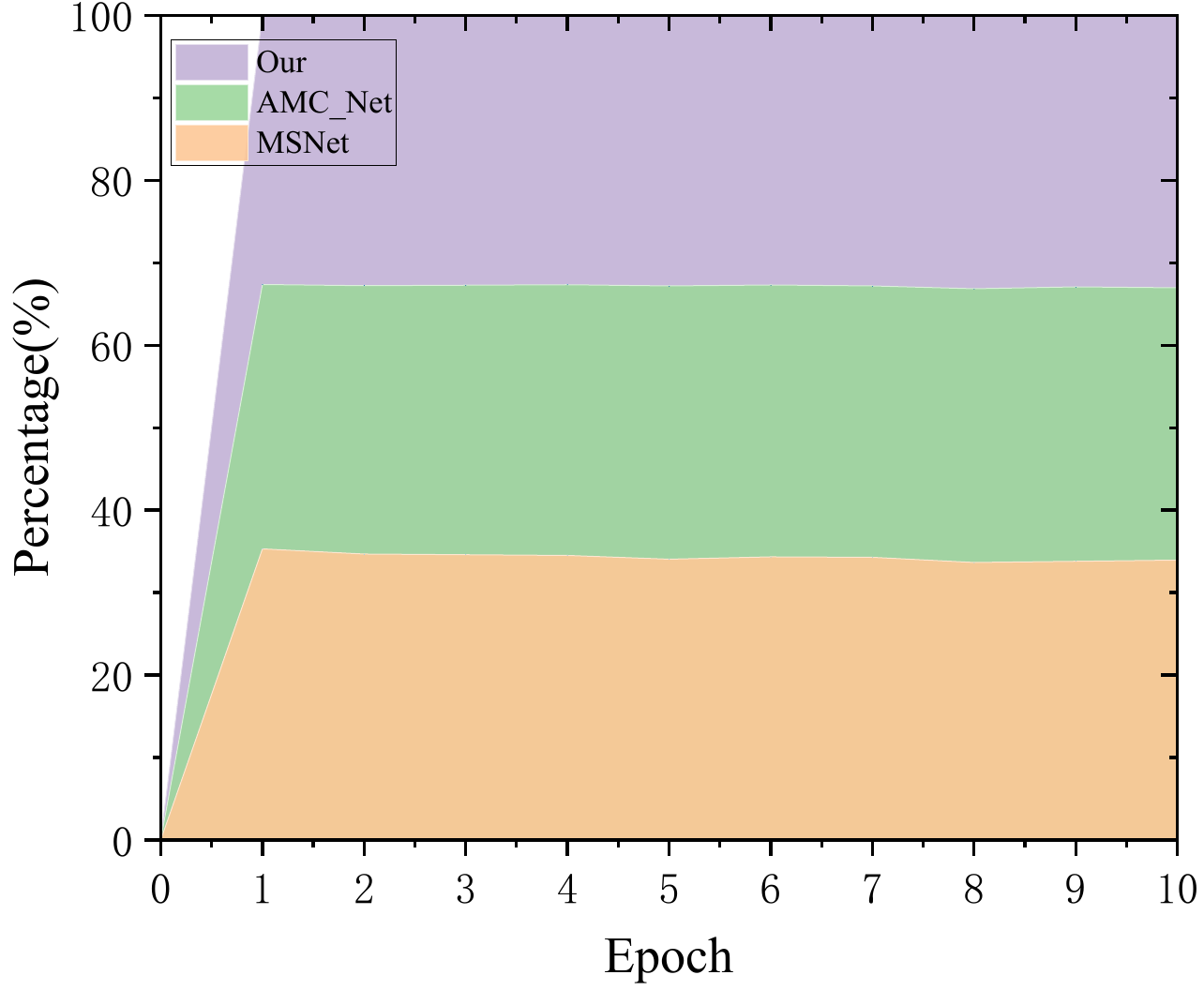}
        \caption{The accuracy curve on RML2016.10A dataset.}
        \label{fig:conv_amc}
    \end{subfigure}
    \begin{subfigure}{0.21\textwidth}
        \centering
        \includegraphics[width=\linewidth]{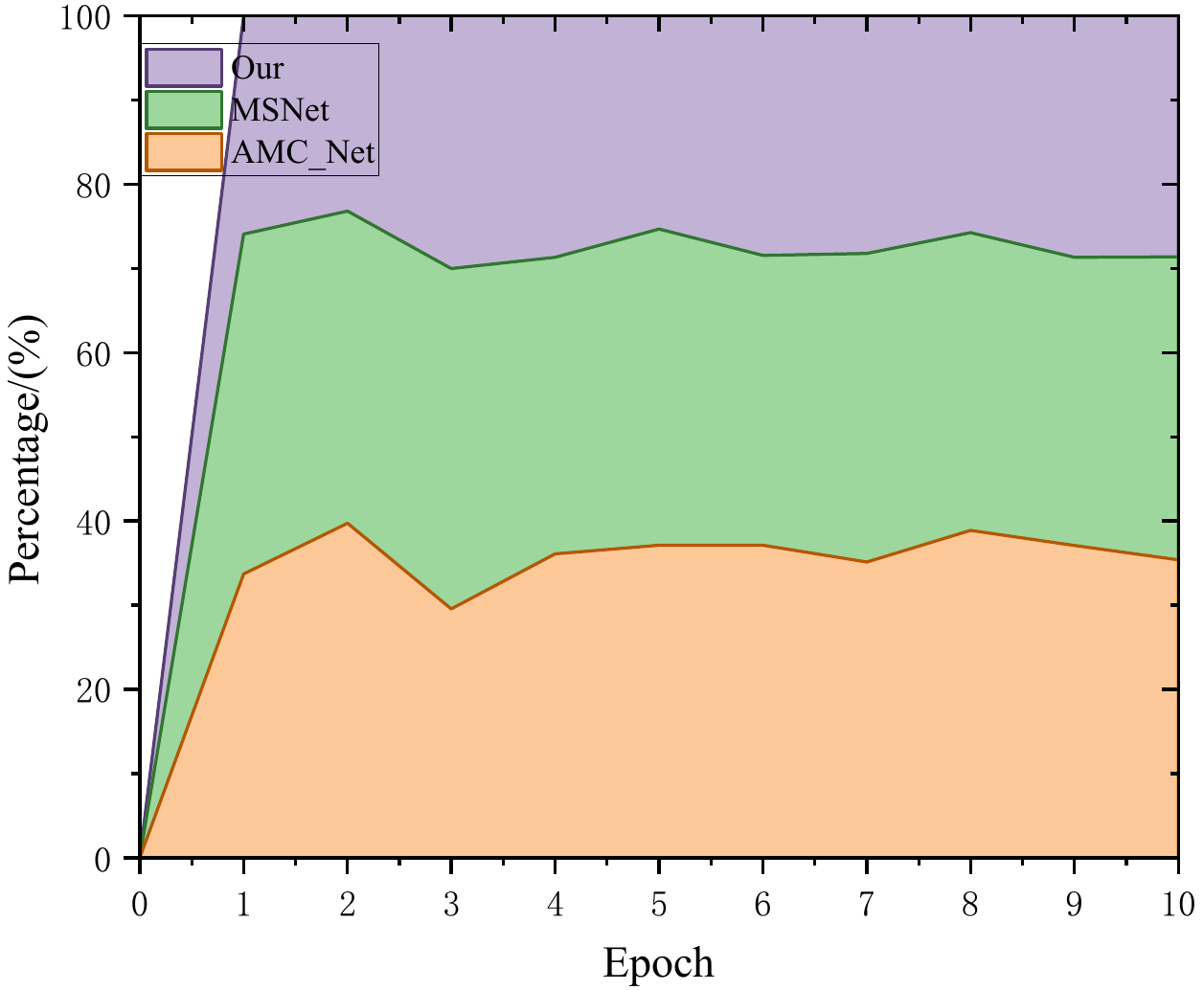}
        \caption{The accuracy curve on TechRec dataset.}
        \label{fig:conv_wtc}
    \end{subfigure}
    \caption{The accuracy curve over epochs.}
    \label{fig:conv}
\end{figure}

To evaluate the learning efficiency of our method, the stacked percentage area chart in Fig. \ref{fig:conv_amc} illustrates the accuracy trends of three models, including our method, MSNet, and AMC\_Net, over ten training epochs. The vertical axis represents accuracy in percentage terms, while the horizontal axis denotes the number of epochs. A key observation from the figure is that our proposed method consistently occupies the largest area in the chart, indicating superior accuracy across all epochs. Furthermore, our method reaches a high accuracy level within the first few epochs, whereas MSNet and AMC\_Net require more epochs to stabilize. The results highlight the exceptional learning efficiency of our model.
\subsection{Wireless Technology Classification Task}
\subsubsection{Dataset}
As detailed in Section \ref{sec:data_collection_and_processing_stage}, the TechRec dataset comprises data captured from various locations including Merelbeke, iGent, and Rabot, which are utilized for pre-training our model. Meanwhile, data collected from the remaining sites, including UZ, Reep, and Gentbrugge, are reserved for fine-tuning and testing, ensuring a robust evaluation of our model performance across different environments.
\subsubsection{Baselines}
We leverage AMC\_Net, Transformer, MSNet, CGDNN, DAE, MCNet, ResNet, VGG, CNN2 and GRU2 described in Section~\ref{sec:baselines} as baseline models.
\subsubsection{Performance Comparison}
\begin{table}[htbp]
    \centering
    \caption{Comparison of Precision, Recall and F1-score in the WTC Task.}
    \label{tab:comparison}
    \begin{tabular}{c|c|c|c}
    \hline
    Models  & Precision & Recall & F1 Score \\  \hline
    ResNet & 0.6723 & 0.6950 & 0.6764 \\ \hline
    VGG & 0.7542 & 0.7404 & 0.7356 \\ \hline
    CNN2 & \underline{0.7851} & 0.7509 & 0.7407 \\ \hline
    AMC\_Net & 0.6492 & 0.6472 & 0.6405 \\ \hline
    DAE & 0.6826 & 0.6735 & 0.6487 \\ \hline
    CGDNN & 0.7815 & 0.7488 & 0.7397 \\ \hline
    MCNet & 0.7210 & 0.7320 & 0.7116 \\ \hline
    GRU2 & 0.7286 & 0.7429 & 0.7225 \\ \hline
    MSNet & 0.7696 & \underline{0.7546} & \underline{0.7481} \\ \hline
    Transformer & 0.7617 & 0.7463 & 0.7415 \\ \hline 
    Our & \textbf{0.8270} & \textbf{0.8253} & \textbf{0.8226} \\ \hline
    \end{tabular}\end{table}
\begin{figure*}[t]
    \centering
    \begin{minipage}[t]{0.23\textwidth}
        \centering
        \includegraphics[width=\linewidth]{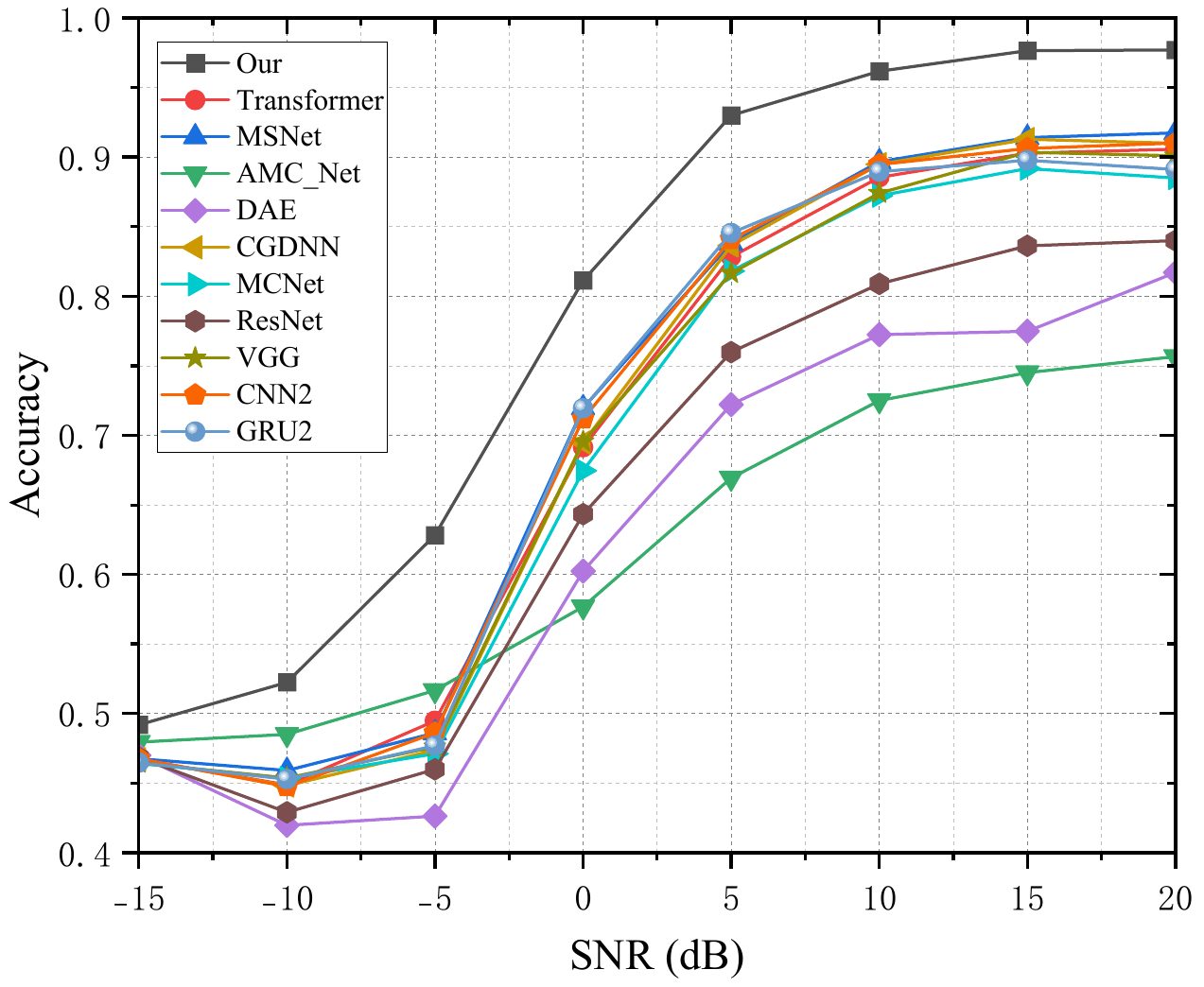}  
        \caption{The accuracy of models on different SNR conditions on the TechRec dataset.}
        \label{fig:accuracy}  
    \end{minipage}
    \begin{minipage}[t]{0.23\textwidth}
        \centering
        \includegraphics[width=\linewidth]{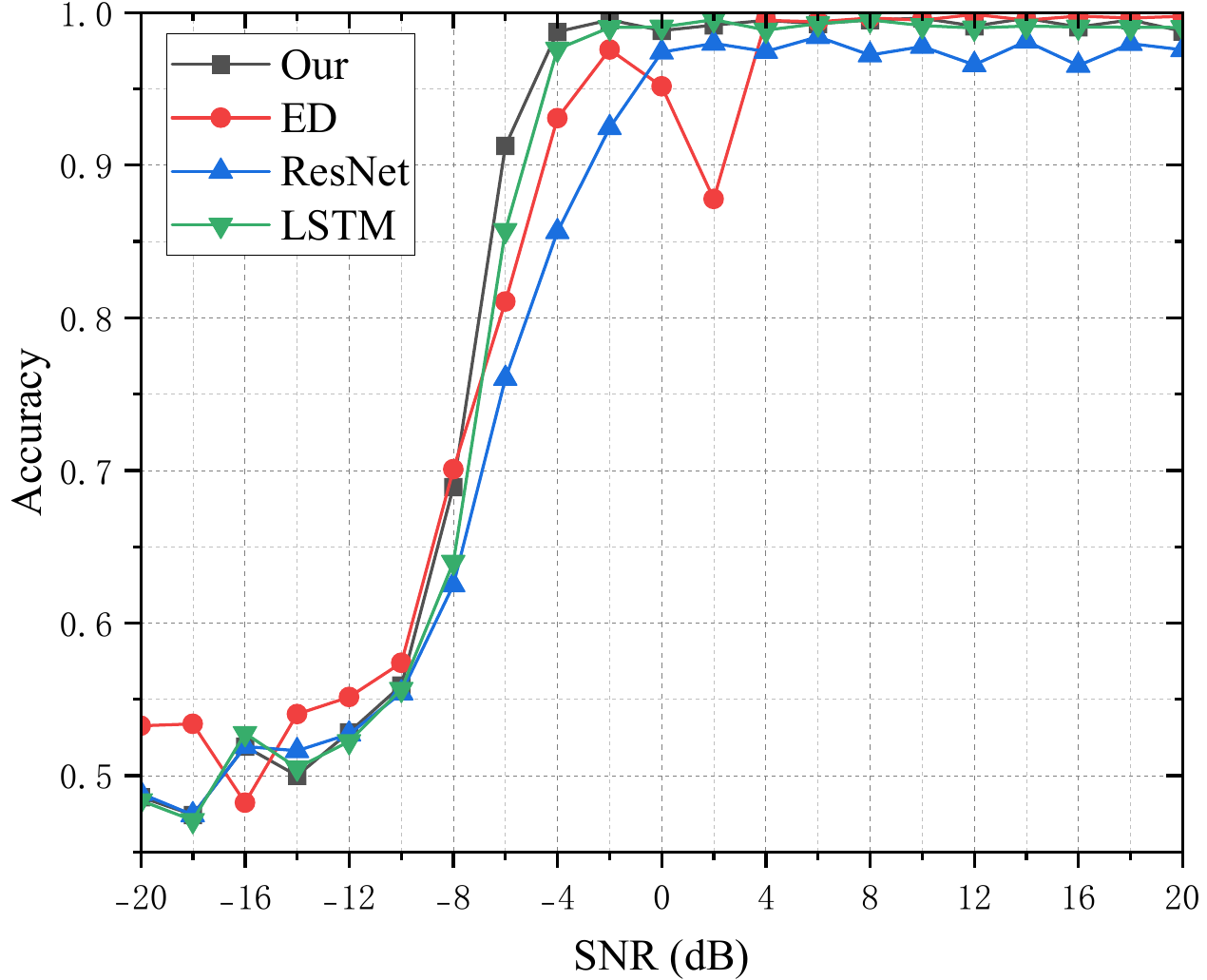}  
        \caption{Sensing accuracy of our model and baseline models at various SNR level.}
        \label{fig:snr_ss}  
    \end{minipage}
    \begin{minipage}[t]{0.23\textwidth}
        \centering
        \includegraphics[width=\linewidth, height=1.35in]{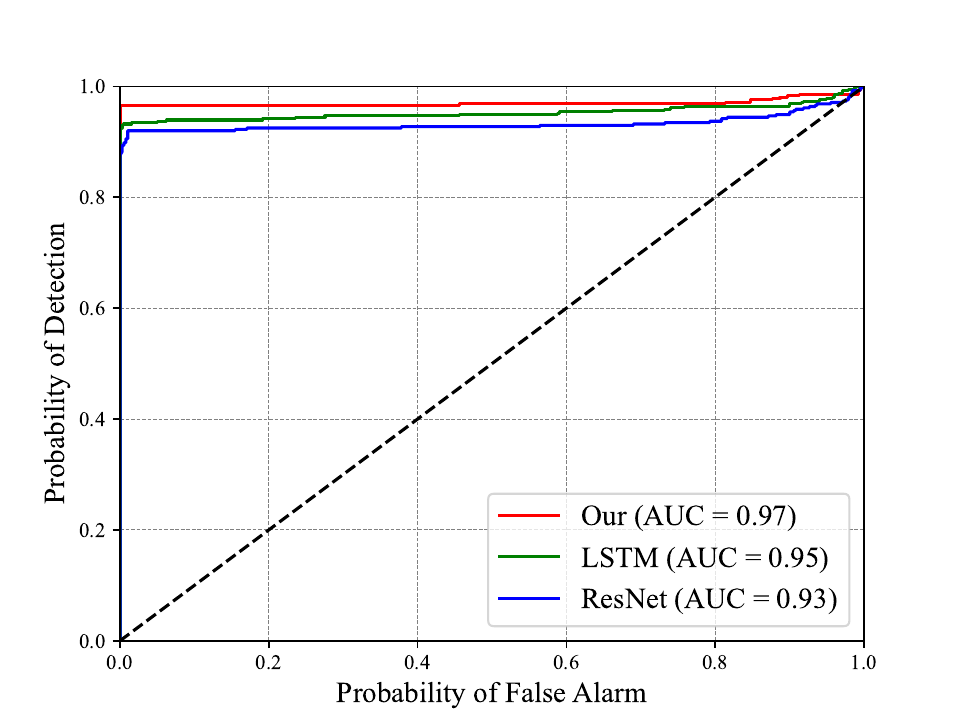}
        \caption{ROC curve of various spectrum sensing method at -4 dB SNR level.}
        \label{fig:ROC_curve_sensing_-4db}
    \end{minipage}
    \begin{minipage}[t]{0.23\textwidth}
        \centering
        \includegraphics[width=\linewidth, height=1.35in]{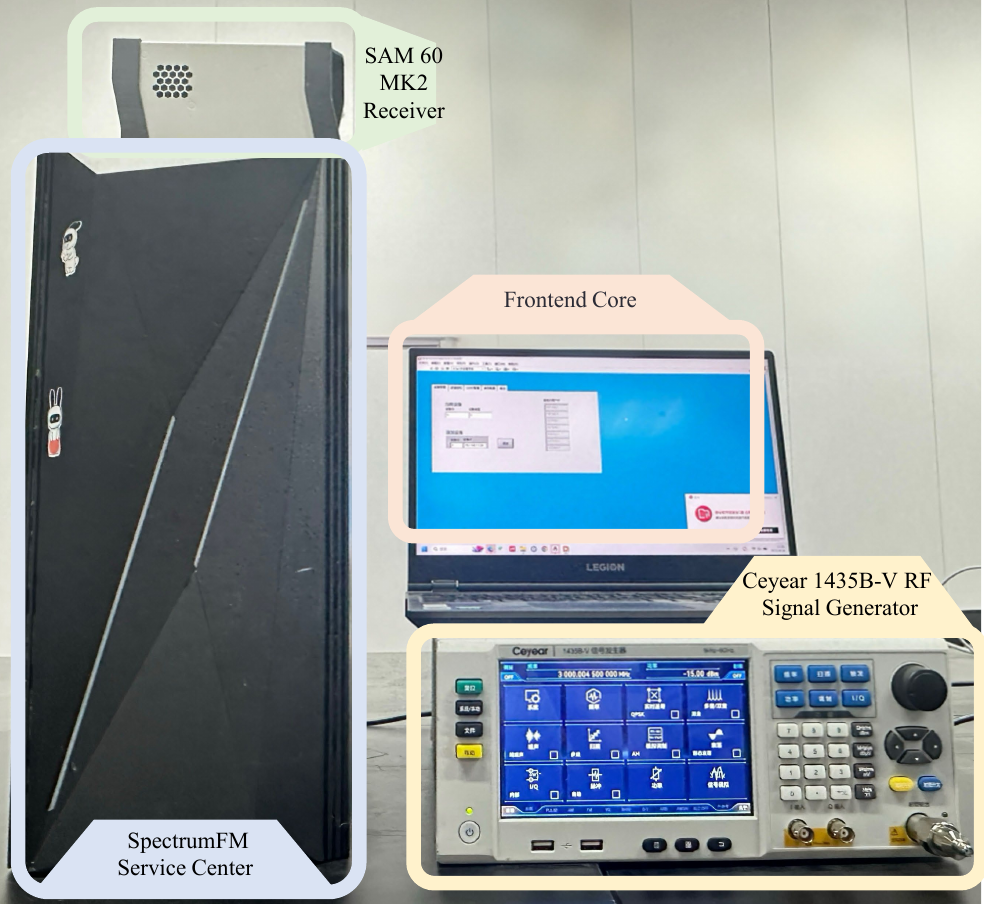}
        \caption{The data acquisition and SpectrumFM deployment platform.}
        \label{fig:pingtai}
    \end{minipage}
\end{figure*}
The experimental results presented in TABLE \ref{tab:comparison} demonstrate the superior performance of our proposed model compared to existing methods in the WTC task. Specifically, our model achieves the highest precision (0.8270), recall (0.8253), and F1-score (0.8226), significantly outperforming the baseline methods. Notably, compared to well-established architectures such as ResNet and VGG, our model exhibits substantial improvements, particularly in Recall and F1-score, indicating its enhanced ability to correctly classify instances while maintaining robustness against misclassification. 

The accuracy of various models under different SNR conditions is illustrated in Fig. \ref{fig:accuracy}. The figure demonstrates that our proposed method achieves superior performance across a wide range of SNR levels. Specifically, our method consistently outperforms other methods. At SNR levels from -15 dB to 0 dB, our method exhibits a higher accuracy rate compared to the other models. As the SNR increases beyond 0 dB, all methods show improved accuracy, but our method maintains a notable lead. The robust performance underscores the effectiveness of our method in both challenging and favorable SNR conditions.

As shown in Fig.~\ref{fig:small_sample_wtc}, our proposed method consistently outperforms MSNet and AMC\_Net across all training set ratios (10\%, 20\%, 30\%, and 100\%). Notably, the model exhibits strong stability as the training data ratio decreases, maintaining high accuracy even under limited data conditions. Furthermore, Fig.~\ref{fig:100_wtc} demonstrates that the model generalizes well to unseen data with 100 samples per class, highlighting the effective few-shot learning capability and reliable generalization of our model.

As evidenced in the percentage area stacked plot in Fig.~\ref{fig:conv_wtc}, our proposed model not only converges more rapidly but also achieves a higher accuracy rate across all epochs, demonstrating its superior learning efficiency compared to AMC\_Net and MSNet.
\subsection{Spectrum Sensing Task}
\subsubsection{Dataset}
The RML2018.01A dataset is a widely used benchmark for AMC. It consists of IQ signal samples representing various digital and analog modulation schemes under different SNR conditions. In this study, pure noise is added to the dataset to enhance its applicability for spectrum sensing task.
\subsubsection{Baselines}
We leverage energy detection (ED), along with the ResNet and LSTM models mentioned in Section~\ref{sec:baselines}, as baseline methods. Energy detection is a fundamental spectrum sensing method that determines the presence or absence of a signal by comparing the received signal energy to a predefined threshold. Due to its low computational complexity and independence from prior signal knowledge, it serves as a widely used benchmark in spectrum sensing research.
   
\subsubsection{Performance Comparison}
Fig. \ref{fig:snr_ss} illustrates the performance of different spectrum sensing methods at various SNR levels. All methodologies exhibit poor performance when the SNR is between -10 dB and -4 dB. However, within the range of SNR from -10 dB to 0 dB, our proposed method demonstrates a higher accuracy rate. Beyond 0 dB, all methodologies achieve nearly perfect sensing accuracy.

Fig. \ref{fig:ROC_curve_sensing_-4db} shows the  receiver operating characteristic (ROC) curve of various spectrum sensing method at -4 dB SNR level. It highlights that our method achieves an AUC of 0.97 at -4 dB SNR, surpassing both LSTM and ResNet, demonstrating its superior accuracy in signal detection. At the same false alarm rate, our method also exhibits better detection performance, further underscoring its effectiveness in the SS task.
\subsection{Anomaly Detection Task}
\begin{figure*}[htbp]
    \centering
    \begin{subfigure}{0.24\textwidth}
        \includegraphics[width=\linewidth]{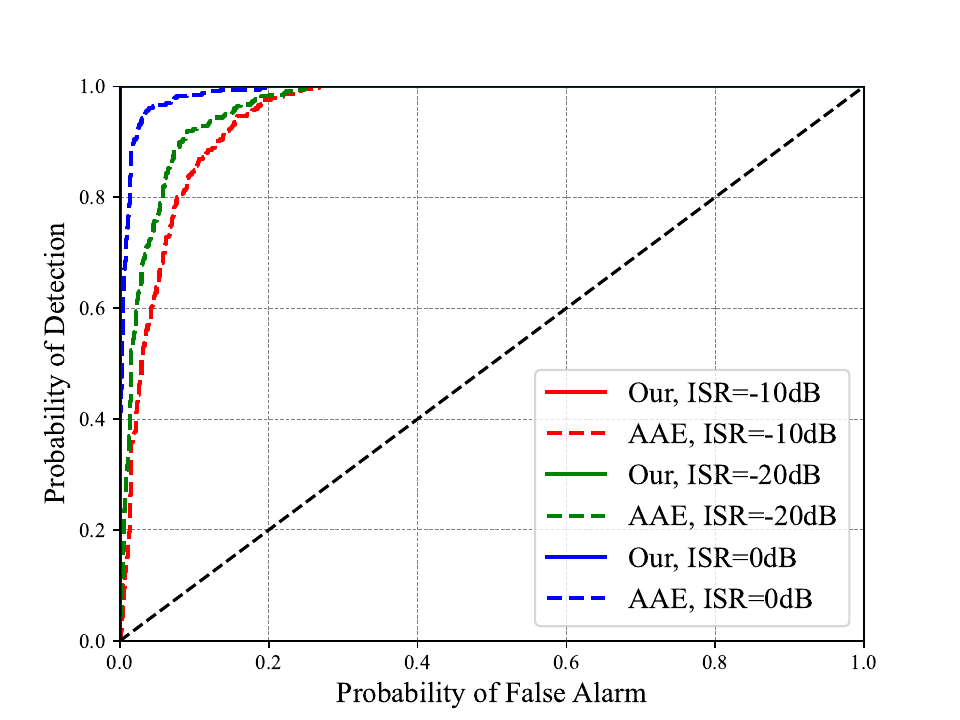}
        \caption{ROC curve for single-tone interference.}
        \label{fig:ROC_curve_singletone}
    \end{subfigure}
    \begin{subfigure}{0.24\textwidth}
        \includegraphics[width=\linewidth]{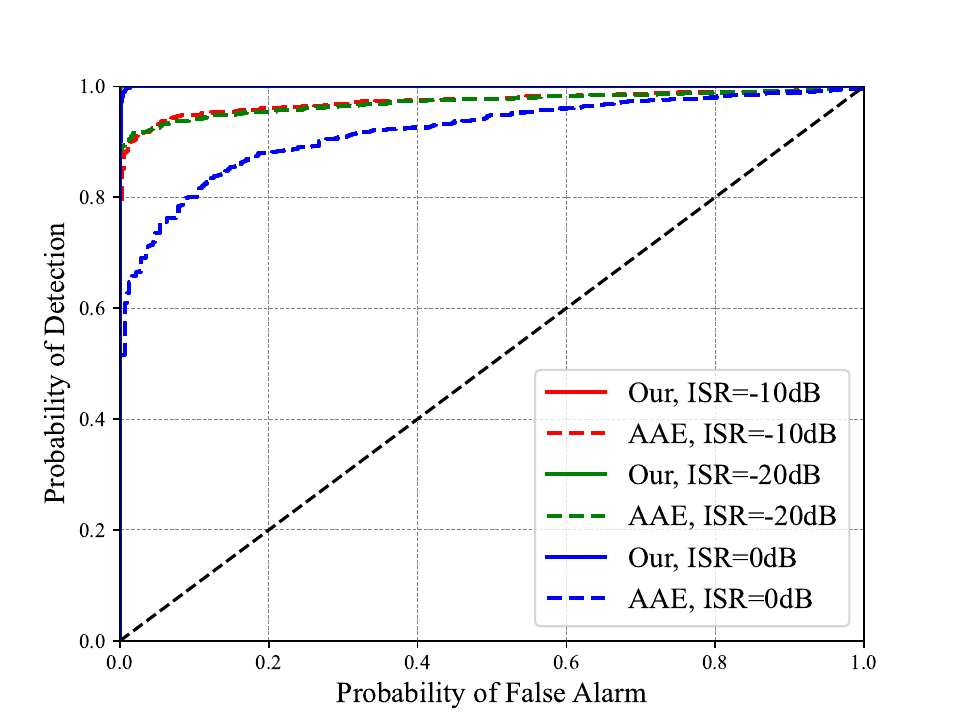}
        \caption{ROC curve for multi-tone interference.}
        \label{fig:ROC_curve_multitone}
    \end{subfigure}
    \begin{subfigure}{0.24\textwidth}
        \includegraphics[width=\linewidth]{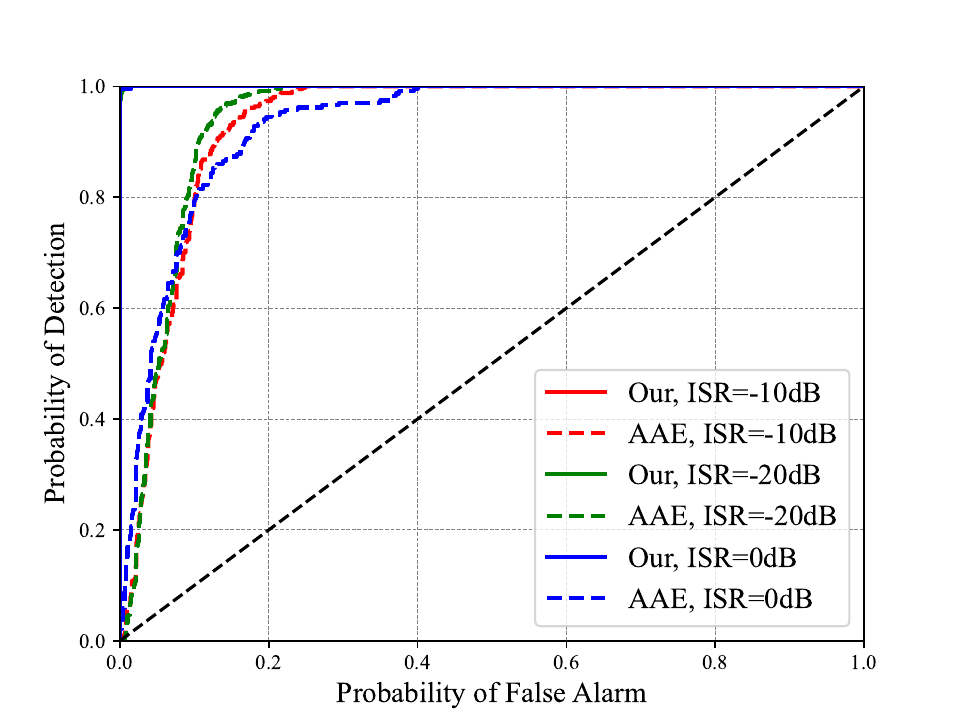}
        \caption{ROC curve for 10M aliased-signal interference.}
        \label{fig:ROC_curve_10M}
    \end{subfigure}
    \begin{subfigure}{0.24\textwidth}
        \includegraphics[width=\linewidth]{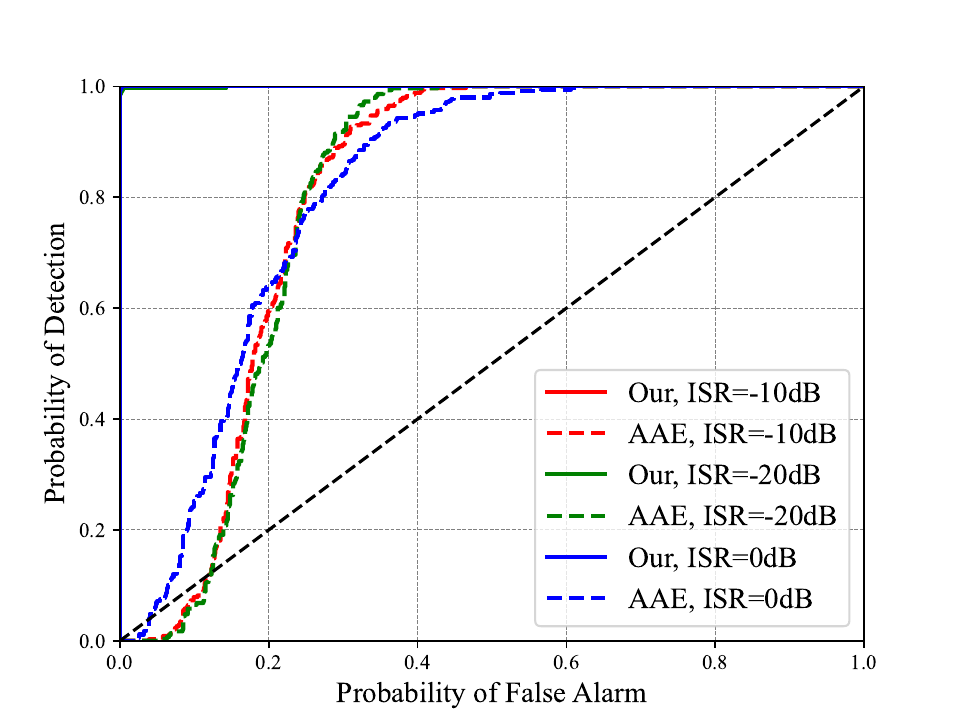}
        \caption{ROC curve for 30M aliased-signal interference.}
        \label{fig:ROC_curve_30M}
    \end{subfigure}
    \caption{Comparison of anomaly detection performance among various types of interference.}
    \label{fig:ROC_curve}
\end{figure*}
\subsubsection{Datasets}
The dataset used for AD task is collected over a two-day period using an over-the-air (OTA) platform, shown in Fig. \ref{fig:pingtai}, which is designed to simulate real-world wireless communication environments. A ceyear 1435B-V RF signal generator is employed to transmit the PU signal, which is modulated using QPSK with a carrier frequency of 2.9 GHz and a bandwidth of 20 MHz. To introduce interference, a ceyear 1465B-V RF signal generator is positioned approximately two meters away, generating abnormal signals intended to disrupt the normal transmission. A SAM 60 MK2 receiver, placed ten meters from both transmitters, is used to capture the received signals.
To replicate realistic interference scenarios, three types of anomalies are introduced. The single-tone interference consisted of a sinusoidal signal sweeping across 2.89 GHz to 2.91 GHz, while the multi-tone interference involved multiple sinusoids sweeping within the same frequency range. The aliased-signal interference is created by generating random signals with bandwidths of 10 MHz and 30 MHz, appearing at 2.885 GHz to 2.915 GHz, thereby causing aliasing effects with the normal QPSK-modulated signal. To further increase the complexity of the aliased interference, these random signals are also QPSK-modulated, making them more challenging to distinguish from the normal transmission.
\subsubsection{Baseline}
We leverage adversarial autoencoder (AAE) \cite{aae} as the baseline model for AD task. The AAE is a generative model that integrates an autoencoder with a discriminator to enforce a specific feature distribution in the latent space. In AAE, the autoencoder consists of an encoder-decoder pair, where the encoder maps input data to a latent representation, and the decoder reconstructs the original input from this representation. Furthermore, a discriminator is employed to guide the encoder in shaping the latent space to follow a predefined distribution, such as a Gaussian or other structured priors. 
\subsubsection{Performance Comparison}
Fig. \ref{fig:ROC_curve} presents the ROC curves for four types of interference under varying interference-to-signal ratios (ISR), comparing the performance of our proposed method with the AAE. Across all scenarios, our method consistently achieves an AUC of 1.0 at -20 dB and -10 dB ISR, showcasing an excellent balance between detection probability and false alarm rate.  Even at 0 dB ISR, our method maintains near-optimal performance, with AUC values significantly outperforming those of AAE, indicating superior robustness and detection capability. Moreover, in the case of aliased-signal interference, the baseline AAE method exhibits a significant decline in performance. The aliased-signal interference is particularly challenging to detect due to its close resemblance to normal signals. Specifically, for 30 MHz aliased-signal interference at 0 dB ISR, the AAE AUC decreases to 0.84, markedly lower than its performance on other less complex interference types. In contrast, our proposed method consistently achieves an AUC of 1.0 across all ISR for both 10 MHz and 30 MHz aliased-signal interferences.  All the findings underscore the robustness, high detection rate, and strong generalization ability of our model.
\subsection{Joint Spectrum Sensing and Automatic Modulation Classification Task}
SpectrumFM is further evaluated in a unified classification setting that combines spectrum sensing and automatic modulation classification~\cite{xing2024joint}. 
The joint task is formulated by augmenting the modulation label space with an additional class representing the idle state, 
so that the model directly predicts either ``idle'' or one of the active modulation types. 
This formulation allows spectrum occupancy and modulation type to be inferred simultaneously with a single classifier. 
As shown in Fig.~\ref{fig:sen_amc}, SpectrumFM achieves a competitive performance across various SNR levels, outperforming baseline methods in most regimes, particularly at moderate to high SNRs. The confusion matrix at 0 dB (Fig.~\ref{fig:sen_amc_confu}) shows that SpectrumFM effectively distinguishes between the idle state and active modulation types, including OOK, 4ASK, and 8ASK, with low misclassification rates. These results demonstrate the feasibility and effectiveness of the unified classification framework for simultaneous spectrum sensing and modulation recognition.

\subsection{Ablation Study}
\begin{figure*}[htbp]
    \centering
    \begin{minipage}[t]{0.23\textwidth}
        \centering
        \includegraphics[width=\linewidth]{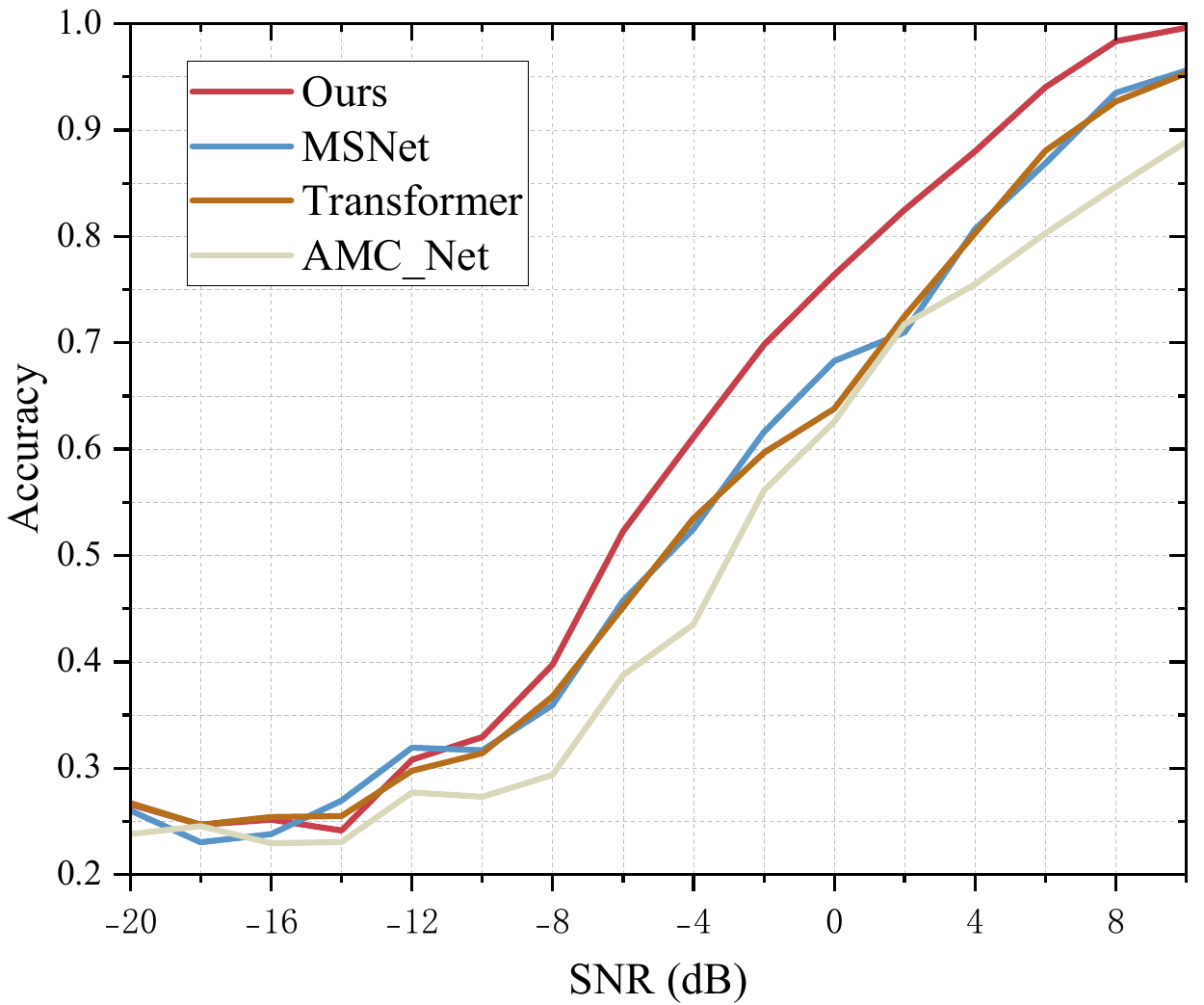}
        \caption{Accuracy of the joint spectrum sensing and automatic modulation classification task.}
        \label{fig:sen_amc}
    \end{minipage}
    \begin{minipage}[t]{0.23\textwidth}
        \centering
        \includegraphics[width=\linewidth]{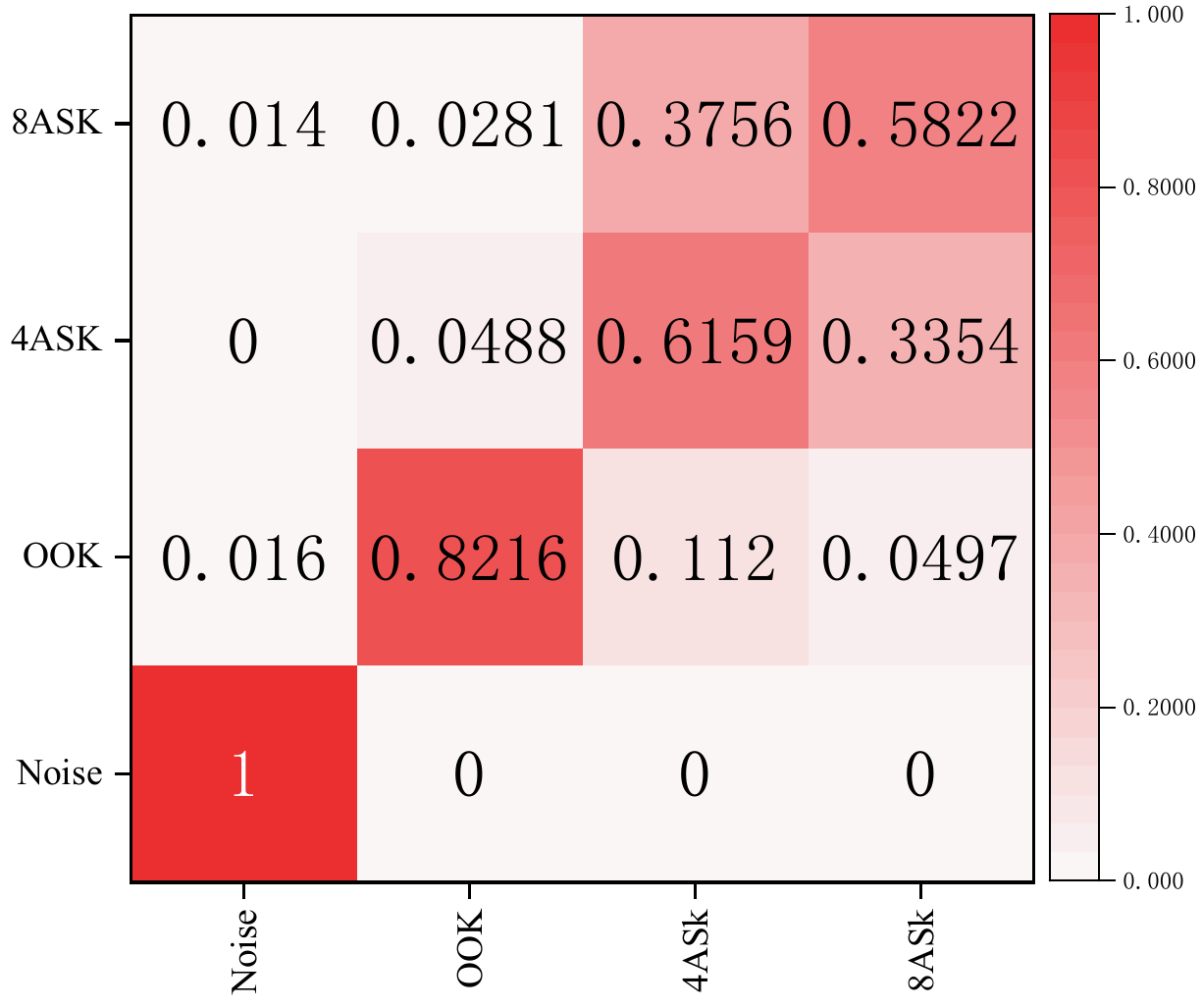}
        \caption{Confusion matrix of SpectrumFM for the joint task at 0 dB SNR.}
        \label{fig:sen_amc_confu}
    \end{minipage}
    \begin{minipage}[t]{0.23\textwidth}
        \centering
        \includegraphics[width=\linewidth]{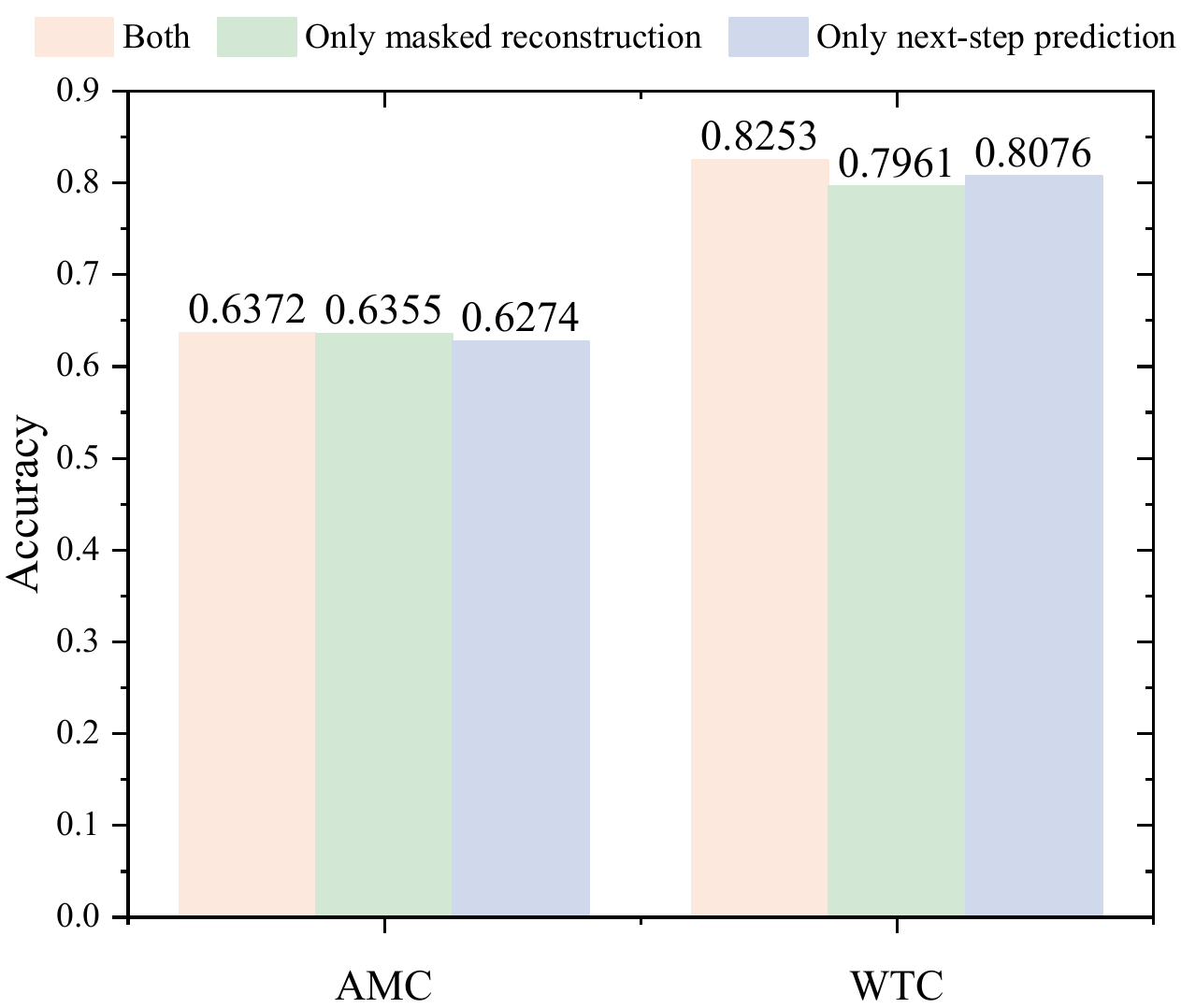}
        \caption{Ablation study on the joint optimization of the masked reconstruction loss and the next-slot signal prediction loss.}
        \label{fig:ablation}
    \end{minipage}
    \begin{minipage}[t]{0.23\textwidth}
        \centering
        \includegraphics[width=\linewidth]{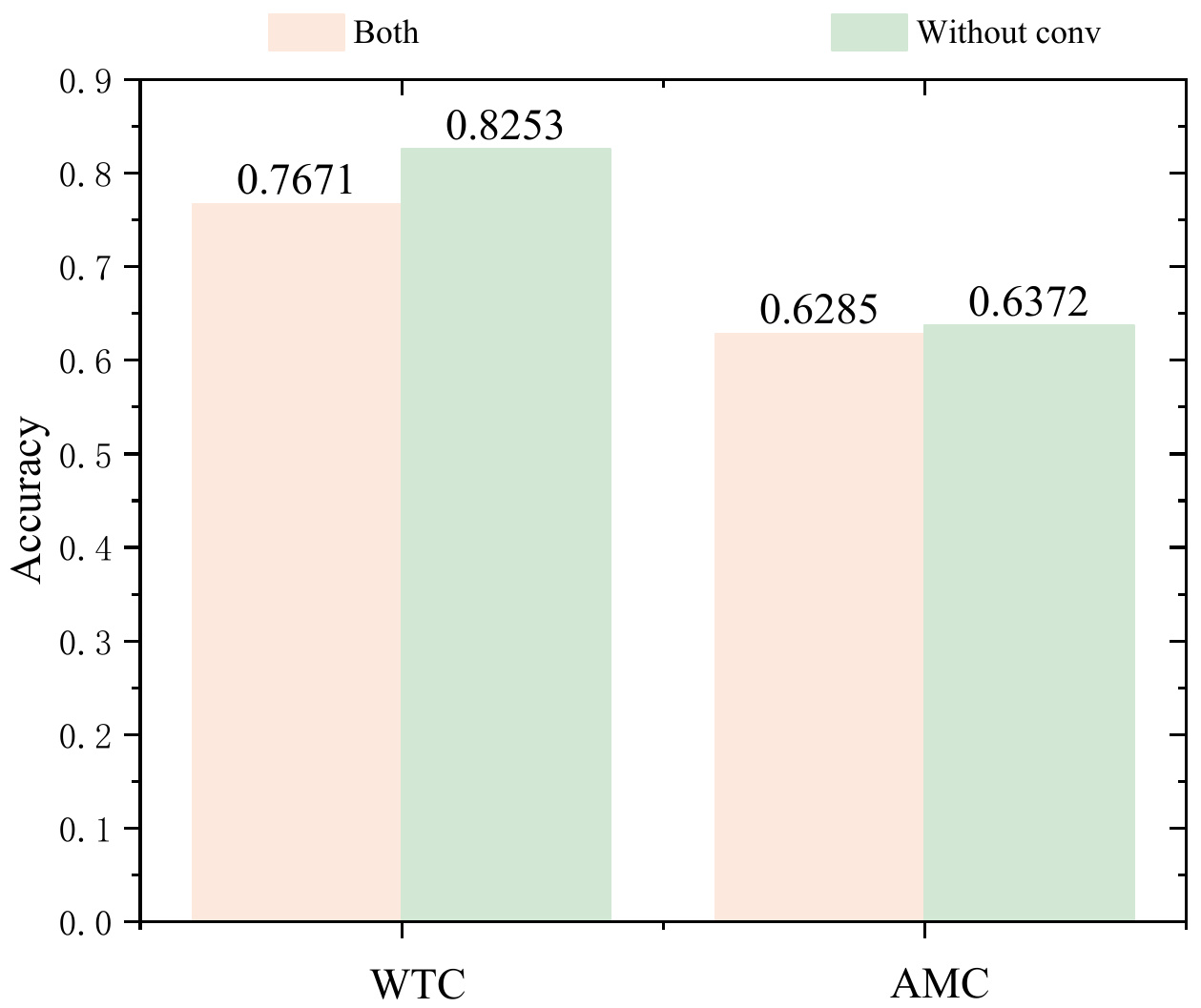}
        \caption{Ablation study on the convolution module.}
        \label{fig:ablation_conv}
    \end{minipage} 
\end{figure*}
To better understand the contribution of each pretraining objective, we conduct an ablation study by separately evaluating the masked reconstruction loss and the next-slot signal prediction loss, in comparison with their joint optimization. The results are shown in Fig.~\ref{fig:ablation}.

When trained with only masked reconstruction loss, the model achieves a reasonable performance, reaching 0.6355 on AMC and 0.7961 on WTC, demonstrating the ability to capture local structural patterns and to enhance robustness against corrupted inputs. Training with only next-slot signal prediction loss yields 0.6274 on AMC and 0.8076 on WTC, showing that this objective effectively models sequential dependencies but does not consistently outperform masked reconstruction across different tasks. Most importantly, the combination of both objectives consistently provides the best results, indicating a clear complementary effect. Specifically, the masked reconstruction loss encourages the model to learn robust feature representations, while the next-slot signal prediction loss enforces the temporal consistency, and their synergy leads to stronger generalization and overall performance gains.

To further evaluate the architectural design, we also conduct an ablation experiment on the convolution module. As shown in Fig.~\ref{fig:ablation_conv}, when this component is removed, the performance on downstream tasks drops notably, with AMC decreasing from 0.6372 to 0.6285 and WTC decreasing from 0.8253 to 0.7671. These results highlight the crucial role of convolutional layers in capturing fine-grained local spectral features, which complement the global dependencies modeled by MHSA and jointly contribute to the superior performance of SpectrumFM.

\subsection{Discussion}
In summary, the superior performance of our model across AMC, WTC, SS, and AD tasks can be attributed to its capacity to effectively learn and generalize spectral features through large-scale pre-training. By exploiting MHSA and convolutional modules, the model captures fine-grained signal characteristics and learns universal spectral patterns, enabling it to perform effectively in diverse, unseen environments. Large-scale pre-training facilitates the development of transferable representations, which enhances the model generalization ability and makes it highly effective even under low SNR or noisy conditions.
Additionally, the leveraging of masked reconstruction tasks during pre-training bolsters the model robustness to signal distortions and interference, fostering enhanced resilience in challenging real-world conditions. The next-slot signal prediction task, in particular, improves the model capacity to anticipate future signal behaviors based on both current and past information, thereby enabling it to navigate dynamic spectrum environments with greater accuracy. These tasks, collectively, significantly augment the model predictive power and robustness, making it particularly well-suited for complex and unpredictable spectrum scenarios.

These advancements offer substantial benefits for wireless communication systems. The improved modulation recognition ensures efficient signal classification, which is critical for adaptive transmission strategies. Technology classification, in turn, enhances interoperability in complex spectrum environments. Furthermore, spectrum sensing facilitates dynamic spectrum access, while anomaly detection strengthens security by mitigating interference and detecting unauthorized transmissions. 
\subsection{Limitations}

Despite its strong performance, SpectrumFM has several limitations that need further investigation. First, the current pre-training dataset and model size, while sufficient for the demonstrated tasks, remain relatively limited, which constrains the model generalization ability and limits its performance in zero-shot scenarios. Second, although SpectrumFM has been evaluated on key downstream tasks such as AMC, WTC, SS, and AD, the range of supported tasks remains limited, and expanding the diversity of downstream applications could further validate its versatility. Third, the current model primarily operates on single-modality spectrum data. Extending SpectrumFM to support multi-modal inputs would enhance its applicability in more complex and heterogeneous wireless environments.

\section{Conclusion}
\label{sec:conclusion}
In this paper, we introduced SpectrumFM, a foundation model designed for spectrum management, representing a novel paradigm in this field. To support its training, a diverse spectrum dataset was collected and processed from multiple sources, including publicly available open-source datasets and self-collected real-world data.
To pre-train SpectrumFM, an encoder architecture exploiting CNNs with MHSA mechanisms was designed to enhance both high-level and fine-grained feature extraction as well as representation learning. Additionally, two novel self-supervised tasks, namely masked reconstruction and next-slot signal prediction, were proposed to enable the model to learn generalizable and foundational spectrum representations. Furthermore, a parameter-efficient fine-tuning strategy was proposed to enable SpectrumFM to effectively adapt to diverse downstream tasks. Extensive experiments conducted across four key spectrum management tasks, including AMC, WTC, SS, and AD, highlight its remarkable efficiency and performance gains over state-of-the-art methods. Specifically, SpectrumFM achieves up to 12.1\% accuracy improvement in AMC, 9.3\% in WTC, an AUC of 0.97 in SS at -4 dB SNR, and an AUC gain exceeding 10\% in AD, underscoring its superior capability in spectrum management. For future research, we will focus on expanding the data scale and model size to improve performance and generalization. Additionally, we believe SpectrumFM can also facilitate dynamic spectrum access, spectrum resource allocation, and secure spectrum sharing.

\bibliographystyle{IEEEtran} 
\bibliography{ref}

@ARTICLE{meaning,
  author={Sun, Jiachen and Chen, Jin and Ding, Guoru and Lin, Fandi and Song, Yehui},
  journal={IEEE Trans. Cognit. Commun. Netw.}, 
  title={Spectrum Recommendation in Cognitive \uppercase{I}nternet of \uppercase{T}hings: A Knowledge-Graph-Based Framework}, 
  year={2024},
  volume={10},
  number={1},
  pages={21-34},
}

@inproceedings{o2016convolutional,
  title={Convolutional Radio Modulation Recognition Networks},
  author={O’Shea, Timothy J and Corgan, Johnathan and Clancy, T Charles},
  booktitle={Int. Conf. Eng. Appl. Neural Netw.},
  pages={213--226},
  year={2016},
}

@ARTICLE{9079888,
  author={Huang, Sai and Dai, Rui and Huang, Juanjuan and Yao, Yuanyuan and Gao, Yue and Ning, Fan and Feng, Zhiyong},
  journal={IEEE Internet Things J.}, 
  title={Automatic Modulation Classification Using Gated Recurrent Residual Network}, 
  year={2020},
  volume={7},
  number={8},
  pages={7795-7807},
}

@ARTICLE{9785878,
  author={Liang, Zhi and Tao, Mingliang and Xie, Jian and Yang, Xin and Wang, Ling},
  journal={IEEE Trans. Cognit. Commun. Netw.}, 
  title={A Radio Signal Recognition Approach Based on Complex-Valued \uppercase{CNN} and Self-Attention Mechanism}, 
  year={2022},
  volume={8},
  number={3},
  pages={1358-1373},
  }

@ARTICLE{10496203,
  author={Zhang, Hao and Zhou, Fuhui and Wu, Qihui and Al-Dhahir, Naofal},
  journal={IEEE Trans. Commun.}, 
  title={SSwsrNet: A Semi-Supervised Few-Shot Learning Framework for Wireless Signal Recognition}, 
  year={2024},
  volume={72},
  number={9},
  pages={5823-5836},
}

@INPROCEEDINGS{10097070,
  author={Zhang, Jiawei and Wang, Tiantian and Feng, Zhixi and Yang, Shuyuan},
  booktitle={IEEE Int. Conf. Acoust. Speech Signal Process.}, 
  title={\uppercase{AMC-N}et: An Effective Network for Automatic Modulation Classification}, 
  year={2023},
  volume={},
  number={},
  pages={1-5},
  }

@ARTICLE{10857965,
  author={Du, Mingyang and Pan, Jifei and Bi, Daping},
  journal={IEEE Trans. Wireless Commun.}, 
  title={ A Contrastive Learner for Automatic Modulation Classification}, 
  year={2025},
  note={to be published}

}

@article{FONTAINE2019101881,
title = {Towards low-complexity wireless technology classification across multiple environments},
journal = {Ad Hoc Netw.},
volume = {91},
pages = {101881},
year = {2019},
issn = {1570-8705},
author = {Jaron Fontaine and Erika Fonseca and Adnan Shahid and Maicon Kist and Luiz {A. DaSilva} and Ingrid Moerman and Eli {De Poorter}},
}

@ARTICLE{9637487,
  author={Yuan, Lu and Zhang, Hao and Xu, Ming and Zhou, Fuhui and Wu, Qihui},
  journal={IEEE Internet Things J.}, 
  title={A Multiscale \uppercase{CNN} Framework for Wireless Technique Classification in Internet of Things}, 
  year={2022},
  volume={9},
  number={12},
  pages={10366-10367},
  }

@article{10.1016/j.vehcom.2022.100563,
author = {Girmay, Merkebu and Maglogiannis, Vasilis and Naudts, Dries and Aslam, Muhammad and Shahid, Adnan and Moerman, Ingrid},
title = {Technology recognition and traffic characterization for wireless technologies in \uppercase{ITS} band},
year = {2023},
volume = {39},
journal = {Veh. Commun.},
pages = {1-15},
}

@ARTICLE{10555427,
  author={Peng, Yang and Zhang, Yibin and Huang, Hao and Wang, Yu and Liu, Pengfei and Lin, Yun and Gui, Guan},
  journal={IEEE Internet Things J.}, 
  title={Low-Complexity Wireless Technique Classification With Multifeature Fusion Broad Learning Network}, 
  year={2024},
  volume={11},
  number={21},
  pages={34434-34442},
}

@article{muzaffar2024review,
  title={A review of spectrum sensing in modern cognitive radio networks},
  author={Muzaffar, Muhammad Umair and Sharqi, Rula},
  journal={Telecommu. Syst.},
  volume={85},
  number={2},
  pages={347--363},
  year={2024},
}

@ARTICLE{5723805,
  author={Pawelczak, Przemyslaw and Nolan, Keith and Doyle, Linda and Oh, Ser Wah and Cabric, Danijela},
  journal={IEEE Commun. Mag.}, 
  title={Cognitive radio: Ten years of experimentation and development}, 
  year={2011},
  volume={49},
  number={3},
  pages={90-100},
}

@ARTICLE{8824091,
  author={Gao, Jiabao and Yi, Xuemei and Zhong, Caijun and Chen, Xiaoming and Zhang, Zhaoyang},
  journal={IEEE Wireless Commun. Lett.}, 
  title={Deep Learning for Spectrum Sensing}, 
  year={2019},
  volume={8},
  number={6},
  pages={1727-1730},
  }

@ARTICLE{10509639,
  author={Zhang, Weishan and Wang, Yue and Chen, Xiang and Cai, Zhipeng and Tian, Zhi},
  journal={IEEE Trans. Wireless Commun.}, 
  title={Spectrum Transformer: An Attention-Based Wideband Spectrum Detector}, 
  year={2024},
  volume={23},
  number={9},
  pages={12343-12353},
  }

@ARTICLE{10758375,
  author={Hao, Xiaoyang and Yang, Shuyuan and Liu, Ruoyu and Feng, Zhixi and Peng, Tongqing and Huang, Bincheng},
  journal={IEEE Trans. Wireless Commun.}, 
  title={\uppercase{VSLM}: Virtual Signal Large Model for Few-Shot Wideband Signal Detection and Recognition}, 
  year={2025},
  volume={24},
  number={2},
  pages={909-925},
}

@ARTICLE{9863875,
  author={Kang, Ying and Wu, Hao and Zhao, Zhihua and Li, Yaxing and Meng, Jin},
  journal={IEEE Trans. Cognit. Commun. Netw.}, 
  title={DL-Based Anomaly Detection at the Physical-Layer of Cognitive Radio by Deep Support Vector Data Description}, 
  year={2022},
  volume={8},
  number={4},
  pages={1689-1705},
}

@ARTICLE{10035489,
  author={Zhang, Han and Song, Zihang and Yang, Jian and Gao, Yue},
  journal={IEEE Trans. Cognit. Commun. Netw.}, 
  title={Adversarial Autoencoder Empowered Joint Anomaly Detection and Signal Reconstruction From Sub-\uppercase{N}yquist Samples}, 
  year={2023},
  volume={9},
  number={3},
  pages={618-628},
  }

@INPROCEEDINGS{10765508,
  author={Fang, Mengqing and Ding, Rui and Zhou, Fuhui and Wu, Qihui},
  booktitle={Int. Conf. Commun. Image Signal Process.}, 
  title={A Simultaneous Spectrum Sensing and Anomaly Detection Deep Learning Framework for Dynamic Spectrum Sharing Networks}, 
  year={2024},
  volume={},
  number={},
  pages={107-111},
  }

@ARTICLE{10834497,
  author={Awais, Muhammad and Naseer, Muzammal and Khan, Salman and Anwer, Rao Muhammad and Cholakkal, Hisham and Shah, Mubarak and Yang, Ming-Hsuan and Khan, Fahad Shahbaz},
  journal={IEEE Trans. Pattern Anal. Mach. Intell.}, 
  title={Foundation Models Defining a New Era in Vision: a Survey and Outlook}, 
  year={2025},
  note={to be published}}

@inproceedings{devlin-etal-2019-bert,
    title = "\uppercase{BERT}: Pre-training of Deep Bidirectional Transformers for Language Understanding",
    author = "Devlin, Jacob  and
      Chang, Ming-Wei  and
      Lee, Kenton  and
      Toutanova, Kristina",
    booktitle = "Conf. North Am. Chapter Assoc. Comput. Linguistics", 
    year = "2019",
    pages = "4171--4186",
}

@inproceedings{10.5555/3600270.3602542,
author = {Chen, Ting and Saxena, Saurabh and Li, Lala and Lin, Tsung-Yi and Fleet, David J. and Hinton, Geoffrey},
title = {A unified sequence interface for vision tasks},
year = {2022},
pages = {31333-31346},
booktitle = {Int. Conf. Neural Inf. Process. Syst.},
}

@article{10.1109/TASLP.2021.3122291,
author = {Hsu, Wei-Ning and Bolte, Benjamin and Tsai, Yao-Hung Hubert and Lakhotia, Kushal and Salakhutdinov, Ruslan and Mohamed, Abdelrahman},
title = {HuBERT: Self-Supervised Speech Representation Learning by Masked Prediction of Hidden Units},
year = {2021},
volume = {29},
journal = {IEEE/ACM Trans. Audio, Speech Lang. Process.},
pages = {3451-3460},
}

@ARTICLE{10490262,
  author={Hong, Danfeng and Zhang, Bing and Li, Xuyang and Li, Yuxuan and Li, Chenyu and Yao, Jing and Yokoya, Naoto and Li, Hao and Ghamisi, Pedram and Jia, Xiuping and Plaza, Antonio and Gamba, Paolo and Benediktsson, Jon Atli and Chanussot, Jocelyn},
  journal={IEEE Trans. Pattern Anal. Mach. Intell.}, 
  title={Spectral\uppercase{GPT}: Spectral Remote Sensing Foundation Model}, 
  year={2024},
  volume={46},
  number={8},
  pages={5227-5244},
}

@misc{yang2025wirelessgptgenerativepretrainedmultitask,
      title={Wireless\uppercase{GPT}: A Generative Pre-trained Multi-task Learning Framework for Wireless Communication}, 
      author={Tingting Yang and Ping Zhang and Mengfan Zheng and Yuxuan Shi and Liwen Jing and Jianbo Huang and Nan Li},
      year={2025},
      eprint={2502.06877},
      archivePrefix={arXiv},
      primaryClass={cs.LG},
      url={https://arxiv.org/abs/2502.06877}, 
}

@ARTICLE{8267032,
  author={O’Shea, Timothy James and Roy, Tamoghna and Clancy, T. Charles},
  journal={IEEE J. Sel. Topics Signal Process.}, 
  title={Over-the-Air Deep Learning Based Radio Signal Classification}, 
  year={2018},
  volume={12},
  number={1},
  pages={168-179},
  }

@ARTICLE{8963964,
  author={Huynh-The, Thien and Hua, Cam-Hao and Pham, Quoc-Viet and Kim, Dong-Seong},
  journal={IEEE Commun. Lett.}, 
  title={M\uppercase{CN}et: An Efficient \uppercase{CNN} Architecture for Robust Automatic Modulation Classification}, 
  year={2020},
  volume={24},
  number={4},
  pages={811-815},
  doi={10.1109/LCOMM.2020.2968030}}

@INPROCEEDINGS{resnet,
  author={Liu, Xiaoyu and Yang, Diyu and Gamal, Aly El},
  booktitle={Asilomar Conf. Signals, Syst., Comput.}, 
  title={Deep neural network architectures for modulation classification}, 
  year={2017},
  volume={},
  number={},
  pages={915-919},
  doi={10.1109/ACSSC.2017.8335483}}

@ARTICLE{dae,
  author={Ke, Ziqi and Vikalo, Haris},
  journal={IEEE Trans. Wireless Commun.}, 
  title={Real-Time Radio Technology and Modulation Classification via an \uppercase{LSTM} Auto-Encoder}, 
  year={2022},
  volume={21},
  number={1},
  pages={370-382},
  doi={10.1109/TWC.2021.3095855}}

@ARTICLE{vgg,
  author={O’Shea, Timothy James and Roy, Tamoghna and Clancy, T. Charles},
  journal={IEEE J. Sel. Topics Signal Process.}, 
  title={Over-the-Air Deep Learning Based Radio Signal Classification}, 
  year={2018},
  volume={12},
  number={1},
  pages={168-179},
  doi={10.1109/JSTSP.2018.2797022}}

@INPROCEEDINGS{gru2,
  author={Hong, Dehua and Zhang, Zilong and Xu, Xiaodong},
  booktitle={IEEE Int. Conf. Comput. Commun.}, 
  title={Automatic modulation classification using recurrent neural networks}, 
  year={2017},
  volume={},
  number={},
  pages={695-700},
  doi={10.1109/CompComm.2017.8322633}}

@ARTICLE{cgdnn,
  author={Njoku, Judith Nkechinyere and Morocho-Cayamcela, Manuel Eugenio and Lim, Wansu},
  journal={IEEE Netw. Lett.}, 
    title={\uppercase{CGDNet}: Efficient Hybrid Deep Learning Model for Robust Automatic Modulation Recognition}, 
  year={2021},
  volume={3},
  number={2},
  pages={47-51},
  doi={10.1109/LNET.2021.3057637}}

@article{vaswani2017attention,
  title={Attention is all you need},
  author={Vaswani, Ashish and Shazeer, Noam and Parmar, Niki and Uszkoreit, Jakob and Jones, Llion and Gomez, Aidan N and Kaiser, {\L}ukasz and Polosukhin, Illia},
  journal={Int. Conf. Neural Inf. Process. Syst.},
  volume={30},
  year={2017}
}

@ARTICLE{msnet,
  author={Yuan, Lu and Zhang, Hao and Xu, Ming and Zhou, Fuhui and Wu, Qihui},
  journal={IEEE Internet Things J.}, 
  title={A Multiscale CNN Framework for Wireless Technique Classification in Internet of Things}, 
  year={2022},
  volume={9},
  number={12},
  pages={10366-10367},
  doi={10.1109/JIOT.2021.3132652}}

@ARTICLE{aae,
  author={Zhang, Han and Song, Zihang and Yang, Jian and Gao, Yue},
  journal={IEEE Trans. Cognit. Commun. Netw.}, 
  title={Adversarial Autoencoder Empowered Joint Anomaly Detection and Signal Reconstruction From Sub-\uppercase{N}yquist Samples}, 
  year={2023},
  volume={9},
  number={3},
  pages={618-628},
  doi={10.1109/TCCN.2023.3241186}}

@ARTICLE{10488747,
  author={Gao, Ang and Wang, Qinyu and Wang, Yongze and Du, Chengyuan and Hu, Yansu and Liang, Wei and Ng, Soon Xin},
  journal={IEEE Trans. Veh. Technol.}, 
  title={Attention Enhanced Multi-Agent Reinforcement Learning for Cooperative Spectrum Sensing in Cognitive Radio Networks}, 
  year={2024},
  volume={73},
  number={7},
  pages={10464-10477},
  doi={10.1109/TVT.2024.3384393}}

@ARTICLE{10685064,
  author={Li, Junjie and Yang, Liang and Wu, Qingqing and Lei, Xianfu and Zhou, Fuhui and Shu, Feng and Mu, Xidong and Liu, Yuanwei and Fan, Pingzhi},
  journal={IEEE J. Sel. Areas Commun.}, 
  title={Active \uppercase{RIS}-Aided NOMA-Enabled Space-Air-Ground Integrated Networks With Cognitive Radio}, 
  year={2025},
  volume={43},
  number={1},
  pages={314-333},
  doi={10.1109/JSAC.2024.3460067}}

@ARTICLE{10667001,
  author={Xing, Huijun and Zhang, Xuhui and Chang, Shuo and Ren, Jinke and Zhang, Zixun and Xu, Jie and Cui, Shuguang},
  journal={IEEE Trans. Wireless Commun.}, 
  title={Joint Signal Detection and Automatic Modulation Classification via Deep Learning}, 
  year={2024},
  volume={23},
  number={11},
  pages={17129-17142},
  doi={10.1109/TWC.2024.3450972}}

@ARTICLE{10598350,
  author={Vlădeanu, Călin and Al-Dulaimi, Omer Mohammed Khodayer and Marţian, Alexandru and Popescu, Dimitrie C.},
  journal={IEEE Trans. Veh. Technol.}, 
  title={Average Energy Detection With Adaptive Threshold for Spectrum Sensing in Cognitive Radio Systems}, 
  year={2024},
  volume={73},
  number={11},
  pages={17222-17230},
  doi={10.1109/TVT.2024.3427664}}

@ARTICLE{10589480,
  author={Qi, Peihan and Jiang, Tao and Xu, Jiabo and He, Jinyang and Zheng, Shilian and Li, Zan},
  journal={IEEE Internet Things J.}, 
  title={Unsupervised Spectrum Anomaly Detection With Distillation and Memory Enhanced Autoencoders}, 
  year={2024},
  volume={11},
  number={24},
  pages={39361-39374},
  doi={10.1109/JIOT.2024.3424837}}

@ARTICLE{9463441,
  author={Zhang, Hao and Zhou, Fuhui and Wu, Qihui and Wu, Wei and Hu, Rose Qingyang},
  journal={IEEE Trans. Cognit. Commun. Netw.}, 
  title={A Novel Automatic Modulation Classification Scheme Based on Multi-Scale Networks}, 
  year={2022},
  volume={8},
  number={1},
  pages={97-110},
  doi={10.1109/TCCN.2021.3091730}}

@ARTICLE{10173618,
  author={Liu, Chunyu and Wu, Wei and Wu, Siyu and Yuan, Lu and Ding, Rui and Zhou, Fuhui and Wu, Qihui},
  journal={IEEE Trans. Knowl. Data Eng.}, 
  title={Social-Enhanced Explainable Recommendation With Knowledge Graph}, 
  year={2024},
  volume={36},
  number={2},
  pages={840-853},
  doi={10.1109/TKDE.2023.3292504}}

@ARTICLE{10680609,
  author={Chamberlain, Jonathan and Starobinski, David and Johnson, Joel T.},
  journal={IEEE J. Sel. Areas Commun.}, 
  title={Facilitating Spectrum Sharing With Passive Satellite Incumbents}, 
  year={2024},
  volume={42},
  number={12},
  pages={3719-3733},
  doi={10.1109/JSAC.2024.3459034}}

@ARTICLE{9479864,
  author={Qu, Yuben and Dai, Haipeng and Wang, Haichao and Dong, Chao and Wu, Fan and Guo, Song and Wu, Qihui},
  journal={IEEE J. Sel. Areas Commun.}, 
  title={Service Provisioning for \uppercase{UAV}-Enabled Mobile Edge Computing}, 
  year={2021},
  volume={39},
  number={11},
  pages={3287-3305},
  doi={10.1109/JSAC.2021.3088660}}

@ARTICLE{10845212,
  author={Xu, Yongqing and Li, Yong and Quek, Tony Q. S.},
  journal={IEEE J. Sel. Areas Commun.}, 
  title={\uppercase{RIS}-Enhanced Cognitive Integrated Sensing and Communication: Joint Beamforming and Spectrum Sensing}, 
  year={2025},
  volume={43},
  number={3},
  pages={795-810},
  doi={10.1109/JSAC.2025.3531531}}

@ARTICLE{10602484,
  author={Ji, Haipeng and Zhang, Tao and Qiao, Xiaoqiang and Wu, Hao and Gui, Guan},
  journal={IEEE Commun. Lett.}, 
  title={\uppercase{TFAM-AAE-Uk}: A Dual-Metric Spectrum Anomaly Detection Algorithm}, 
  year={2024},
  volume={28},
  number={11},
  pages={2638-2642},}

@ARTICLE{10841938,
  author={Song, Shezheng and Li, Xiaopeng and Li, Shasha and Zhao, Shan and Yu, Jie and Ma, Jun and Mao, Xiaoguang and Zhang, Weimin and Wang, Meng},
  journal={IEEE Trans. Knowl. Data Eng.}, 
  title={How to Bridge the Gap between Modalities: Survey on Multimodal Large Language Model}, 
  year={2025},
  note={to be published}
  }

@ARTICLE{10876763,
  author={Huang, Zhongzhan and Zhong, Shanshan and Zhou, Pan and Gao, Shanghua and Zitnik, Marinka and Lin, Liang},
  journal={IEEE Trans. Pattern Anal. Mach. Intell.}, 
  title={A Causality-Aware Paradigm for Evaluating Creativity of Multimodal Large Language Models}, 
  year={2025},
  volume={47},
  number={5},
  pages={3830-3846},
  doi={10.1109/TPAMI.2025.3539433}}

@article{gulati2020conformer,
  title={Conformer: Convolution-augmented transformer for speech recognition},
  author={Gulati, Anmol and Qin, James and Chiu, Chung-Cheng and Parmar, Niki and Zhang, Yu and Yu, Jiahui and Han, Wei and Wang, Shibo and Zhang, Zhengdong and Wu, Yonghui and others},
  journal={arXiv preprint arXiv:2005.08100},
  year={2020}
}

@article{xing2024joint,
  title={Joint signal detection and automatic modulation classification via deep learning},
  author={Xing, Huijun and Zhang, Xuhui and Chang, Shuo and Ren, Jinke and Zhang, Zixun and Xu, Jie and Cui, Shuguang},
  journal={IEEE Trans. Wireless Commun.},
  volume={23},
  number={11},
  pages={17129-17142},
  year={2024},
  publisher={IEEE}
}










\begin{IEEEbiography}
[{\includegraphics[width=1in,height=1.25in,clip,keepaspectratio]{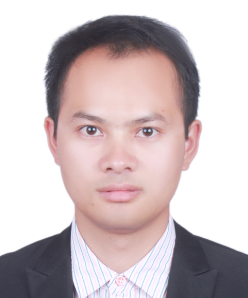}}]
{Fuhui Zhou} (Senior Member, IEEE) is currently a
Full Professor with Nanjing University of Aeronautics and Astronautics, Nanjing, China, where he is
also with the Key Laboratory of Dynamic Cognitive
System of Electromagnetic Spectrum Space. His
research interests include cognitive radio, cognitive
intelligence, knowledge graph, edge computing, and
resource allocation.
Prof. Zhou has published over 200 papers in
internationally renowned journals and conferences in
the field of communications. He has been selected
for 1 ESI hot paper and 13 ESI highly cited papers. He has received 4
Best Paper Awards at international conferences such as IEEE Globecom and
IEEE ICC. He was awarded as 2021 Most Cited Chinese Researchers by
Elsevier, Stanford World's Top 2\% Scientists, IEEE ComSoc Asia-Pacific
Outstanding Young Researcher and Young Elite Scientist Award of China and
URSI GASS Young Scientist. He serves as an Editor of  IEEE Wireless Communication
Letters, IEEE Access and Physical Communications.

\end{IEEEbiography}
\begin{IEEEbiography}
[{\includegraphics[width=1in,height=1.25in,clip,keepaspectratio]{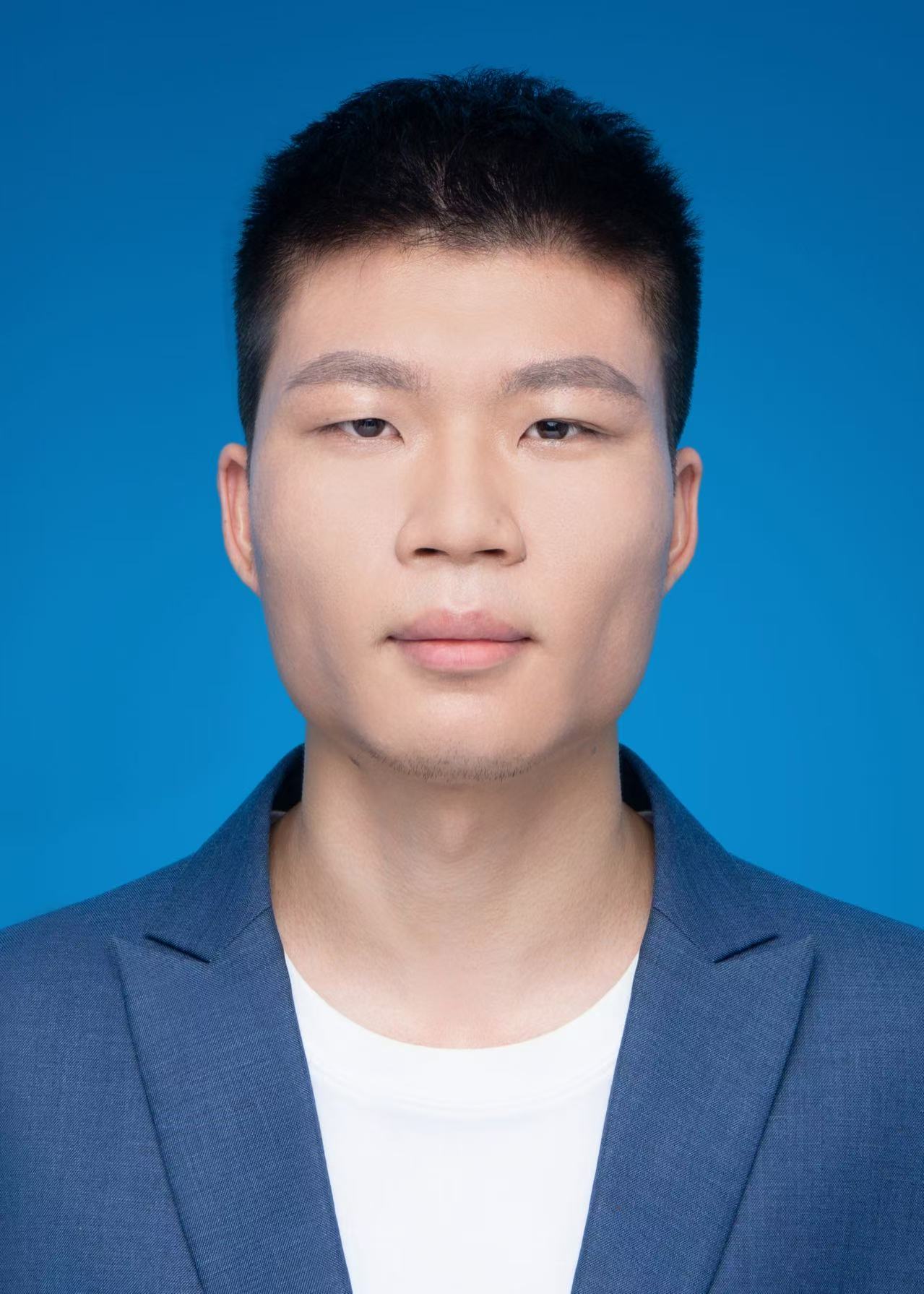}}]
{Chunyu Liu} (Graduate Student Member, IEEE) is currently pursuing the Ph.D degree in electronic and information engineering with the School of Electronic and Information Engineering, Nanjing University of Aeronautics and Astronautics. His research interests include spectrum foundation models, knowledge graph, and wireless communication.
\end{IEEEbiography}
\begin{IEEEbiography}
    [{\includegraphics[width=1in,height=1.25in,clip,keepaspectratio]{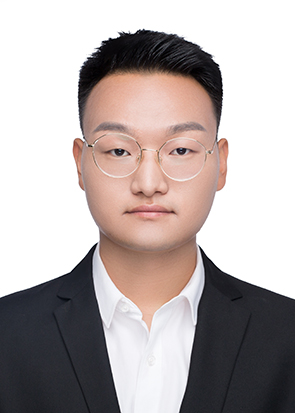}}]
    {Hao Zhang} (Member, IEEE) received the B.E.
and M.Eng. from Nanchang University, China, in
2017 and 2020, respectively, and the Ph.D. degree
from the College of Electronic and Information
Engineering, Nanjing University of Aeronautics and
Astronautics, Nanjing, China, in 2025. He is now
a postdoctoral fellow with the College of Artificial Intelligence, Nanjing University of Aeronautics and Astronautics, Nanjing, China. He was a
visiting Ph.D. student at the School of Electrical
\& Electronic Engineering, Nanyang Technological
University, Singapore, in 2024. His research interests focus on deep learning,
foundation models, wireless communication, wireless signal processing, and
spectrum cognition.
    \end{IEEEbiography}
\begin{IEEEbiography}
        [{\includegraphics[width=1in,height=1.25in,clip,keepaspectratio]{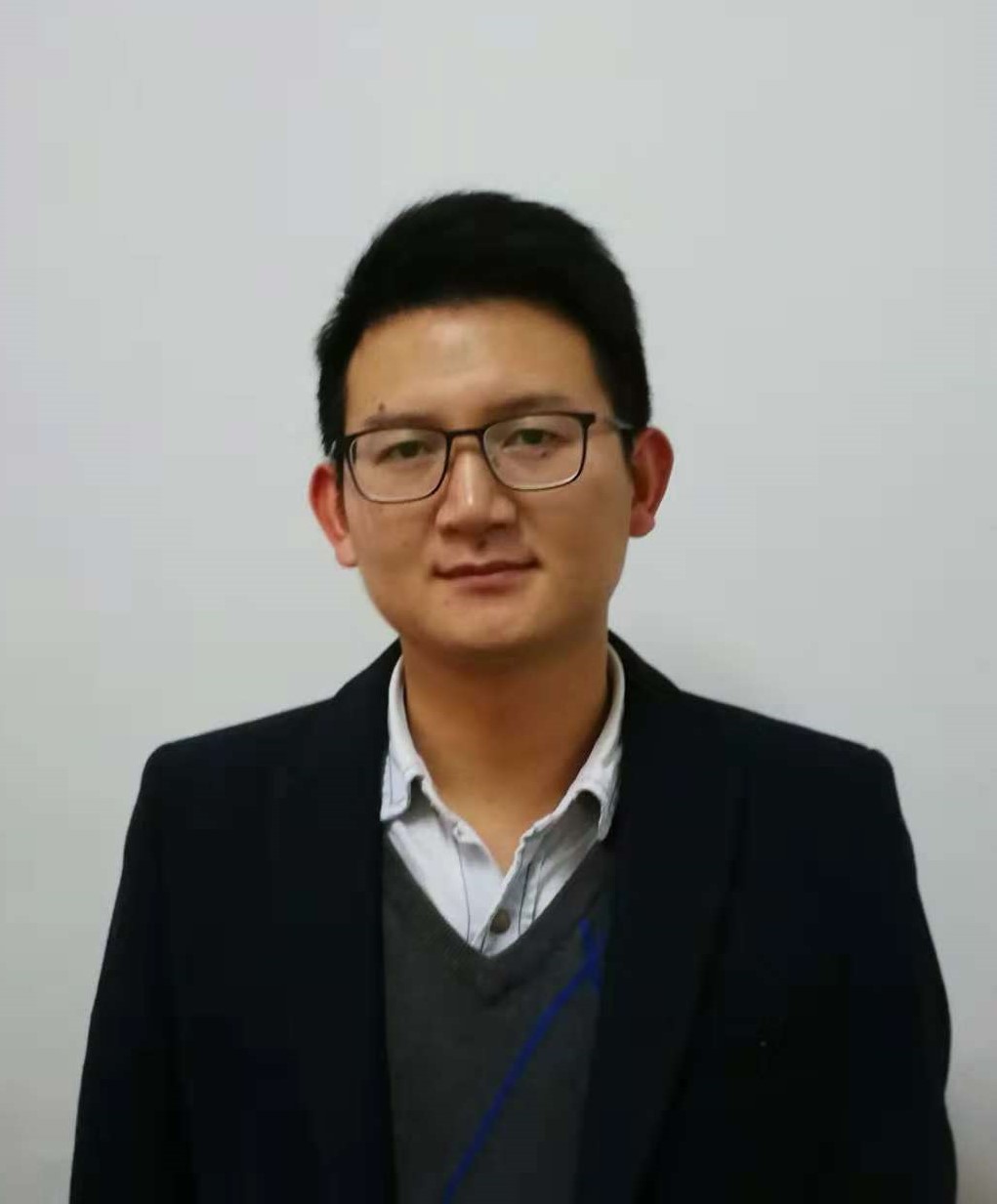}}]
        {Wei Wu} (Member, IEEE) is currently an Full professor of the Nanjing University of Posts and
    Telecommunications. His research interests include
    spectrum sharing, semantic communication, knowledge
    graph, and physical layer security. He was the
    Technical Program Committee member for many international
    conferences, such as IEEE GLOBECOM
    and IEEE ICC. He is the editor of Physical Communications.
        \end{IEEEbiography}
   
\begin{IEEEbiography}
[{\includegraphics[width=1in,height=1.25in,clip,keepaspectratio]{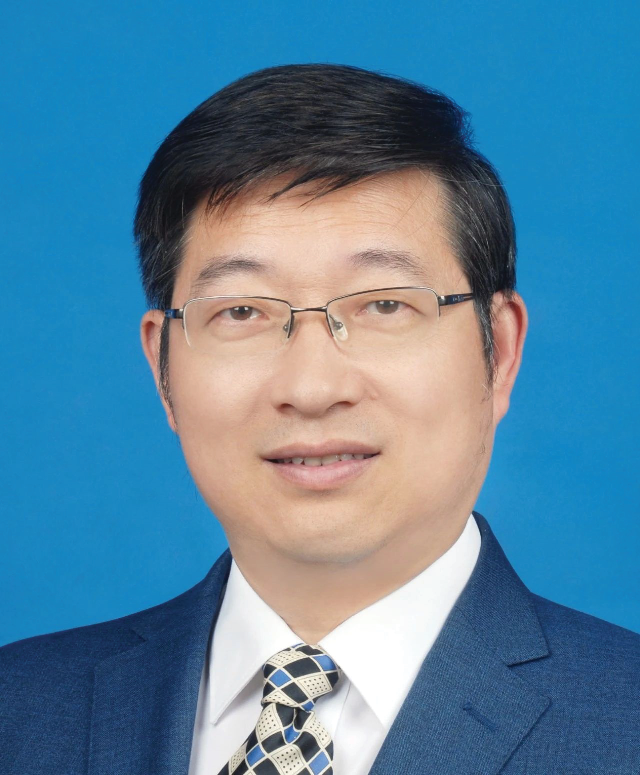}}]
{Qihui Wu} (Fellow, IEEE) received the B.S. degree in communications engineering and the M.S. and Ph.D. degrees
in communications and information systems from the Institute of Communications Engineering, Nanjing, China, in 1994, 1997, and 2000, respectively.
Since May 2016, he has been a Full Professor with the College of Electronic and
Information Engineering, Nanjing University of Aeronautics and Astronautics,
Nanjing. From March 2011 to September 2011, he was an Advanced Visiting
    Scholar with the Stevens Institute of Technology, Hoboken, NJ, USA. His
    current research interests include the areas of wireless communications and
    statistical signal processing, with emphasis on system design of softwaredefined radio, cognitive radio, and smart radio.
\end{IEEEbiography}
\begin{IEEEbiography}
[{\includegraphics[width=1in,height=1.25in,clip,keepaspectratio]{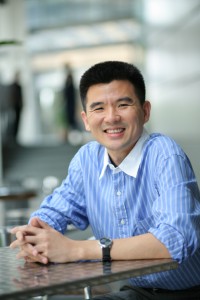}}]
{Tony Q.S. Quek} (Fellow, IEEE) received the B.E. and M.E. degrees in electrical and electronics engineering from
the Tokyo Institute of Technology in 1998 and 2000, respectively, and the Ph.D. degree in electrical engineering and computer science from the Massachusetts Institute of Technology in 2008. Currently, he is the Cheng Tsang Man Chair Professor with Singapore University of Technology and Design (SUTD) and ST Engineering Distinguished
Professor. He also serves as the Director of the Future Communications R\&D Programme, the Head
of ISTD Pillar, and the Deputy Director of the SUTD-ZJU IDEA. His current
research topics include wireless communications and networking, network
intelligence, non-terrestrial networks, open radio access network, and 6G. Dr.
Quek has been actively involved in organizing and chairing sessions, and
has served as a member of the Technical Program Committee as well as
symposium chairs in a number of international conferences. He is currently
serving as an Area Editor for the IEEE TRANSACTIONS ON WIRELESS
COMMUNICATIONS. Dr. Quek was honored with the 2008 Philip Yeo
Prize for Outstanding Achievement in Research, the 2012 IEEE William R.
Bennett Prize, the 2015 SUTD Outstanding Education Awards - Excellence
in Research, the 2020 IEEE Stephen O. Rice Prize, the 2020 Nokia Visiting
Professor, and the 2022 IEEE Signal Processing Society Best Paper Award. He
is a Fellow of IEEE and a Fellow of the Academy of Engineering Singapore.
\end{IEEEbiography}
\begin{IEEEbiography}
[{\includegraphics[width=1in,height=1.25in,clip,keepaspectratio]{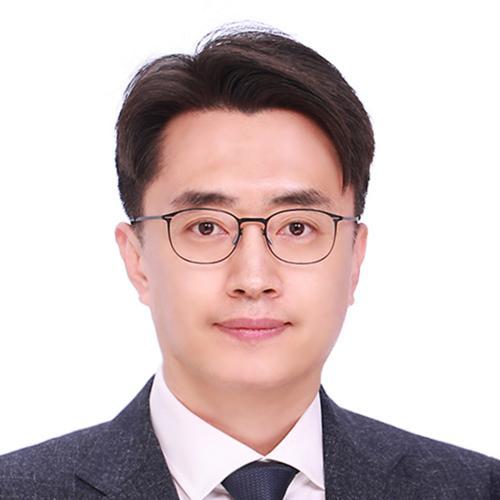}}]
{Chan-Byoung Chae} (Fellow, IEEE) received the
Ph.D. degree in electrical and computer engineering from The University of Texas at Austin (UT),
USA in 2008, where he was a member of wireless
networking and communications group (WNCG).
Prior to joining UT, he was a Research Engineer
at the Telecommunications R\&D Center, Samsung
Electronics, Suwon, South Korea, from 2001 to
2005. He is currently an Underwood Distinguished
Professor and Lee Youn Jae Fellow (Endowed Chair
Professor) with the School of Integrated Technology,
Yonsei University, South Korea. Before joining Yonsei, he was with Bell Labs,
Alcatel-Lucent, Murray Hill, NJ, USA, from 2009 to 2011, as a Member of
Technical Staff, and Harvard University, Cambridge, MA, USA, from 2008
to 2009, as a Post-Doc., Fellow and Lecturer.
Dr. Chae was a recipient/co-recipient of the Korean Ministry of ICT and
Science Award in 2024, the Korean Ministry of Education Award in 2024,
the KICS Haedong Scholar Award in 2023, the CES Innovation Award in
2023, the IEEE ICC Best Demo Award in 2022, the IEEE WCNC Best
Demo Award in 2020, the Best Young Engineer Award from the National
Academy of Engineering of Korea (NAEK) in 2019, the IEEE DySPAN
Best Demo Award in 2018, the IEEE/KICS Journal of Communications and
Networks Best Paper Award in 2018, the IEEE INFOCOM Best Demo Award
in 2015, the IEIE/IEEE Joint Award for Young IT Engineer of the Year in
2014, the KICS Haedong Young Scholar Award in 2013, the IEEE Signal
Processing Magazine Best Paper Award in 2013, the IEEE ComSoc AP
Outstanding Young Researcher Award in 2012, and the IEEE VTS Dan. E.
Noble Fellowship Award in 2008.
Dr. Chae has held several editorial positions, including Editor-in-Chief of
the IEEE TRANSACTIONS ON MOLECULAR, BIOLOGICAL, AND MULTISCALE COMMUNICATIONS, Senior Editor of the IEEE WIRELESS COMMUNICATIONS LETTERS, and Editor of the IEEE COMMUNICATIONS MAGAZINE, IEEE TRANSACTIONS ON WIRELESS COMMUNICATIONS, and IEEE
WIRELESS COMMUNICATIONS LETTERS. He was an IEEE ComSoc Distinguished Lecturer from 2020 to 2023 and is an IEEE VTS Distinguished
Lecturer from 2024 to 2025. He is an elected member of the National
Academy of Engineering of Korea.

\end{IEEEbiography}
\end{document}